\documentclass[sigconf]{acmart}
\usepackage{enumitem}
\usepackage{cleveref}
\usepackage{subcaption}
\usepackage{stfloats}

\copyrightyear{2026}
\acmYear{2026}
\setcopyright{cc}
\setcctype{by}
\acmConference[CHI '26]{Proceedings of the 2026 CHI Conference on Human Factors in Computing Systems}{April 13--17, 2026}{Barcelona, Spain}
\acmBooktitle{Proceedings of the 2026 CHI Conference on Human Factors in Computing Systems (CHI '26), April 13--17, 2026, Barcelona, Spain}
\acmDOI{10.1145/3772318.3791741}
\acmISBN{979-8-4007-2278-3/2026/04}

\newcommand{\dtag}[1]{\textsc{#1}}

\newcommand{\myKw}[1]{\hspace{0.2em}{\small\texttt{#1}}\hspace{0.2em}}

\newcommand{\mySubsub}[2][]{%
  \par
  \def\temp{#1}%
  \def\nospace{nospace}%
  \ifx\temp\nospace
    \addvspace{0pt}%
  \else
    \addvspace{0.6\baselineskip plus 2pt minus 0.5pt}%
  \fi
  \noindent{\bfseries #2}
  \par
  \nobreak
  \addvspace{0.15\baselineskip}%
  \noindent
  \ignorespaces 
}

\makeatletter
\newcommand{\myAppSection}[3][]{%
  \par
  \in@{nospacebefore}{#1}%
  \ifin@
    \addvspace{0pt}%
  \else
    \addvspace{3ex \@plus 0.5ex \@minus 0.2ex}%
  \fi
  \refstepcounter{section}%
  \noindent{\normalfont\normalsize\bfseries \thesection \ \ \ #2}%
  \label[appendix]{#3}%
  \par
  \in@{nospaceafter}{#1}%
  \ifin@
    \addvspace{0pt}%
  \else
    \addvspace{1.0ex \@plus 0.2ex}%
  \fi
  \noindent\ignorespaces
}

\newcommand{\myAppSubSection}[3][]{%
  \par
  \in@{nospacebefore}{#1}%
  \ifin@
    \addvspace{0pt}%
  \else
    \addvspace{2.5ex \@plus 0.5ex \@minus 0.2ex}%
  \fi
  \refstepcounter{subsection}%
  \noindent{\normalfont\normalsize\bfseries \thesubsection \ \ \ #2}%
  \label[appendix]{#3}%
  \par
  \in@{nospaceafter}{#1}%
  \ifin@
    \addvspace{0pt}%
  \else
    \addvspace{0.8ex \@plus 0.2ex}%
  \fi
  \noindent\ignorespaces
}
\makeatother

\newcounter{myfinding}
\newcounter{mysubfinding}[myfinding]
\Crefname{finding}{Finding}{Findings}

\renewcommand{\themyfinding}{\arabic{myfinding}}

\makeatletter
\newcommand{\myFinding}[2]{%
  \refstepcounter{myfinding}%
  \par\addvspace{0.75\baselineskip plus 1pt minus 0.1pt}
  \noindent%
  {\bfseries\Large Finding \themyfinding: #1}%
  \label[finding]{#2}%
  \par
  \nobreak
  \addvspace{0.25\baselineskip}
  \noindent
  \ignorespaces
}
\newcommand{\mySubFinding}[3][]{%
  \par
  \def\temp{#1}
  \def\nospace{nospace}
  \ifx\temp\nospace
    \addvspace{0pt}%
  \else
    \addvspace{0.5\baselineskip}%
  \fi
  \noindent%
  \refstepcounter{mysubfinding}%
  \textit{\sffamily\arabic{mysubfinding}. #2.}%
  \label[finding]{#3}%
  \enspace%
  \ignorespaces
}
\makeatother

\newcommand{\myDiscussion}[2][]{%
  \par
  \def\temp{#1}%
  \def\nospace{nospace}%
  \ifx\temp\nospace
    \addvspace{0pt}%
  \else
    \addvspace{0.9\baselineskip plus 0pt minus 1pt}%
  \fi
  \noindent{\bfseries\large #2}%
  \par
  \nobreak
  \addvspace{0.15\baselineskip}%
  \noindent
  \ignorespaces
}

\definecolor{MetaBlue}{RGB}{0, 100, 224}
\definecolor{unboundedLink}{HTML}{D67A2C} 

\newcommand{\imgwidth}{0.096\textwidth}%
\newcommand{\textwidthvar}{0.08999\textwidth}%
\newcommand{\pdfwidth}{0.093\textwidth}%
\newcommand{\imgsep}{1.5pt}%

\AtBeginDocument{%
  }

\newcounter{qquote}
\newcommand{\q}[1]{(Q\ref{q:#1})}
\newcommand{\qs}[1]{Q\ref{q:#1}}
\newcommand{\Q}[3]{%
  \refstepcounter{qquote}%
  \label{q:#2}%
  \par\noindent
  {\textbf{Q\theqquote}\ \textit{#1:}\enspace ``#3''}%
  \par
}

\begin{document}

\title{Unbounded: Object--Boundary Interaction in Mixed Reality}
\author{Zhuoyue Lyu}
\affiliation{%
 \institution{Department of Engineering\\University of Cambridge}
 \city{Cambridge}
 \country{United Kingdom}}
  \email{zl536@cam.ac.uk}

\author{Per Ola Kristensson}
\affiliation{%
 \institution{Department of Engineering\\University of Cambridge}
 \city{Cambridge}
 \country{United Kingdom}}
  \email{pok21@cam.ac.uk}

\begin{abstract}
Boundaries such as walls, windows, and doors are ubiquitous in the physical world, yet their potential in mixed reality (MR) remains underexplored. We present Unbounded, a Research through Design inquiry into object--boundary interaction (OBI). Building on prior work, we articulate a design space aimed at providing a shared language for OBI. To demonstrate its potential, we design and implement eight examples across productivity and art exploration scenarios, showcasing how OBIs can enrich and reframe everyday interactions. We further engage with six MR experts in one-on-one feedback sessions, using the design space and examples as design probes. Their reflections broaden the conceptual scope of OBI, reveal new possibilities for how the framework may be applied, and highlight implications for future MR interaction design.
\end{abstract}



\begin{teaserfigure}
\vspace{-1em}
  \includegraphics[width=\textwidth]{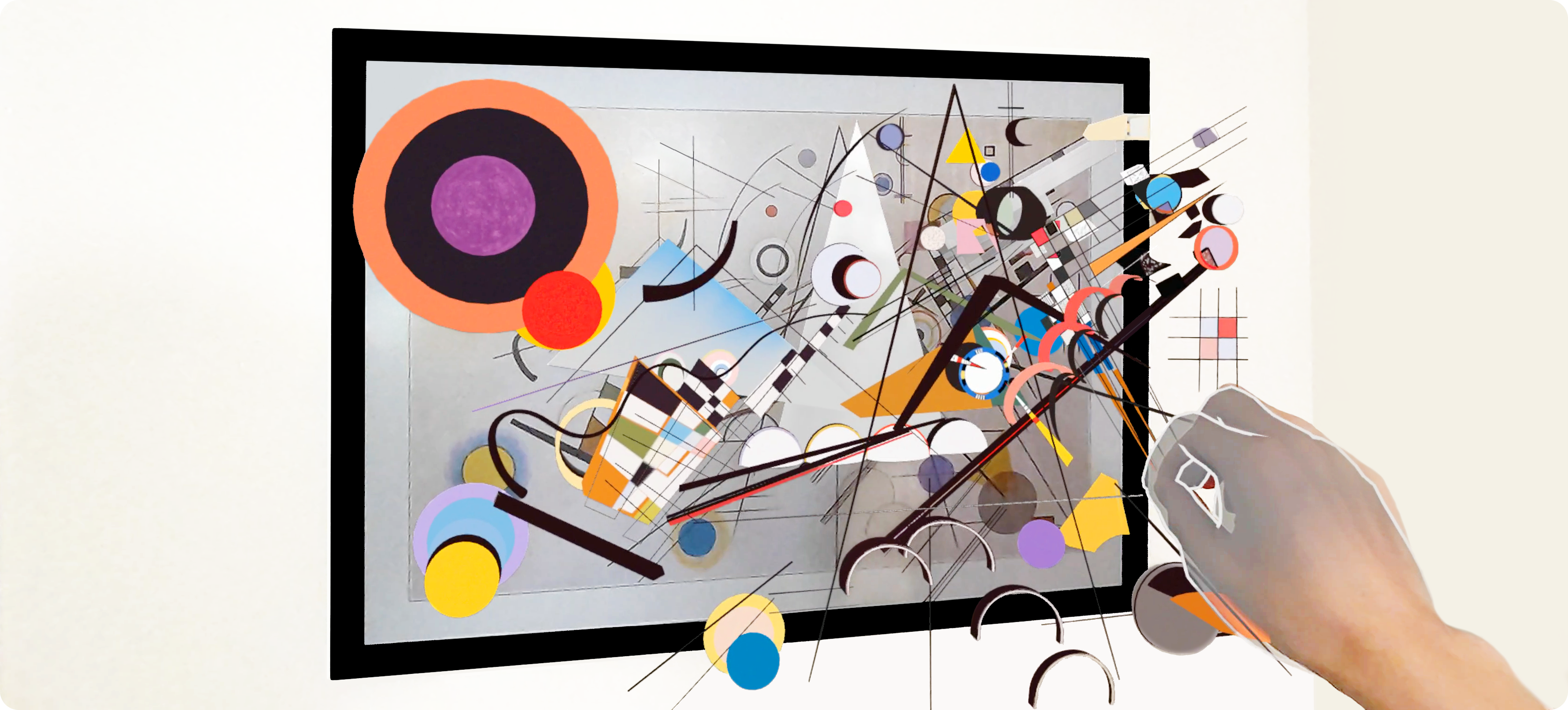}
  \vspace{-2em}
  \caption{Interactive Painting is one of our eight examples exploring the potential of object--boundary interaction. It invites the viewer to become a creative participant by reimagining Kandinsky's \textit{Composition VIII}.}
  \Description{A first-person perspective showing a hand pinching and pulling outward a framed abstract painting on a white wall. Instead of remaining flat on the canvas, numerous colorful geometric elements from the artwork---such as concentric circles, semi-circles, grids, and intersecting lines---emerge into three-dimensional space in the foreground.}
  \label{fig:teaser}
\end{teaserfigure}

\maketitle
\section{Introduction and Context} \label{sec:introduction}
Boundaries such as floors and walls are ubiquitous in the physical world and are typically understood as restrictions: floors constrain objects to their surface, while walls prevent direct passage. Prior work has challenged this by enabling objects to pass through boundaries via portals, thereby affording novel interactions~\cite{nakagaki_disappearables_2022}. We were intrigued by how such behaviors could be more fluidly realized in mixed reality (MR), where objects are no longer bound by physical laws~\cite{kari_scene_2023,lyu_objestures_2026} and can traverse boundaries without mechanical intervention. This ontological shift creates affordances difficult or impossible in the physical world.

Prior MR work has explored specific boundaries, such as pulling objects out of 2D desktop screens~\cite{rau_traversing_2025}, interacting through mirrors~\cite{zhou_reflected_2024}, or making walls transparent to access hidden objects~\cite{lilija_augmented_2019}. These highlight possibilities but remain tied to particular contexts. What is lacking is a broader perspective that treats the object--boundary relationship itself as a design material, extending beyond any single boundary or object.

We therefore present a Research through Design (RtD)~\cite{zimmerman_research_2007} inquiry into object--boundary interaction (OBI) in MR. We first articulate a design space for OBI, drawing inspiration from and reflecting on prior work (\Cref{sec:framework}). We then demonstrate its descriptive and generative potential by applying it to existing interactions (\Cref{fig:existing}) and by designing and implementing eight examples\footnote{\href{https://www.zhuoyuelyu.com/unbounded}{zhuoyuelyu.com/unbounded}} across Productivity and Art Exploration scenarios that use all its elements (\Cref{sec:demonstrations}). Finally, using the framework and examples as design probes~\cite{wallace_making_2013}, we conducted one-on-one feedback sessions with six MR design experts to elicit critique, reflection, and insights (\Cref{sec:expert-sessions}). The findings broaden the scope of OBI, reveal ways the framework may be applied and enriched, and suggest implications for future MR design (\Cref{sec:discussion}). 

While MR definitions vary~\cite{speicher_what_2019}, we build on Milgram and Kishino's continuum~\cite{paul_milgram_taxonomy_1994}, ``in which real world and virtual world objects are presented together,'' and treat MR as ``Strong AR''~\cite{speicher_what_2019}, where digital content not only appears but also interacts with both users and the physical world. Although we focus on MR, we also draw on augmented reality (AR) and virtual reality (VR) from the broader notion of extended reality (XR), as well as tangible user interfaces (TUIs) and other paradigms that explore similar boundary phenomena. Together, these inform how OBI can be reimagined through MR.

\subsection{Grounding the Transcendent}
Prior XR work has explored interactions that transcend physical constraints between humans and their environments. Some give users superhuman abilities such as flying~\cite{liu_virtual_2022}, walking at giant-like speeds~\cite{abtahi_im_2019}, or slowing down time~\cite{li_benefits_2025}. Others reimagine the environment itself: making a real chair disappear through illusions~\cite{kari_scene_2023}, changing the colors of everyday items (e.g., ties, plates)~\cite{rixen_exploring_2021}, or ad-blocking in real life (e.g., replacing street billboards with art)~\cite{katins_ad-blocked_2025}. The line between these two conceptual lenses can be blurred. For example, letting users see through a wall to insert a USB stick behind it~\cite{lilija_augmented_2019} can be interpreted either as empowering users with superhuman vision, or as reimagining the wall itself as transparent. 

We position our work in this relational in-between space, where transcendence is not an attribute of either human or environment but, as Merleau-Ponty argues, emerges through their intertwining~\cite{merleau-ponty_visible_1968}. These works also reveal a tension: transcendence often breaks physical causality, which can undermine perceptual trust. Rather than unexplained illusions (e.g., objects vanishing without cause~\cite{lee_diminishar_2025, cheng_towards_2022}), we preserve perceptual integrity (form, location, and color) that matches human expectations of perceptual causality~\cite{michotte_perception_2017} and object permanence~\cite{piaget_construction_2013}, utilizing ``pre-existing knowledge of the everyday, non-digital world'' in reality-based principles~\cite{jacob_reality-based_2008}. While prior work examined implications (e.g., whether VR flying relates to empowerment~\cite{liu_virtual_2022}), we focus on how such interactions can be envisioned and articulated.

Prior work has specifically explored boundary-related transcendence. Some utilize properties of environmental boundaries, such as the ceiling's position for throwing and catching~\cite{lin_throwio_2023}, the reflectiveness of mirrors for improvisational dance~\cite{zhou_here_2023}, or the diffuse reflectance of everyday surfaces for projection-based interactions~\cite{kim_perspective_2023,jones_illumiroom_2013,muller_baselase_2015}. Others exploit boundaries afforded by computing devices, such as transferring content between computer screens and AR~\cite{rau_traversing_2025}, sliding fingers on both sides of a tablet~\cite{wigdor_lucid_2007}, or traversing virtual walls and doors in VR~\cite{ogawa_you_2020,van_gemert_doorways_2024}. Yet, few center on OBI; most touch on boundary phenomena to demonstrate technical or conceptual novelty. We therefore provide a design space inspired by these scattered examples, emphasizing not only boundary affordances but broader relationships among humans, objects, and boundaries (\Cref{fig:final-design-space}). To avoid repetition, we describe more OBI instances in prior work throughout the design space dimensions (\Cref{sec:dimensions}).

Tangible Bits~\cite{ishii_tangible_1997} envisioned coupling digital information with graspable physical objects that transcend the mundane. Decades later, adoption remains limited, as \citet{holmquist_bits_2023} argued they are ``significantly more expensive to create; to control; to modify; to maintain; and to mass-produce and distribute than GUI-based systems'', prompting the idea of liberated pixels (LPs): pixels that exist in 3D and are directly visible to the naked eye. Although current MR still requires headsets, it offers a compelling way to prototype this vision. LPs, however, lack tactile feedback. Recent work addresses this by combining tangible and visual elements: creating illusions that preserve tactile-visual consistency, such as a chair that disappears visually but reappears upon touch~\cite{kari_scene_2023}; turning everyday surfaces into touchable interfaces~\cite{dupre_tripad_2024}; and transforming everyday objects into input devices~\cite{gong_affordance-based_2023, monteiro_teachable_2023}. These efforts share a common philosophy: interaction should not confine people within digital devices but reconnect them with the physical world~\cite{lyu_objestures_2026}. By letting hands engage with tables, walls, and objects, users experience the physical presence not as a mere backdrop for digital overlays but as a companion in interaction---a perspective we also embody.

\subsection{Anchoring the Speculative}
Design spaces can organize and anchor novel, often speculative~\cite{dunne_speculative_2013}, elements. Prior work has constructed such spaces with different emphases on how they articulate envisioned futures. Some foreground properties of the artifacts or environments they consider while leaving interactions implicit: geometrical parameters of haptic wearables~\cite{han_parametric_2023}; power sources, display technologies, and communication mechanisms in situated displays~\cite{grosse-puppendahl_exploring_2016}; and shapes, sewing methods, materials, and compositions in smart textiles~\cite{ma_sensequins_2022}.
Some focus on interaction possibilities more directly, presenting properties mainly as contextual background rather than focal elements: 22 gait gestures for on-the-move input~\cite{tsai_gait_2024}, 12 joint interactions between smartphones and AR headsets~\cite{zhu_bishare_2020}, and five interaction types spanning object-based and mid-air gestures, yielding 85 uni- and bimanual interactions~\cite{lyu_objestures_2026}.
Others entwine properties and interactions: presentations and interactions for 3D visualization view management~\cite{liu_datadancing_2023}, interactions and kinetic parameters in shape-changing interfaces~\cite{rasmussen_shape-changing_2012}, and appearance and reappearance effects in actuated TUIs across nine types of portals~\cite{nakagaki_disappearables_2022}.

We align with the last view, with slightly greater emphasis on design than on system considerations. We believe both are essential to anchor the speculative, while acknowledging that current technologies (e.g., the Meta Quest~3) impose limitations that we aim not to let constrain imagination. 
We also observe that prior frameworks often present elements only through text and symbols, which can limit expressiveness, as describing new interactions requires entirely new illustrations. In response, we introduce a visual design language (\Cref{sec:language}) that encodes elements to support the composition of new symbols and interaction illustrations. Beyond aesthetics, prior research shows that diagrammatic representations (e.g., causal pathway diagrams~\cite{zhong_ai-assisted_2024}) can facilitate creative exploration and communication. Our visual language serves a different purpose but may similarly support designers' thinking and collaboration. 

Many RtD works incorporate artistic or speculative elements, often through the creation of artifacts such as supernumerary robotic limbs that enable arm swapping~\cite{yamamura_social_2023}, a radio that supports experiences of living with music from one's past~\cite{odom_design_2018}, and a design fiction film envisioning hybrid virtual funerals~\cite{uriu_designing_2025}. They often provide deep design reflection, yet perhaps due to their conceptual nature, rarely offer files or tools for reproduction or extension. Conversely, non-RtD works that focus on system and engineering, such as making objects disappear in MR~\cite{kari_scene_2023,lindlbauer_remixed_2018} or enabling giant-like walking in VR~\cite{abtahi_im_2019}, offer tools and technical details but often lack design reflection. We therefore bring these two worlds together by offering an RtD inquiry for designers and a prototype for developers, enabling both to build upon our work.

\vspace{1em}
\noindent Together, we hope our work serves as an anchor for imagination and expression. As John Dewey wrote in \emph{Art as Experience} (1934), \emph{``[\ldots{}] meanings are actually embodied in a material which thereby becomes the medium for their expression [\ldots{}] Its imaginative quality dominates, because meanings and values that are wider and deeper than the particular here and now in which they are anchored are realized by way of expressions''}~\cite[p.~273]{dewey_art_1980}.

\section{Design Space}\label{sec:framework}

We developed the design space in two stages. This section focuses on the first stage, which includes the initial iterations (\Cref{sec:iterations}), the resulting dimensions (\Cref{sec:dimensions}), and a visual design language for communication (\Cref{sec:language}). The second stage involved expert feedback sessions, detailed later in \Cref{sec:expert-sessions}, which led to the final design space (\Cref{fig:final-design-space}).

\subsection{Initial Iterations}\label{sec:iterations}
\begin{figure*}
\centering
\includegraphics[width=\textwidth]{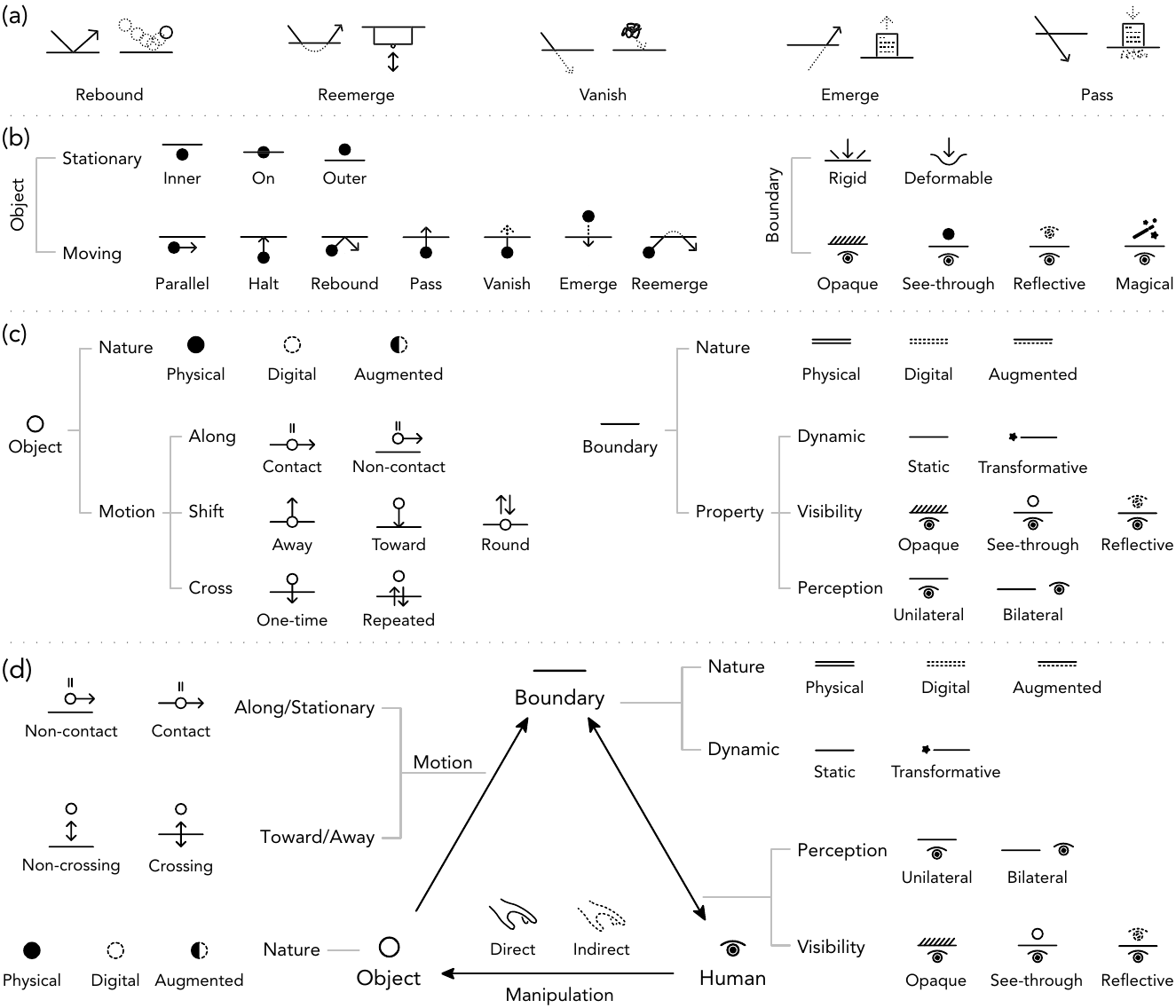}
\vspace{-2em}
\caption{Initial iterations of the design space (a--d; details in \Cref{sec:iterations}). Part (d) shows the resulting space, defined in \Cref{sec:dimensions} and its illustration explained in \Cref{sec:language}. Through expert feedback sessions (\Cref{sec:expert-sessions}), we refined (d) and present the final version in \Cref{fig:final-design-space}. The iterations shown here remain unchanged to document the RtD process.}
\Description{A four-part diagram showing the evolution of a design space. Part (a) shows basic interaction sketches like Rebound, Vanish, and Pass. Part (b) organizes these into Object states (Stationary, Moving) and Boundary types (Rigid, Deformable, Opaque, etc.). Part (c) turns this into a hierarchical tree, dividing Object into Nature (Physical, Digital, Augmented) and Motion (Along, Shift, Cross), and Boundary into Nature and Property (Dynamic, Visibility, Perception). Part (d) condenses this into a triangular relationship diagram connecting Object, Boundary, and Human, with lists of their respective properties branching off each node.}
\label{fig:design-space}
\end{figure*}

In this first stage, the design space (\Cref{fig:design-space}) was iteratively developed by the first author with feedback from the coauthor. The first author presented the iterations twice to ten HCI and design researchers during our group's weekly discussion slot (30--60 minutes), where ideas were refined through collective critique. Because feedback was informal, we did not record attendance or conduct formal analysis as in \Cref{sec:expert-sessions}, but meeting notes and audio recordings were preserved using a note-taking application (Notability), and relevant comments were quoted where appropriate.

We began by sketching object motion patterns (\Cref{fig:design-space}a): \dtag{Rebound}---bouncing off (a ball bouncing off a tabletop); \dtag{Reemerge}---crossing through and returning (a virtual drawer in a wall); \dtag{Vanish}---crossing through without returning (discarding a paper ball into a tabletop); \dtag{Emerge}---emerging from the boundary (retrieving a faxed document from a tabletop slit); or \dtag{Pass}---passing through with visibility on both sides (shredding a document).

We presented these sketches to the group through in-situ enactments (e.g., mimicking throwing a paper ball at the meeting table or pulling an imaginary drawer from the wall). Feedback highlighted its potential---especially using boundaries to hide digital information---while noting that the five categories were limited. The group suggested various boundary possibilities such as ``extended mirror realities \ldots{} where you have something living in the mirror world and something living in this world,'' and ``[shredding a document] through a window or a sheet of glass.''

Consequently, we expanded the design space (\Cref{fig:design-space}b): an object could also move parallel to a boundary (\dtag{Parallel}), move toward and rest on it (\dtag{Halt}), or remain stationary, directly on the boundary (\dtag{On}), or on either side (\dtag{Inner}, \dtag{Outer}). Boundaries themselves could be \dtag{Rigid} or \dtag{Deformable}, \dtag{Opaque}, \dtag{See-through}, \dtag{Reflective}, or \dtag{Magical} (e.g., melting~\cite{lee_icy_2024}).

Reflecting on this process, we recognized that early feedback reinforced an implicit assumption that objects are digital and boundaries are physical. This framing overlooked how materiality shapes interaction; for instance, a physical object colliding with a soft physical boundary may produce a visible deformation, whereas a digital object does not, even though their motion patterns are similar. We therefore introduced a \dtag{Nature} dimension---\dtag{Digital}, \dtag{Physical}, or \dtag{Augmented}---to capture these distinctions. For conceptual coherence and visual clarity, we merged \dtag{Stationary} into \dtag{Motion}, added \dtag{Perception}, and combined \dtag{Deformable} and \dtag{Magical} into \dtag{Transformative}. Although later expert feedback highlighted the value of distinguishing properties such as deformability (\Cref{sec:visble-forms-material}), we expanded these details in \Cref{fig:final-design-space}.

In \Cref{fig:design-space}c, \dtag{Cross} refers to object motions that cross the boundary, either \dtag{One-time} or \dtag{Repeated}. \dtag{Shift} and \dtag{Along} do not involve crossing: \dtag{Shift} changes the distance between the object and the boundary (\dtag{Away}, \dtag{Toward}, or both, as in \dtag{Round}); \dtag{Along} maintains the same distance to the boundary, either in \dtag{Contact} or \dtag{Non-contact}, and may be stationary or moving along it. Definitions of elements shared with \Cref{fig:design-space}d are provided in \Cref{sec:dimensions}.

We presented this version to our research group, whose feedback challenged our object--boundary focus. For example, \dtag{Perception} was seen as less a fixed property of the boundary, since ``people can also move around [to change perception],'' and different OBIs could shape users' sense of agency---for instance, whether a thrown ball passes through or bounces off a wall---which led us to include \dtag{Human} (\Cref{fig:design-space}d). There were also comments on interacting with multiple boundaries at once and on objects accelerating toward boundaries, but other colleagues noted that ``it is [already] comprehensive, probably a bit too comprehensive.'' As a result, we did not add further elements and instead regrouped the existing ones for clarity; for example, the distinctions between \dtag{One-time} and \dtag{Repeated} and between \dtag{Round} and \dtag{Away}/\dtag{Toward} were deemed too granular, thus seven motions were consolidated into four. Definitions of all elements are provided in \Cref{sec:dimensions}.

This led us to define OBI as the dynamic relationship in which the object engages with the boundary through human action and perception. Here, ``interaction'' refers not only to contact or motion between object and boundary but also to the intentionality through which the human initiates, observes, and interprets these relations. 


Throughout the iterations, we gathered representative, inspirational prior work to inform the dimensions. We searched the ACM Digital Library, focusing on full papers from CHI, UIST, and DIS using the keywords \myKw{object}, \myKw{boundary}, \myKw{mixed reality}, \myKw{augmented reality}, \myKw{extended reality}, and \myKw{virtual reality}. Because the notion of OBI has not been explicitly theorized, such searches often yielded weakly related results. We thus included \myKw{mirror}, \myKw{disappear}, \myKw{tangible}, \myKw{shader}, and \myKw{occlusion}, with iterative snowballing through Google Scholar---reviewing the top results, retaining those strongly related, and conducting forward/backward citation chasing. We selected 31 papers that: (1) involved object--boundary phenomena regardless of paradigm (e.g., XR or TUI), and (2) could be realized or reimagined in MR. (The limitations in excluding invisibles are discussed in \Cref{sec:invisible-main}). These papers are referenced throughout \Cref{sec:dimensions} and listed in \Cref{list-papers}.

\subsection{Dimensions}\label{sec:dimensions}
\mySubsub[nospace]{\dtag{Object}}
\dtag{Physical}: Objects that exist in the physical world, such as ping-pong balls that bounce on a table~\cite{ishii_pingpongplus_1999}, or situated displays placed on a window~\cite{grosse-puppendahl_exploring_2016}.
\dtag{Digital}: Virtual objects unconstrained by physical form, such as MR flowers that bloom on walls for breathing exercises~\cite{soler-dominguez_arcadia_2024}, or virtual human replicas for exploring body perception~\cite{dollinger_are_2023}. Note that we define \dtag{Digital} as virtual overlays (e.g., via projection or XR), distinct from hardware devices such as phones, tablets, and computers.
\dtag{Augmented}: Physical objects enhanced with digital overlays, such as everyday objects with digitally adjustable opacity~\cite{cheng_towards_2022}, or physical pots containing digitally rendered plants~\cite{lyu_objestures_2026}. 

\mySubsub{\dtag{Object} $\longrightarrow$ \dtag{Boundary}}
\dtag{Along/Stationary}: Movements that are parallel or static relative to the boundary. These can occur in \dtag{Contact} with the boundary, such as digital cursors moving across everyday surfaces for extended desktops~\cite{kim_perspective_2023}, or graspable objects sliding on tabletops~\cite{fitzmaurice_bricks_1995}; or in \dtag{Non-contact}, such as mid-air objects suspended from ceilings~\cite{yu_aerorigui_2023} or humans interacting with mirrors while seated~\cite{rajcic_mirror_2020}.

\dtag{Toward/Away}: Movements that change distance relative to the boundary. These include \dtag{Non-crossing}, where movements stay on one side of the boundary, such as objects transitioning from 2D screens into 3D~\cite{rau_traversing_2025}, throwable objects sticking to and returning from ceilings~\cite{lin_throwio_2023}, or ping-pong balls bouncing off tables~\cite{ishii_pingpongplus_1999}. \dtag{Crossing} involves movements across both sides, such as two-wheeled robots passing through actuated wall portals~\cite{nakagaki_disappearables_2022}, users walking through virtual doorways~\cite{van_gemert_doorways_2024}, or collaborators entering shared MR whiteboard spaces to work face-to-face~\cite{gronbaek_blended_2024}. 

\mySubsub{\dtag{Boundary}}
\dtag{Nature}: Similar to objects, boundaries can be \dtag{Physical}, such as walls, tables~\cite{han_corobos_2025}, or windows~\cite{benford_understanding_1998}; \dtag{Digital}, such as virtual grading gates~\cite{ogawa_you_2020} or portals~\cite{van_gemert_doorways_2024}; or \dtag{Augmented}, such as floors with laser projections for interactive play~\cite{muller_baselase_2015}, or whiteboards with MR canvases~\cite{gronbaek_blended_2024}. We define \dtag{Boundary} as visual, spatial, or material divisions rather than conceptual or social constructs; later expert feedback expands this definition (\Cref{sec:reimagining-boundaries}).

\dtag{Dynamic}: Boundaries can be \dtag{Static}, meaning stable and fixed in form or position, such as walls~\cite{lilija_augmented_2019} or floors~\cite{branzel_gravityspace_2013}; or \dtag{Transformative}, meaning capable of changing form, structure, or function over time, such as portals enabling robots to emerge from underground~\cite{nakagaki_disappearables_2022}, movable partitions that reconfigure workspaces for privacy and interaction~\cite{onishi_waddlewalls_2022}, or ice interfaces that gradually melt to alter form and experience~\cite{lee_icy_2024}.

\mySubsub{\dtag{Human} $\longleftrightarrow$ \dtag{Boundary}}
\dtag{Perception}: Refers to the user's positional relationship to boundaries. \dtag{Unilateral} when users view the boundary from only one side, for example, standing in front of walls and reaching behind~\cite{lilija_augmented_2019}, or interacting on tabletops that conceal electronic components underneath~\cite{hendriks_undertable_2024}. \dtag{Bilateral} when users are positioned at or near the boundary plane and can perceive both sides simultaneously, such as bookshelf planks revealing objects above and below~\cite{cheng_towards_2022}, or visual cut planes that reveal both sides of a divided cake~\cite{pohl_integrated_2024}.

\dtag{Visibility}: Refers to how the boundary mediates visual access for the user. It is \dtag{Opaque} when the boundary obstructs sight, such as walls separating front and backstage in a theater~\cite{nakagaki_disappearables_2022}; \dtag{See-through}, as with pseudo-transparent screens that support back-of-device interaction~\cite{wigdor_lucid_2007}, or digital lenses that alter the style of underlying images~\cite{riche_ai-instruments_2025}; and \dtag{Reflective}, such as augmented mirrors used in dance to explore multi-layered presence~\cite{zhou_here_2023}, or to foster connection between self and the world~\cite{jacobs_performative_2019}. 

\mySubsub{\dtag{Human} $\longrightarrow$ \dtag{Object}}
Manipulations can be \dtag{Direct}, where the object is acted upon by the human's hand (e.g., touching~\cite{wigdor_lucid_2007} or grabbing~\cite{rau_traversing_2025}); or \dtag{Indirect}, where the object is influenced by the human via an intermediary (e.g., automation~\cite{nakagaki_disappearables_2022}, controllers~\cite{kim_perspective_2023}, or proxies~\cite{pohl_poros_2021,liu_reality_2025}).

\subsection{Visual Design Language}\label{sec:language}
We aimed to create a minimal yet expressive symbol set to encode design space attributes and illustrate interactions. The design iteration began with simple metaphors---a ball for \dtag{Object}, a line for \dtag{Boundary}---to abstract physical concepts. For \dtag{Nature}, we used solid {(conveys tangibility)}, dashed {(virtuality)}, and half-solid/half-dashed {(hybridity)} styles to represent \dtag{Physical}, \dtag{Digital}, and \dtag{Augmented}, respectively. For \dtag{Object}, the styles are applied as fills. \dtag{Boundary} uses double parallel lines to denote planar quality, with the same style variations. For \dtag{Dynamic}, \dtag{Static} is a plain line, while \dtag{Transformative} adds a star symbol, associated with change, spark, or magic. \dtag{Perception} is shown by the eye symbol's position relative to the boundary. For \dtag{Visibility}, \dtag{Opaque} uses hatching to indicate obstruction; \dtag{See-through} shows the target on the opposite side for transparency; and \dtag{Reflective} renders a duplicate, evoking mirrors producing virtual images. \dtag{Motion} employs symbol placement for spatial relations, arrows for movement, and the ``\textbf{|\,|}'' for \dtag{Stationary}. For \dtag{Manipulation}, \dtag{Direct} is drawn with a solid hand, denoting immediacy; \dtag{Indirect} uses a dashed outline, signaling mediation.

\subsubsection*{Compositional Principles} Symbols can be composed into interaction illustrations by (1) placing the eye at the human's viewpoint, and the hand on the actual side or opposite side depending on visual balance; (2) styling the object and boundary to indicate their \dtag{Nature} and \dtag{Dynamic}, adding a star for \dtag{Transformative}; and (3) depicting \dtag{Motion} via relative positioning, arrows, and ``\textbf{|\,|}''. 

\subsubsection*{Negotiating (Productive) Ambiguity}\label{para:flexibility-ambiguity} We initially viewed the framework's ability to accommodate atypical cases as productive ambiguity~\cite{gaver_ambiguity_2003}. For instance, for a \dtag{Reflective} boundary with \dtag{Unilateral} perception (e.g., the {Magic Mirror} in \Cref{sec:demonstrations}), we stylized the eye symbol as half-solid and half-dashed. Similarly, for the grid fence in~\cite{ogawa_you_2020} (\Cref{fig:existing}), we illustrated visibility as half \dtag{Opaque} and half \dtag{See-through}, as it partially obstructs vision while allowing visibility through its openings. Likewise, in \Cref{fig:existing}, when interactions occurred primarily between Human and Boundary---such as a person dancing before a mirror~\cite{zhou_here_2023}---we conflated \dtag{Human} and \dtag{Object} into a single concept, since in such cases the \dtag{Object} is essentially the \dtag{Human}'s body that they control. However, expert feedback later revealed the risk of actual ambiguity in this conflation; thus, we illustrate these elements distinctly in the final version (\Cref{fig:final-design-space}).

\section{Demonstration}\label{sec:demonstrations}
\begin{figure*}  
\centering
\begin{tabular}{@{}*{10}{@{\hspace{\imgsep}}c}@{}}
\includegraphics[width=\imgwidth]{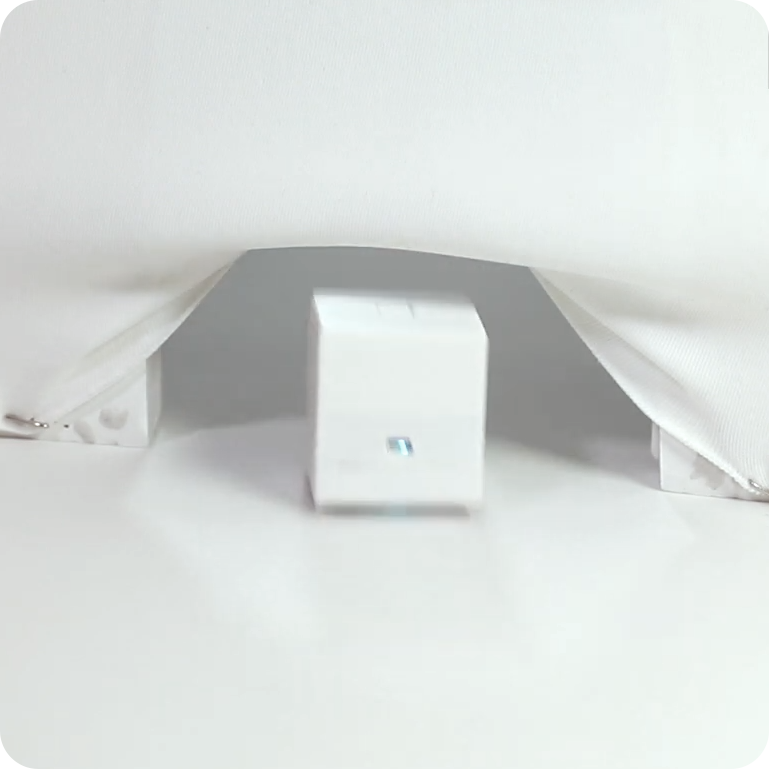} &
\includegraphics[width=\imgwidth]{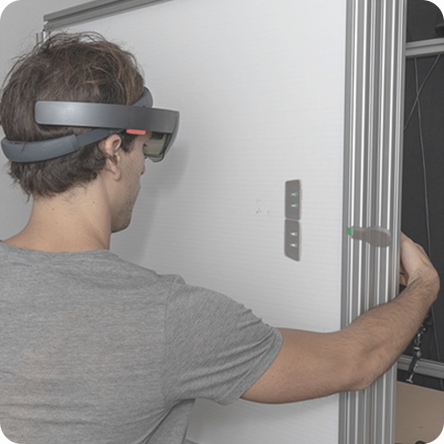} &
\includegraphics[width=\imgwidth]{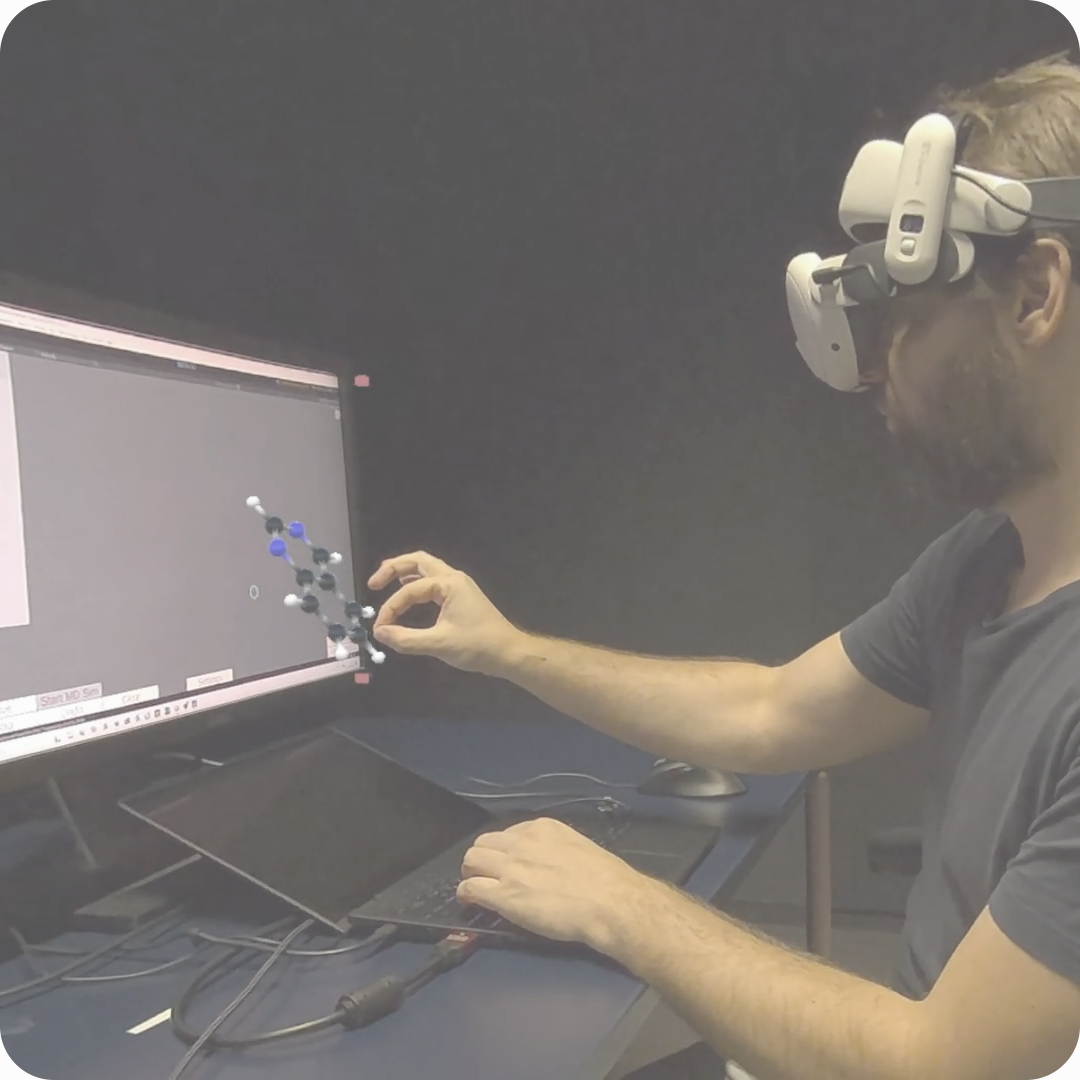} &
\includegraphics[width=\imgwidth]{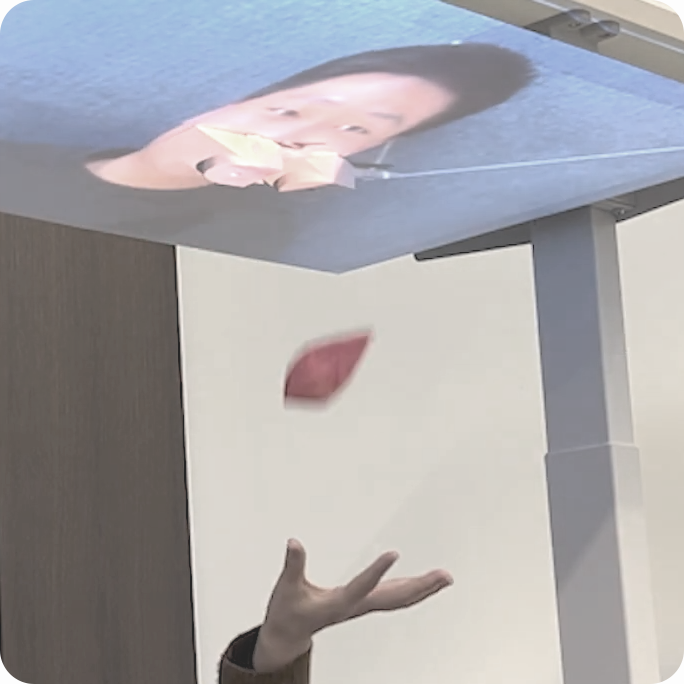} &
\includegraphics[width=\imgwidth]{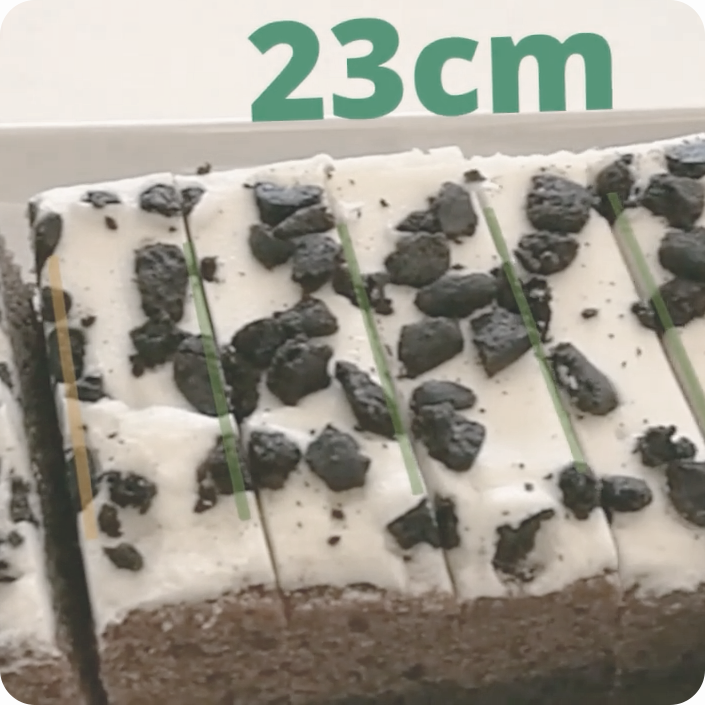} &
\includegraphics[width=\imgwidth]{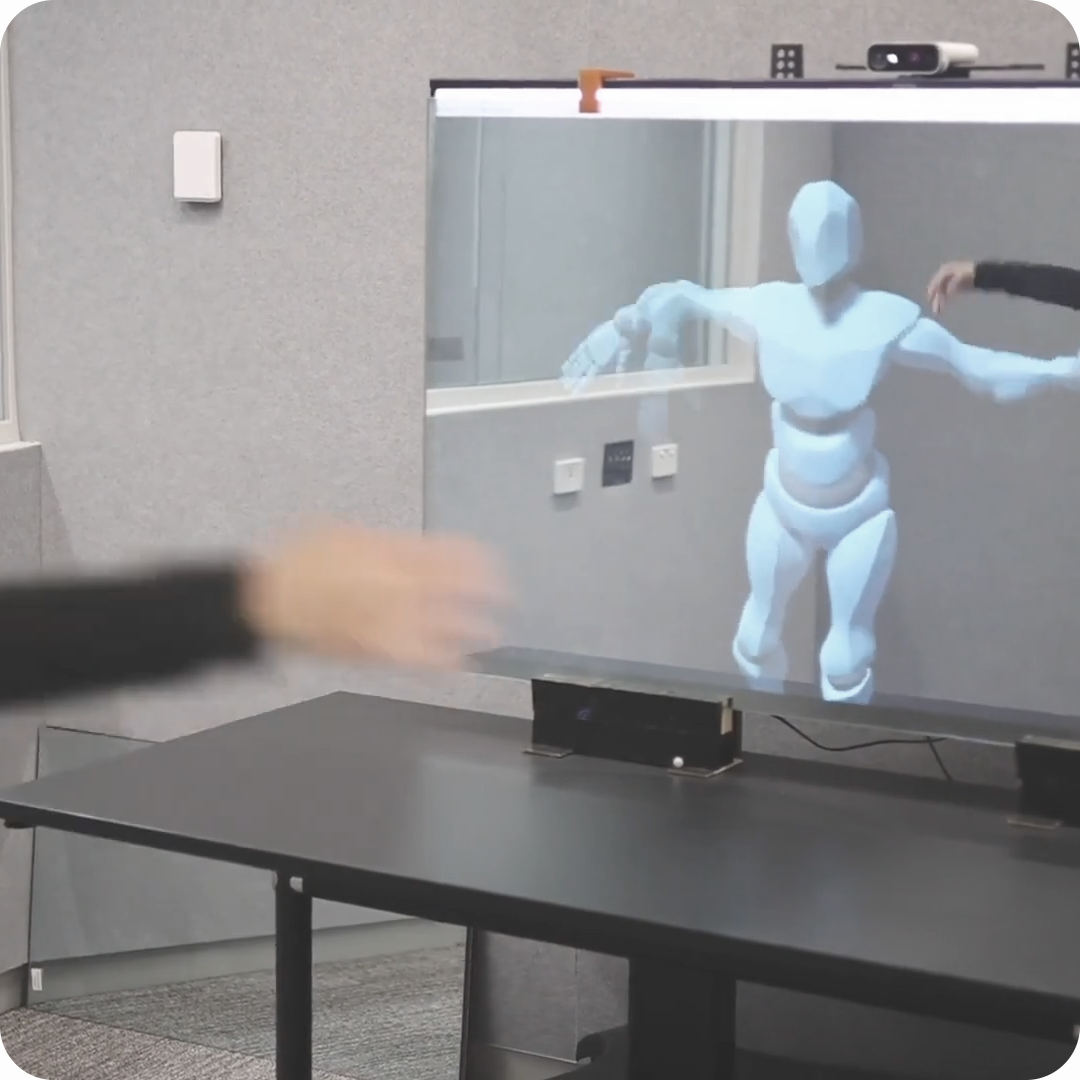} &
\includegraphics[width=\imgwidth]{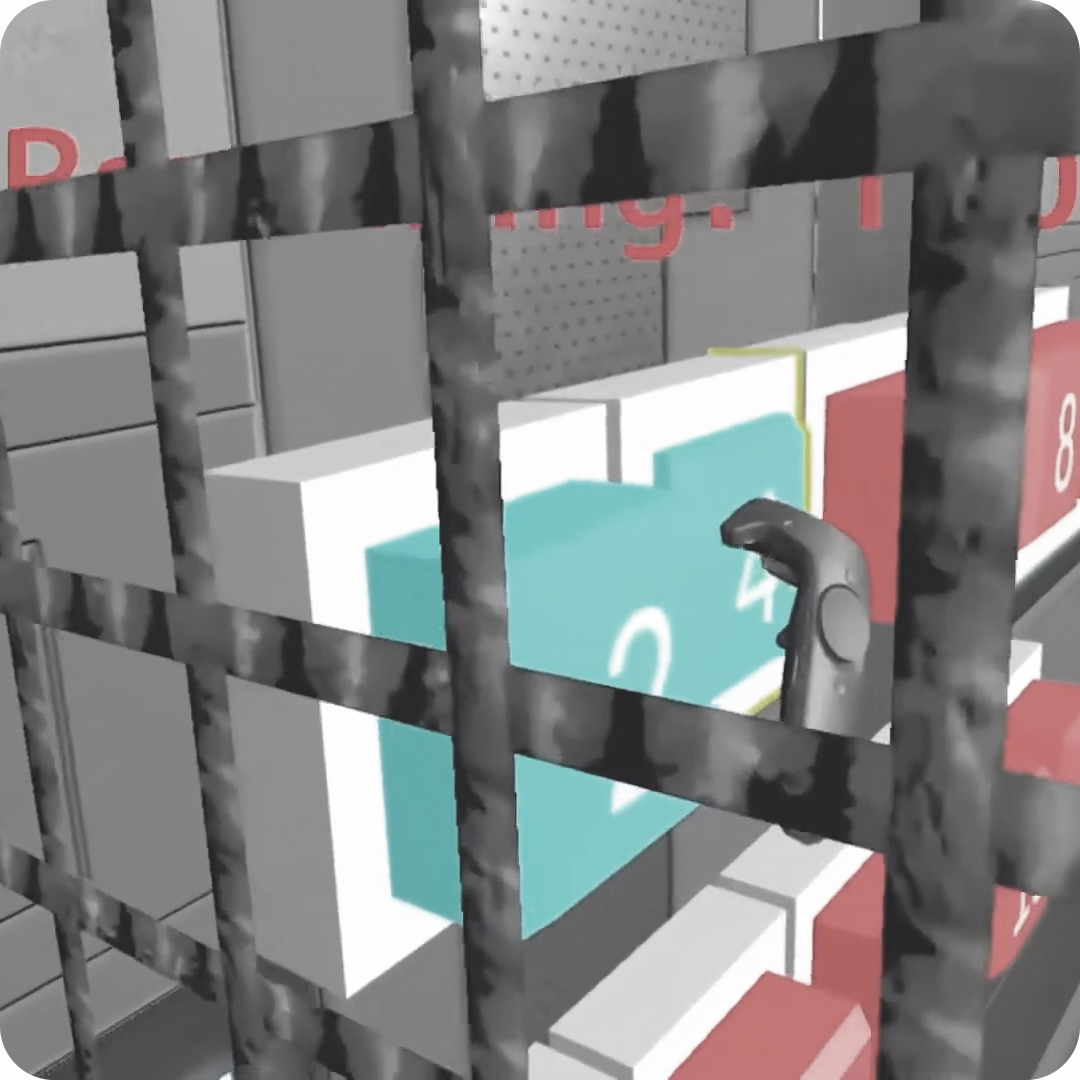} &
\includegraphics[width=\imgwidth]{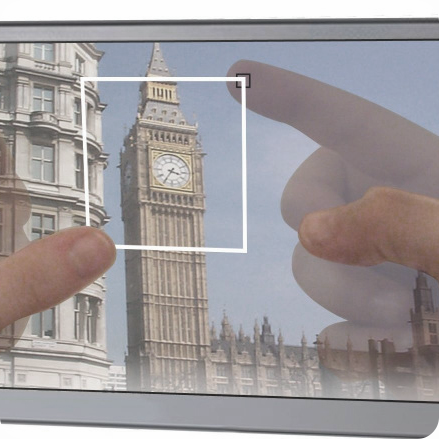} &
\includegraphics[width=\imgwidth]{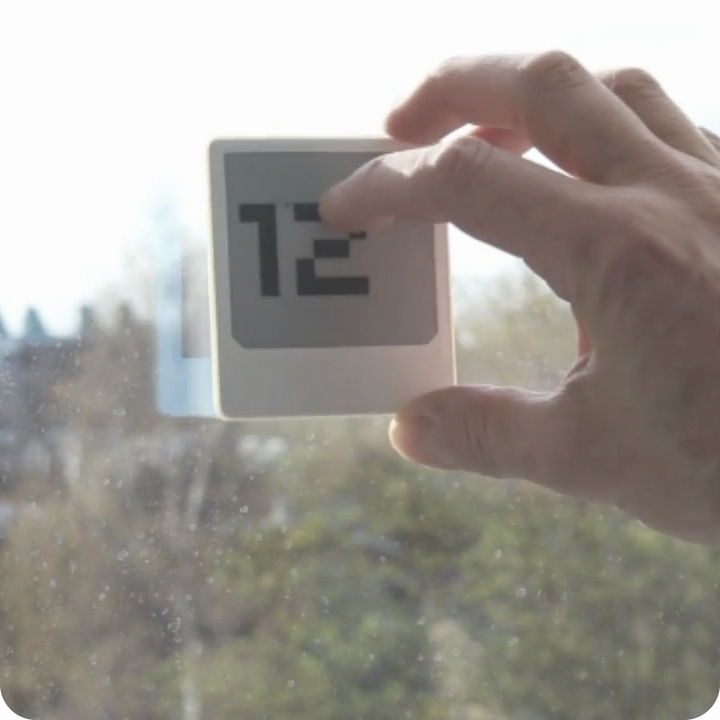} &
\includegraphics[width=\imgwidth]{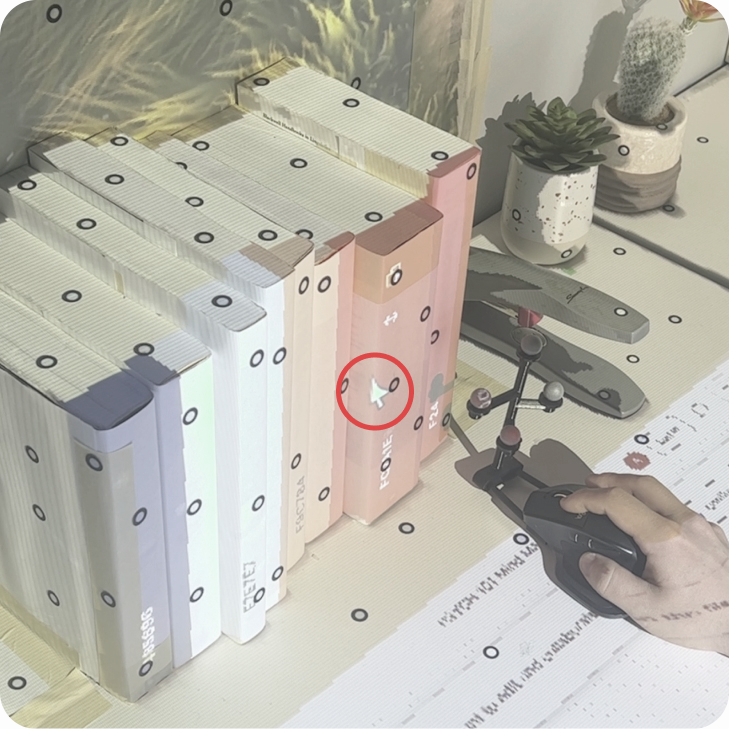} \\[-2pt]

\vtop{\hsize=\textwidthvar \footnotesize\raggedright 
Sending a programmed robot through an actuated wall~\cite{nakagaki_disappearables_2022}.}&
\vtop{\hsize=\textwidthvar \footnotesize\raggedright 
Inserting a USB stick into the back of a wall with visual augmentation~\cite{lilija_augmented_2019}.}&
\vtop{\hsize=\textwidthvar \footnotesize\raggedright 
Pulling a molecular model from a 2D screen into 3D space~\cite{rau_traversing_2025}.}&
\vtop{\hsize=\textwidthvar \footnotesize\raggedright 
Throwing and catching objects via ceiling-integrated robots~\cite{lin_throwio_2023}.}&
\vtop{\hsize=\textwidthvar \footnotesize\raggedright 
Guiding cake division using visually adjustable lines~\cite{pohl_integrated_2024}.}&
\vtop{\hsize=\textwidthvar \footnotesize\raggedright 
Dancing before an augmented mirror with an avatar reflection~\cite{zhou_here_2023}.}&
\vtop{\hsize=\textwidthvar \footnotesize\raggedright 
Reaching through a virtual grid while holding a controller~\cite{ogawa_you_2020}.}&
\vtop{\hsize=\textwidthvar \footnotesize\raggedright 
Sliding fingers on both sides of a see-through screen for selection~\cite{wigdor_lucid_2007}.}&
\vtop{\hsize=\textwidthvar \footnotesize\raggedright 
Attaching a situated display to a window~\cite{grosse-puppendahl_exploring_2016}.}&
\vtop{\hsize=\textwidthvar \footnotesize\raggedright 
Controlling a cursor projected onto everyday surfaces using a mouse~\cite{kim_perspective_2023}.}\\[39pt] 

\includegraphics[width=\pdfwidth]{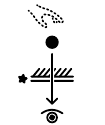} &
\includegraphics[width=\pdfwidth]{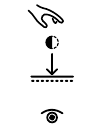} &
\includegraphics[width=\pdfwidth]{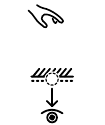} &
\includegraphics[width=\pdfwidth]{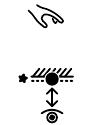} &
\includegraphics[width=\pdfwidth]{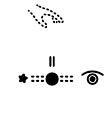} &
\includegraphics[width=\pdfwidth]{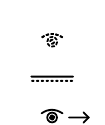} &
\includegraphics[width=\pdfwidth]{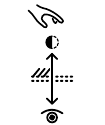} &
\includegraphics[width=\pdfwidth]{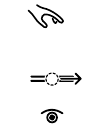} & 
\includegraphics[width=\pdfwidth]{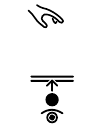} &
\includegraphics[width=\pdfwidth]{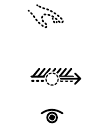} \\
\end{tabular}

\vspace{-0.5em}
\caption{Ten existing OBIs described using our framework. Each includes an interaction scene and description, with its illustration using our visual design language (\Cref{sec:language}). For example, the first illustration depicts a \dtag{Physical} object performing a \dtag{Crossing} through a \dtag{Physical}, \dtag{Transformative} boundary, with \dtag{Unilateral} perception, \dtag{Opaque} visibility, and \dtag{Indirect} manipulation.}
\Description{The figure presents a horizontal grid with ten columns, each showcasing a unique interaction scenario across three aligned rows. The top row features visual captures of users engaging with various physical and digital interfaces. Beneath each capture is a brief text label describing the specific task being performed. The bottom row consists of abstract, symbolic diagrams that map each scenario into standardized visual components: a hand symbol at the top indicating the manipulation method, a central arrangement of geometric shapes (like filled or hollow circles) and lines (solid, dashed, or hatched) combined with arrows depicting the object's movement relative to the boundary, and an eye symbol at the bottom illustrating the user's visual perspective. Through variations in these icons, line textures, and spatial arrangements, the diagrams visually deconstruct the complex interactive relationships shown in the corresponding captures.}
\label{fig:existing}
\end{figure*}

To demonstrate the descriptive power~\cite{beaudouin-lafon_designing_2004} of our framework, we illustrated 10 OBIs from prior work using it (\Cref{fig:existing}). The framework is not limited to these 10; all 31 examples cited in \Cref{sec:dimensions} and others can likewise be represented.

To demonstrate its technical feasibility, we implemented a proof-of-concept prototype using Meta Quest~3 (v79) and Unity (6000.1.4f1) with the Meta XR All-in-One SDK. We developed shaders to achieve different visibility effects (\Cref{fig:materials}) and custom tools to locate objects and fine-tune positions (\Cref{fig:tools}). 

To demonstrate its generative power~\cite{beaudouin-lafon_designing_2004}, we used it to design and implement eight examples across two scenarios---Productivity and Art Exploration---that collectively cover all its elements. They envision how OBI may manifest once MR becomes part of everyday life: one emphasizes practical usage, the other expressive possibilities. All examples serve as functional diegetic prototypes~\cite{kirby_future_2010,uriu_designing_2025} embedded in an overarching narrative, as shown in the accompanying video\footnote{\href{https://www.zhuoyuelyu.com/unbounded}{zhuoyuelyu.com/unbounded}} that begins with an opening scene and continues through two scenarios. 

We present the examples below. For each, we illustrate the interaction, reflect on the design rationale and narrative in the spirit of RtD, describe the framework elements it used, and detail the implementation. An overall reflection on the design process is provided in \Cref{sec:design-implementation}.

\subsection*{Opening Scene}
\mySubsub[nospace]{\large Example 1: Bouncing Ball}
A ball bounces on a tabletop, but instead of rebounding on its final contact, it passes through the surface.
\begin{center}   
\includegraphics[width=0.999\linewidth]{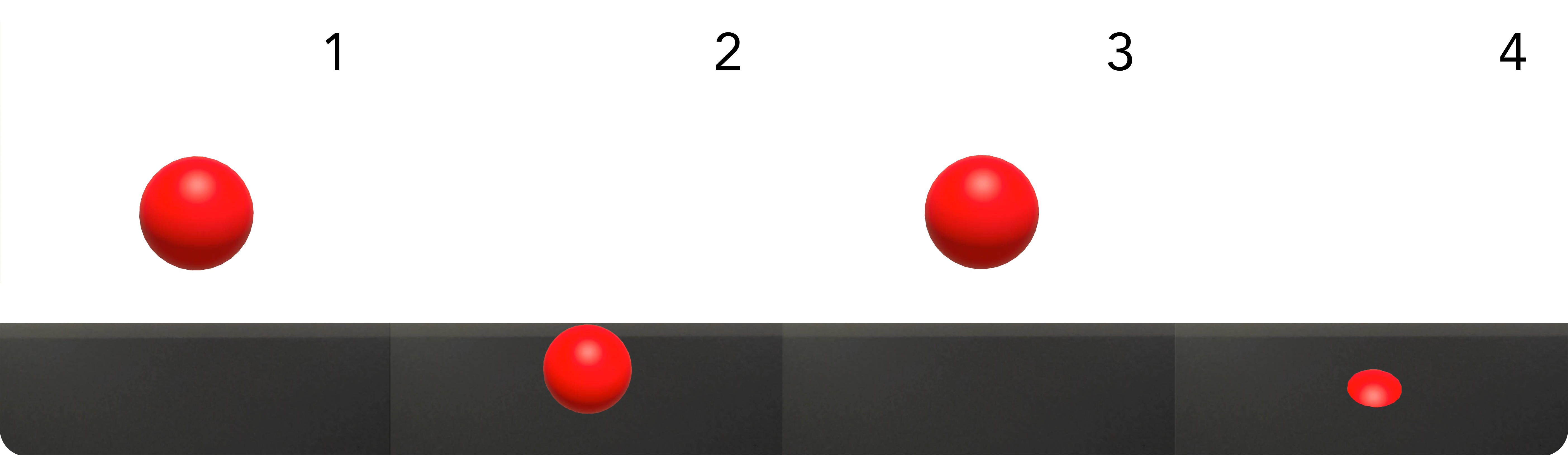}   
\Description{A visual sequence of four frames showing a red 3D sphere interacting with a black horizontal surface. In frame 1, the sphere is suspended in mid-air. In frame 2, it rests directly on the surface. Frame 3 shows the sphere back in mid-air. In the final frame 4, the sphere is partially submerged into the dark surface, with only its top, flattened curve visible above the boundary.}
\end{center}
\vspace{-0.8\baselineskip}
\subsubsection*{Design} This design draws on the iconic red ball in the seminal tangible interface work inFORM~\cite{follmer_inform_2013}, carrying two intertwined layers of meaning. Formally, it begins with the familiar cadence of a plausible bounce---descent, impact, rebound---before breaking the rhythm through sudden disappearance, a small rupture that invites the viewer to pause and reconsider. Symbolically, the red ball represents classic tangibility, here deliberately reimagined to gesture toward affordances unbound by physical law. In both registers, this minimal, recognizable motion becomes a quiet but potent vessel for surprise and reinterpretation.

\vspace{-0.3\baselineskip}
\noindent
\begin{tabular}{@{} p{0.8\linewidth} @{\hspace{0.01\linewidth}} p{0.19\linewidth} @{}}
  \vspace{0pt} 
\subsubsection*{Composition} Human \dtag{Indirect} manipulation of a \dtag{Digital} object (the red ball) that moves \dtag{Toward/Away}, involving both \dtag{Non-crossing} and \dtag{Crossing} a \dtag{Physical}, \dtag{Static} boundary (the tabletop), with \dtag{Opaque} visibility and \dtag{Unilateral} perception.
&
  \vspace{0.3\baselineskip}
\centering
\includegraphics[width=\linewidth]{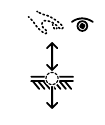}
\Description{The illustration visually encodes the interaction using a dotted-outline hand and a solid eye symbol at the top. Below, a vertical double-headed arrow passes through a dotted circle resting on a double horizontal hatched line, depicting motion crossing the boundary.}
\end{tabular}
\vspace{-0.3\baselineskip}

\subsubsection*{Implementation} The tabletop is wrapped with a custom Cloak of Invisibility material (\Cref{fig:materials}b). Initially, a Box Collider enables precise physical collisions and bouncing of the ball. After a few seconds, the collider is removed, allowing the ball to pass through.

\subsection*{Scenario 1: Productivity}
This scenario envisions how OBIs enabled by our framework could be integrated into future productivity settings, such as offices or home studios. The scenario centers around a desk with common work-related items such as markers, documents, and a cup.

\mySubsub{\large Example 2: Paper Shredder}
To discard an unwanted digital document (e.g., an expired bank statement), users can pick it up and feed it through a horizontal shelf, which shreds it, with fragments visibly dispersing below.
\begin{center}   
\includegraphics[width=0.999\linewidth]{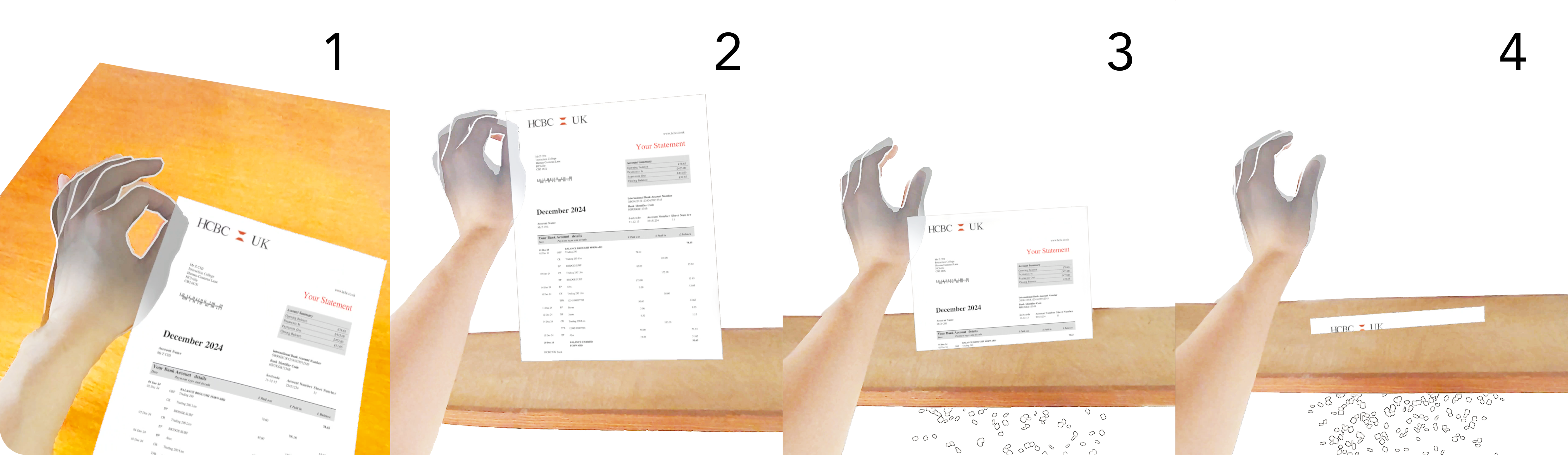}   
\Description{A visual sequence of four frames depicting a user's hand interacting with a digital document and a physical wooden shelf. In frame 1, the hand picks up the document from the desktop. In frame 2, the document is lowered toward the edge. In frame 3, the document begins to pass vertically through the shelf, with small digital fragments visually dispersing beneath it. In frame 4, only the top edge of the document remains visible above the shelf while fragments continue to fall below.}
\end{center}
\vspace{-0.8\baselineskip}
\subsubsection*{Design} Although people inhabit a 3D world, many information artifacts, such as documents, remain rooted in 2D conventions. This persistence is not merely a matter of habit; the planar arrangement of text and images continues to offer efficiency in reading and comprehension. We speculated that, in a future where physical and digital artifacts coexist, digital ``paper'' might inherit familiar physical qualities, such as subtle deformation when lifted, to preserve embodied expectations. Building on this, the act of shredding is reimagined through a bookshelf that reveals both sides of the transformation: the intact sheet above and the stream of confetti below. This grounds the digital act in a familiar material metaphor and follows the spirit of Weiser's vision~\cite{weiser_computer_1999}, where everyday objects are reimagined as computational interfaces.

\vspace{-0.4\baselineskip}
\noindent
\begin{tabular}{@{} p{0.8\linewidth} @{\hspace{0.01\linewidth}} p{0.19\linewidth} @{}}
  \vspace{0pt} 
\subsubsection*{Composition} Human \dtag{Direct} manipulation of a \dtag{Digital} object (the document), performing a \dtag{Crossing} through a \dtag{Physical}, \dtag{Static} boundary (the shelf). The boundary offers \dtag{Opaque} visibility but allows \dtag{Bilateral} perception from the human's position.
&
  \vspace{0.3\baselineskip}
\centering
\includegraphics[width=\linewidth]{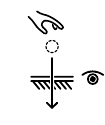}
\Description{Features a solid hand symbol positioned above a dotted circle with a downward-pointing arrow that crosses a double horizontal hatched line. A solid eye symbol is placed to the right of the boundary line, looking horizontally toward the point of intersection.}
\end{tabular}
\vspace{-0.1\baselineskip}
\subsubsection*{Implementation} 
Simulating paper presented a major challenge. While research on paper deformation exists~\cite{narain_folding_2013}, we found no realistic real-time simulator available in Unity. As a workaround, we customized the Obi Cloth simulator~\cite{ObiCloth_Unity} to approximate paper properties (see figure below): applying 0.05 Aerodynamics to make dropped paper fall slowly rather than as a rigid board (1); minimizing Distance and Bend constraints to ensure the paper retains its structural form when placed on objects rather than draping like cloth (2); and adding colliders with high friction to ensure stable placement on flat surfaces while allowing slow sliding on non-flat surfaces (2--4).
\begin{center}   
\includegraphics[width=0.999\linewidth]{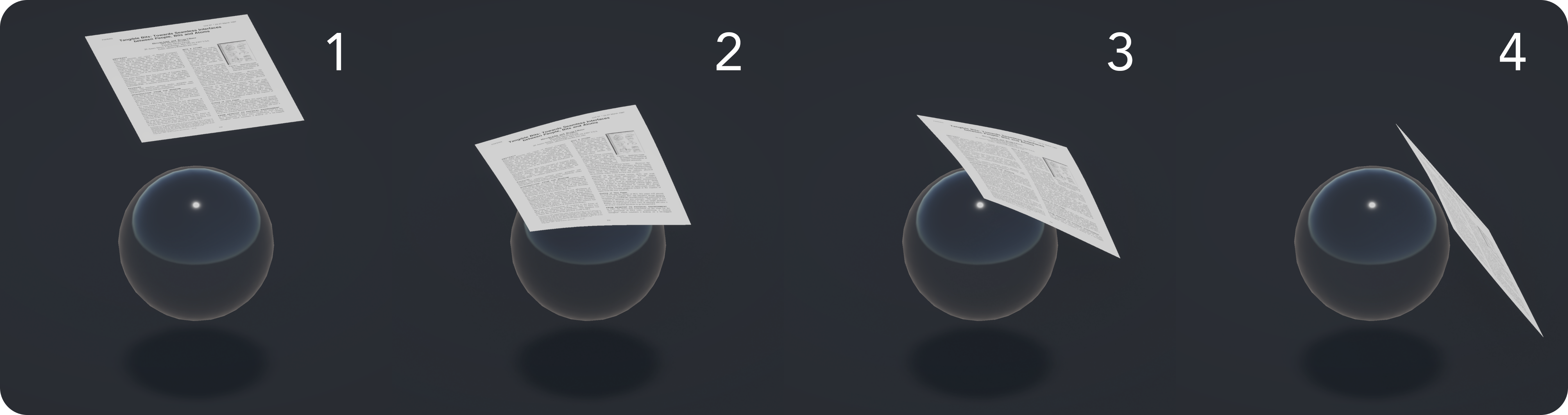}   
\Description{A sequence of four frames demonstrating the physical simulation of a digital sheet of paper interacting with a sphere. In frame 1, the flat rectangular paper falls through the air toward the sphere. In frame 2, it lands and bends slightly upon contact with the sphere. In frames 3 and 4, the paper slides off the sphere while showing slight bends and rotation like a real piece of paper would.}
\end{center}

Paper interaction is handled by anchoring the nearest cloth particles to the fingertip upon pinch detection, with a height threshold and time-based override to filter false positives. Shredding is triggered when particles cross the shelf's top plane; for each crossing particle, a confetti prefab is instantiated via Particle System at $p_i - (0,\delta,0)$ (where $\delta$ is the shelf thickness). To manage performance, active confetti is capped at $K$ using a FIFO queue. Once all particles cross, the cloth object is destroyed. To simulate the document vanishing into the shelf, we applied the Cloak of Invisibility material (\Cref{fig:materials}b) to the space beneath the shelf. Hand overlays are assigned Phasing Object material (\Cref{fig:materials}f) to remain always visible.

\mySubsub{\large Example 3: Trash Can}
To discard office waste (e.g., a dried-out marker), the user moves it to the desk corner, where a circular trash area appears; dropped items disappear through the tabletop.
\begin{center}   
\includegraphics[width=0.999\linewidth]{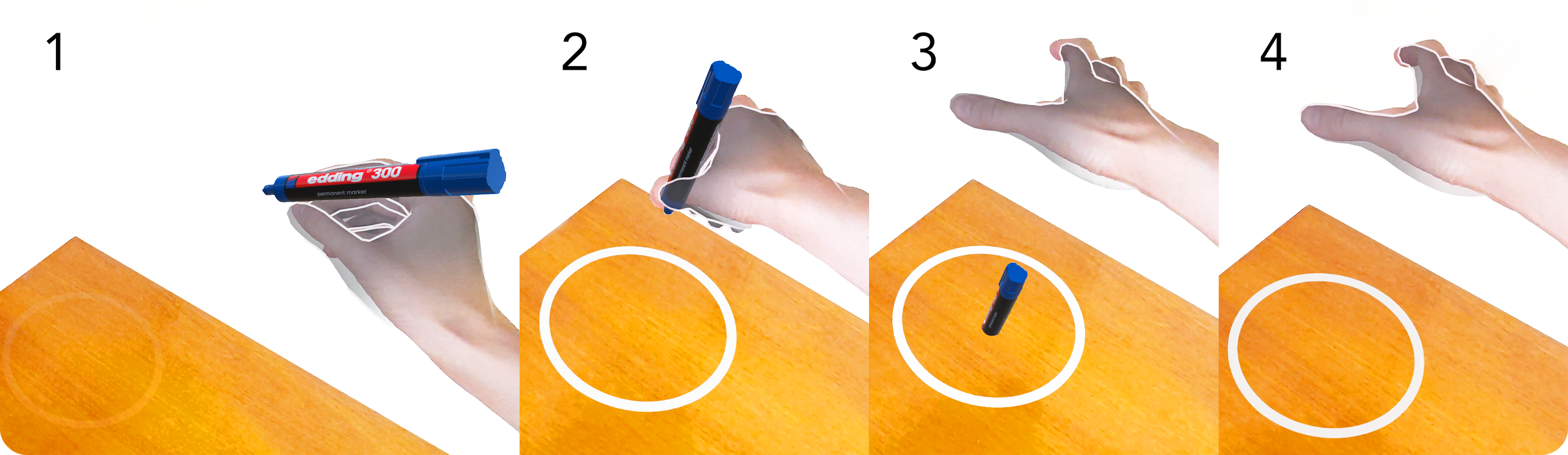} 
    \Description{A visual sequence of four frames showing a user's hand holding a digital marker above a wooden table. In frame 1, the hand hovers near the table, and a faint white circular outline begins to appear. In frame 2, the hand moves the marker directly on top of the now-bright white circle. In frame 3, the hand releases the marker, which is visually depicted sinking halfway through the solid wood. In frame 4, the marker has completely disappeared, leaving only the white circle on the surface.}
\end{center}
\vspace{-0.8\baselineskip}
\subsubsection*{Design} Extending the idea of embedding functions within everyday objects, we reimagine a desk corner as a subtle trash receptacle. In daily life, tossing a paper ball into a bin from a distance carries the small satisfaction of seeing it disappear inside. Here, the hard-to-reach corner, often free from accidental contact, becomes an apt location for discarding objects. The bin is indicated only by a faint circle that becomes more visible as the hand approaches, keeping the affordance unobtrusive until needed. When an item crosses the surface, it vanishes much like the Bouncing Ball, yet here the act moves from speculative play to a functional affordance for disposal.

\vspace{-0.4\baselineskip}
\noindent
\begin{tabular}{@{} p{0.8\linewidth} @{\hspace{0.01\linewidth}} p{0.19\linewidth} @{}}
  \vspace{0pt} 
\subsubsection*{Composition} Human \dtag{Direct} manipulation of a \dtag{Digital} object (the marker) that moves \dtag{Toward} and performs \dtag{Crossing} of a \dtag{Augmented}, \dtag{Static} boundary (the tabletop with a circle overlay), with \dtag{Opaque} visibility and \dtag{Unilateral} perception.
&
  \vspace{0.3\baselineskip}
\centering
\includegraphics[width=\linewidth]{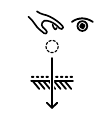}
\Description{Shows a solid hand and a solid eye symbol grouped at the top. Beneath them, a downward-pointing arrow passes through a dotted circle and vertically crosses a dotted horizontal line above a solid hatched horizontal line.}
\end{tabular}
\vspace{-0.3\baselineskip}

\subsubsection*{Implementation}
We render the faint circle on the tabletop using Line Render, whose opacity encodes objects' proximity and whose interior deletes objects that drop sufficiently below the plane. The displayed opacity is computed as $\alpha=\max_j f(d_j)$, where $d_j$ is the distance from object $j$ to the circle center; $f(d)$ equals 1 for $d \le d_{\min}$, 0 for $d \ge d_{\max}$, and varies linearly in between. Colliders are disabled when $d\le d_{\min}$ to allow entry; objects are removed if they fall by a vertical distance $h$; opacity is smoothed each frame (lerp).

\mySubsub{\large Example 4: Marker}
To jot down notes or thoughts (e.g., a to-do list), users can pick up a marker and write on any surrounding wall as if it were a whiteboard.
\begin{center}   
\includegraphics[width=0.999\linewidth]{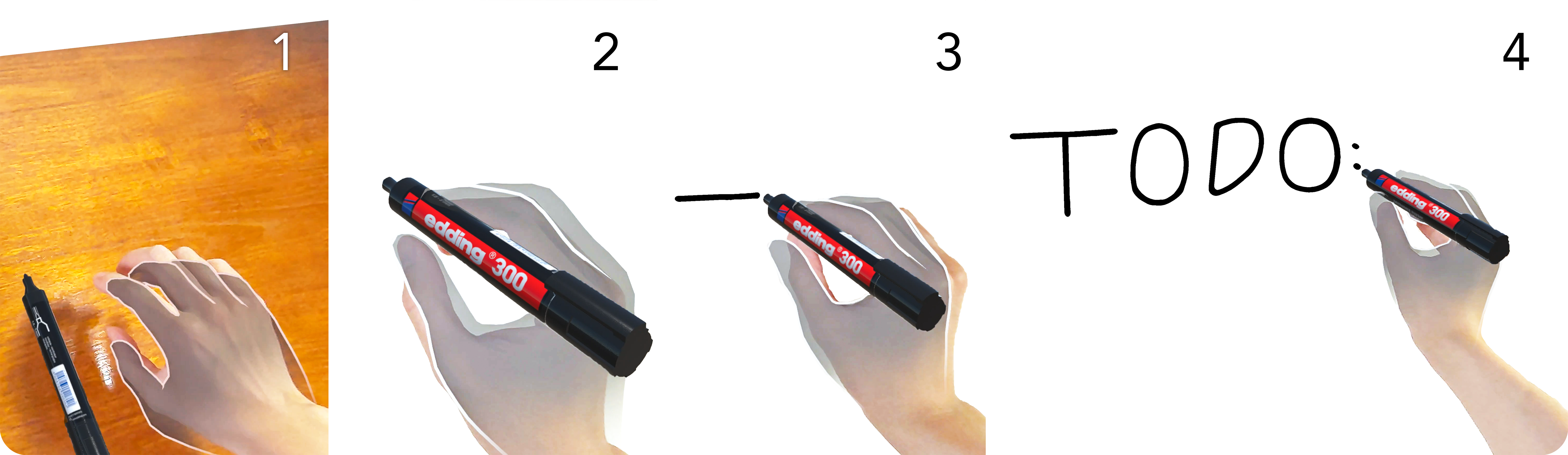}   
    \Description{A visual sequence of four frames showing a user's hand interacting with a digital marker. In frame 1, the hand picks up the black marker from a wooden surface. In frame 2, the hand holds the marker in a ready position against a blank white wall. In frames 3 and 4, the hand uses the marker to write the word "TODO:" in black digital ink on the wall.}
\end{center}
\vspace{-0.8\baselineskip}
\subsubsection*{Design} Walls in domestic or office environments offer vast, underutilized expanses that can inherently resemble the tabula rasa of a whiteboard. Choosing walls over surfaces like desktops leverages these vertical planes to bypass the heterogeneity of surface items and textures, providing a consistent canvas for spontaneous notes or sketches. The marker is not tied to a fixed whiteboard but extends its affordance to any suitable surface, providing a more fluid form of workspace where the dichotomy between architecture and interface dissolves.

\vspace{-0.25\baselineskip}
\noindent
\begin{tabular}{@{} p{0.8\linewidth} @{\hspace{0.01\linewidth}} p{0.19\linewidth} @{}}
  \vspace{0pt} 
\subsubsection*{Composition} Human \dtag{Direct} manipulation of a \dtag{Digital} object (the marker) moving \dtag{Along} with \dtag{Contact} on a \dtag{Physical}, \dtag{Static} boundary (the wall), with \dtag{Opaque} visibility and \dtag{Unilateral} perception.
&
  \vspace{0.25\baselineskip}
\centering
\includegraphics[width=\linewidth]{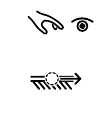}
\Description{Depicts a solid hand and an eye symbol at the top. Below, a dotted circle sits directly on top of a double horizontal hatched line, accompanied by a horizontal arrow pointing to the right to illustrate sliding motion along the surface.}
\end{tabular}
\vspace{-1.2\baselineskip}
\subsubsection*{Implementation} The marker is a 3D asset with no built-in logic~\cite{Edding300Marker_TheLostSock_2022}; all functionality was custom-implemented. The drawing plane is defined by sampling three points on the physical wall with the controller (\Cref{fig:tools}). To avoid view-dependent distortions caused by line renderers, we generate strokes by rapidly instantiating small solid circle prefabs on the plane when the tip is within a threshold distance. Drawing is gated by asymmetric thresholds to account for plane penetration: $\leq1.5$~cm on the front, where overshoot is conspicuous, and $\leq5$~cm behind, where penetration is less perceptible. The projected tip is lightly smoothed and resampled at $\geq1$~mm spacing to ensure stroke fluidity.

\mySubsub{\large Example 5: Acoustic Partition}
When outdoor noise becomes distracting, users can pull out an acoustic partition from a seam along the windowsill to block sound, creating a quieter workspace on their side.
\begin{center}   
\includegraphics[width=0.999\linewidth]{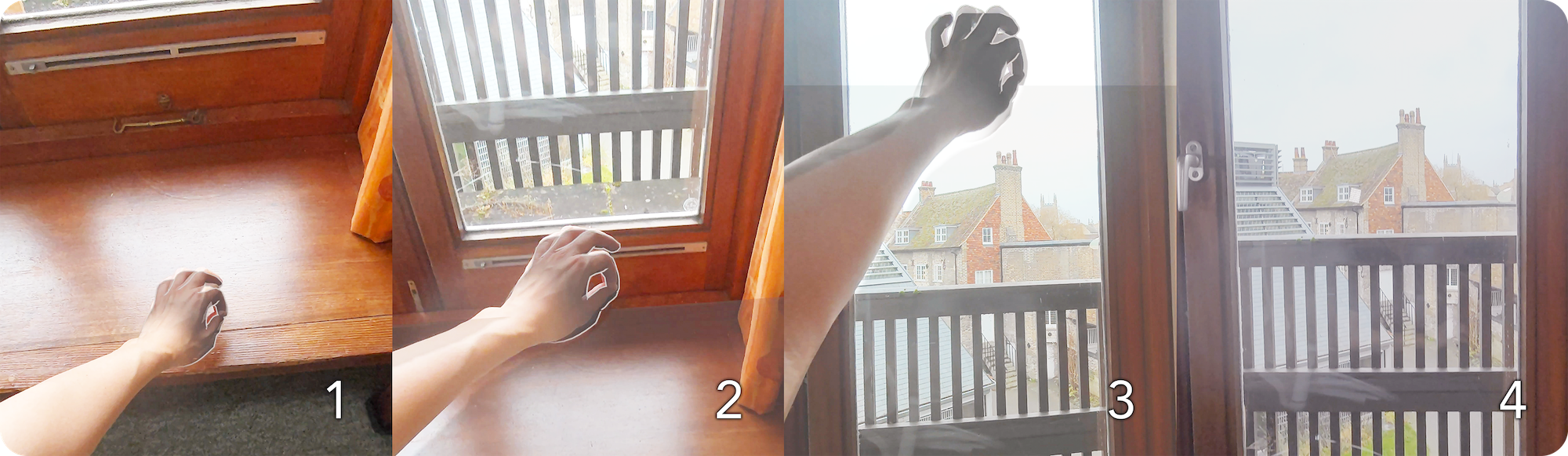}  
    \Description{A visual sequence of four frames depicting a user's hand interacting with a seam along the wooden windowsill. In frame 1, the hand pinches on top of the seam. In frames 2 and 3, the hand pulls upward from the seam, which draws out a translucent virtual pane. In frame 4, the translucent pane is fully extended upward.}
\end{center}
\vspace{-0.8\baselineskip}
\subsubsection*{Design} The idea stems from a common experience in shared study spaces---the desire to block surrounding noise without signaling irritation to others. In physical settings, erecting a barrier is cumbersome and socially conspicuous; here, a semi-transparent digital boundary provides acoustic isolation while maintaining a subtle visual presence, conveying ``something is there'' without feeling like an opaque wall. The partition unfurls from a narrow seam along the windowsill, as if the frame itself stores and releases the barrier. Sound reduction is mapped to its height: raising the boundary increases attenuation, mirroring the real-world sensation of closing a window and hearing the outside fade. 

\vspace{-0.5\baselineskip}
\noindent
\begin{tabular}{@{} p{0.8\linewidth} @{\hspace{0.01\linewidth}} p{0.19\linewidth} @{}}
  \vspace{0pt} 
\subsubsection*{Composition} 
(1) The act of pulling out the partition: Human \dtag{Direct} manipulation of an \dtag{Augmented} object (the seam extending into a partition) that moves \dtag{Away} with \dtag{Non-crossing} on a \dtag{Physical}, \dtag{Static} boundary (the windowsill), with \dtag{Opaque} visibility and \dtag{Unilateral} perception. 
&
  \vspace{0.5\baselineskip}
\centering
\includegraphics[width=\linewidth]{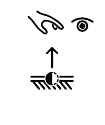}
\Description{Displays a solid hand and an eye symbol at the top. Below them, an upward-pointing arrow originates from a circle that is half-solid black and half-white, resting on a double horizontal hatched line.}
\end{tabular}

\vspace{-0.5\baselineskip}
\newlength{\mySavedIndent}
\setlength{\mySavedIndent}{\parindent}
\noindent
\begin{tabular}{@{} p{0.8\linewidth} @{\hspace{0.01\linewidth}} p{0.19\linewidth} @{}}
  \vspace{0pt} 
\setlength{\parindent}{\mySavedIndent}
\vspace{0pt}
(2) Once extended, the partition becomes a \dtag{Digital}, \dtag{Transformative} boundary; the human (as \dtag{Object}) remains \dtag{Non-contact} relative to it, with \dtag{See-through} visibility and \dtag{Unilateral} perception.
&
  \vspace{-0.25\baselineskip}
\centering
\includegraphics[width=\linewidth]{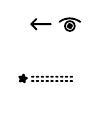}
\Description{This graphic contains no hand or object icons. A solid eye symbol appears at the top right, with a leftward-pointing arrow positioned to its left. Below them is a double horizontal dashed boundary line preceded by a star symbol.}
\end{tabular}
\vspace{-1.6\baselineskip}
\subsubsection*{Implementation}
A translucent plane is pulled up from a windowsill covered by the Cloak of Invisibility material (\Cref{fig:materials}b). Its height is proportional to the reduction of street noise from an audio file, smoothed with a lerp function. We capped the maximum height to match a seated user's reachable range, and accordingly limited the noise reduction to 2\% of the original volume.

\subsection*{Scenario 2: Art Exploration}\label{sec:art}
This scenario envisions how our framework could support artistic experiences, such as in museums or galleries. The scenario centers on a painting---Kandinsky's \textit{Composition VIII}---with a drawer and mirror as part of the experience.

\mySubsub{\large Example 6: Interactive Painting}
We envision a more interactive form of art appreciation. Viewers can pull elements of a painting out of the 2D frame into 3D space (1--3) and participate in the creative process by adding or rearranging elements (see Magic Mirror).
\begin{center}   
\includegraphics[width=0.999\linewidth]{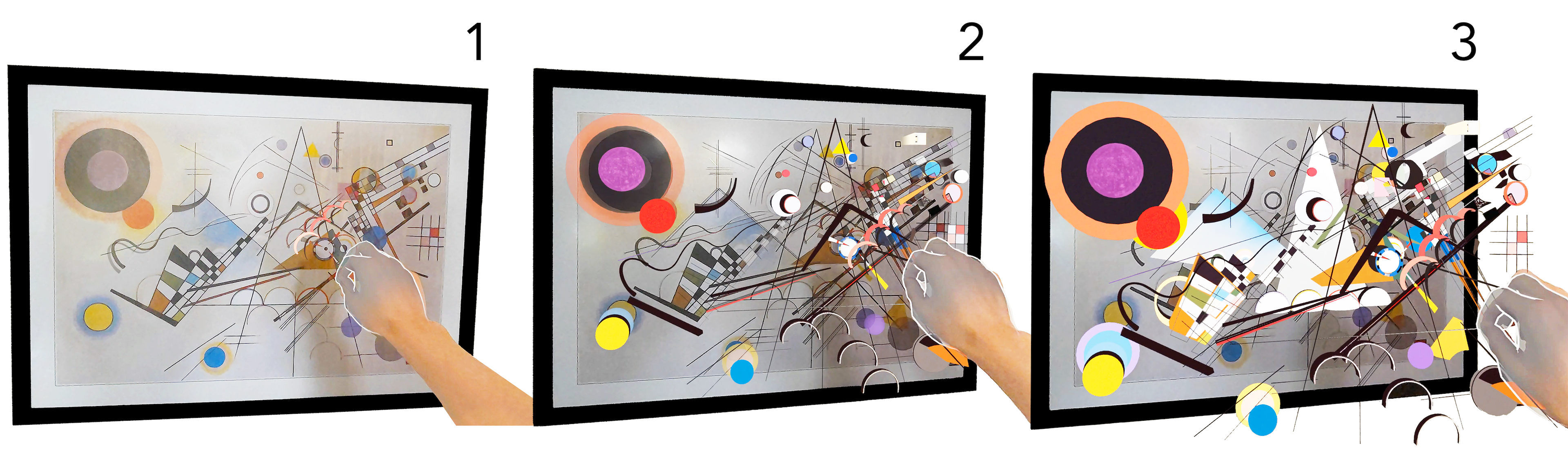}
\Description{A sequence of three frames showing a user's hand interacting with a framed abstract painting. In frame 1, the hand pinches in front of the artwork. In frame 2, the hand pinches and pulls a portion of the artwork out from the canvas into 3D space. In frame 3, all the shapes and lines from the painting are now pulled into 3D space in front of the frame.}
\end{center}
\vspace{-0.8\baselineskip}
\subsubsection*{Design} This example explores three intertwined ideas. First, traditional gallery experiences position the viewer as a passive recipient, relying on wall labels or audio commentary while adhering to ``do not touch'' protocols. MR transcends these limits by reclaiming the in-between space for interaction. Thereby, the viewer becomes a creative participant, able to interact with the work and engage in a cross-temporal dialogue with its original artist.

Second, for centuries, painters have negotiated spatial relationships on 2D surfaces, from the sfumato of \textit{Mona Lisa} to the spatial illusion of \textit{Las Meninas} and the shifting vantage points of the Chinese masterpiece \textit{Dwelling in the Fuchun Mountains} (1350)~\cite{huang1350fuchun}. True three-dimensionality was reserved for sculpture. This design asks: could there be a medium that bridges painting and sculpture? Seen head-on, the piece reads as 2D; when pulled outward, its lines, shapes, and structures unfold into spatial relationships that invite new interpretations. 

Third, the act of pulling with the hand creates a point of connection between the viewer and the work. Unlike an automatic transformation, this gesture lets the viewer control the speed, depth, and rhythm of the unfolding, making the transition not only a visual shift but a moment of agency and intimacy with the piece.

\vspace{-0.4\baselineskip}
\noindent
\begin{tabular}{@{} p{0.8\linewidth} @{\hspace{0.01\linewidth}} p{0.19\linewidth} @{}}
  \vspace{0pt} 
\subsubsection*{Composition} Human \dtag{Direct} manipulation of an \dtag{Augmented} object (the physical painting with 3D augmentation) that moves \dtag{Away} with \dtag{Non-crossing} on a \dtag{Physical}, \dtag{Static} boundary (the painting's surface), with \dtag{Opaque} visibility and \dtag{Unilateral} perception.
&
  \vspace{0.5\baselineskip}
\centering
\includegraphics[width=\linewidth]{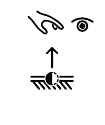}
\Description{Visually identical to the first diagram of Example 5 (Acoustic Partition), it shows a top-aligned solid hand and eye symbol, with an upward-pointing arrow emerging from a half-solid, half-white circle resting on a double horizontal hatched line.}
\end{tabular}
\vspace{-0.25\baselineskip}
\subsubsection*{Implementation} We used a framed reproduction of Kandinsky's \textit{Composition VIII} (A1 size) and a corresponding 3D asset~\cite{patrix_Composition8_2023}. To enable interaction with individual elements from the asset's single mesh, we developed a MeshSplitter that separates connected components via breadth-first search on triangle adjacency (defined by a vertex-pair proximity tolerance). Each component was processed into a sub-mesh with preserved UVs and duplicated materials, then assembled into a parent prefab. We manually regrouped related components and added hand grab, free transform, and occlusion using the Meta SDK. Two reference points were marked using the controllers for alignment (\Cref{fig:tools}).

\mySubsub{\large Example 7: Drawer}
Below the painting, a small circular knob invites the user to pull open a drawer embedded in the wall. Inside are several everyday objects: a Rubik's cube, a seashell, and a pair of headphones.
\begin{center}   
\includegraphics[width=0.999\linewidth]{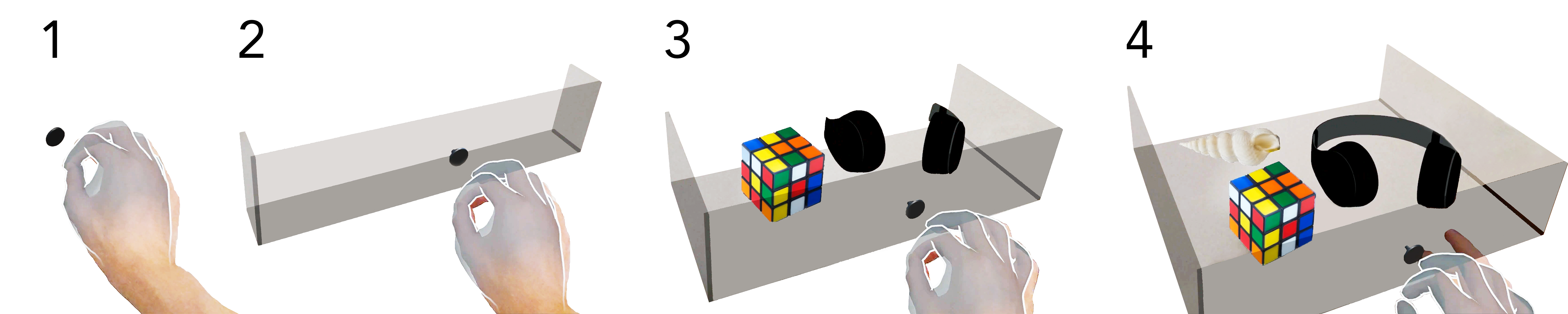}  
    \Description{A four-frame sequence depicts a user interacting with a virtual drawer that is attached to a physical wall. In frame 1, a hand pinches a dark knob on the wall. In frame 2, pulling the knob outward reveals a semi-transparent rectangular drawer. In frame 3, the open drawer reveals 3D digital objects inside: a Rubik's cube and black headphones. In frame 4, a white seashell is also visible inside the drawer.}
\end{center}
\vspace{-0.8\baselineskip}
\subsubsection*{Design} We were intrigued by the idea of hiding information within boundaries. Here, it takes the form of a drawer that pulls out from the wall, a minimal unit of the concept that could scale to reveal an entire hidden bookshelf. Initially, we considered concealing the drawer completely, but realized that without visible cues, users would lack the affordance to discover it; we therefore left the knob exposed. We explored drawer opacity, weighing the clarity of a solid form against the spatial disruption it would cause, and ultimately chose semi-transparency to convey permeability, softening the divide between digital and physical space. 

Kandinsky composed with abstract points, lines, and shapes, viewing his paintings as symphonic arrangements. Here, we begin with concrete objects, intending for them to be later transformed into similar visual primitives (see Magic Mirror) that carry the richness of lived experiences---much as Kandinsky encoded musicality into his art. We selected three objects to span diverse human senses. Headphones represent sound; a Rubik's cube, color and tactile engagement; and a seashell---small, natural, resonant---embodies the sensory breadth of smell, taste, temperature, wind, and the longing for nature.

\vspace{-0.25\baselineskip}
\noindent
\begin{tabular}{@{} p{0.8\linewidth} @{\hspace{0.01\linewidth}} p{0.19\linewidth} @{}}
  \vspace{0pt} 
\subsubsection*{Composition} Human \dtag{Direct} manipulation of a \dtag{Digital} object (the drawer) that moves \dtag{Away} with \dtag{Crossing} on a \dtag{Physical}, \dtag{Static} boundary (the wall), with \dtag{Opaque} visibility and \dtag{Unilateral} perception.
&
  \vspace{0\baselineskip}
\centering
\includegraphics[width=\linewidth]{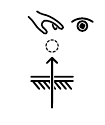}
\Description{Features a solid hand and an eye symbol at the top. Below, an upward-pointing arrow rises from beneath a horizontal hatched line, passes through it, and points toward a dotted circle above.}
\end{tabular}
\vspace{-1\baselineskip}
\subsubsection*{Implementation}
To ensure a sharp visual transition without Meta SDK's occlusion blur, we applied a Cloak of Invisibility material (\Cref{fig:materials}b) to the wall. This allowed the drawer to appear as if emerging directly from the surface. We constrained the drawer's movement to horizontal translation with a fixed maximum extension, setting all contents as child objects. The Rubik's cube~\cite{AntijnvanderGun_RubiksCube_2016}, seashell~\cite{NHMWien_AmaeaMagnifica_2016}, and headphones~\cite{Poum_WirelessHeadphones_2018} were existing 3D assets, while their abstract representations were predominantly custom-designed (detailed in the next example).

\mySubsub{\large Example 8: Magic Mirror}
A rotatable circular mirror sits between the painting and the drawer. Holding an object in front of the mirror reveals an abstracted reflection (1). When rotated sideways (2), placing an object on one side causes its abstraction to appear on the opposite side (3). The mirror supports both digital and physical objects (e.g., a glass cup) (4). Abstracted representations can be extracted, transformed, and added to the painting.
\begin{center}   
\includegraphics[width=0.999\linewidth]{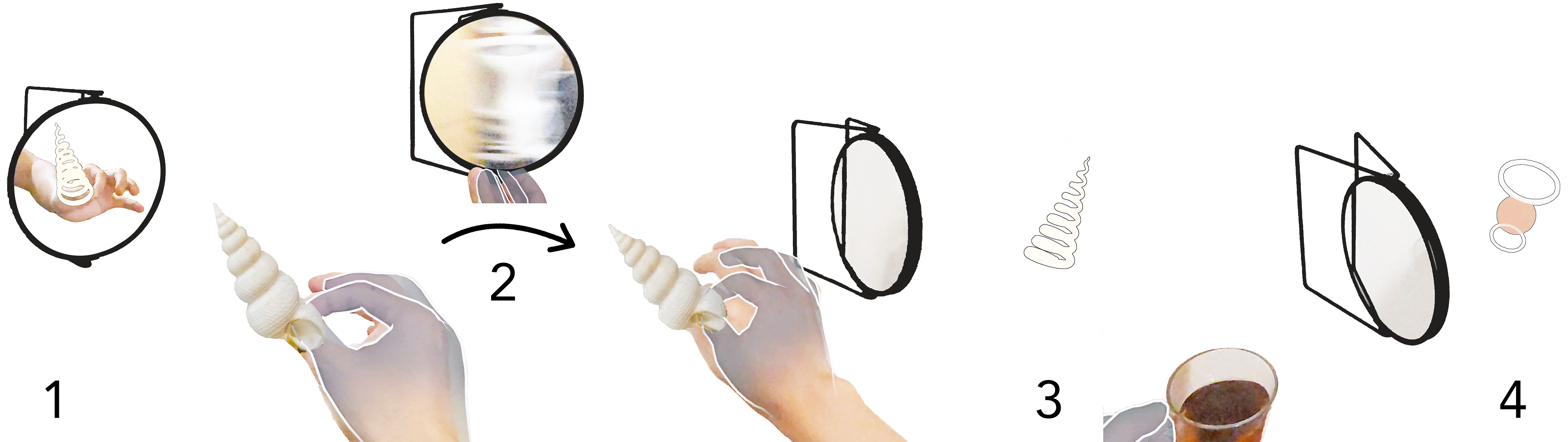}
\Description{A four-frame sequence showing interactions with a circular mirror mounted on the wall. In frame 1, a hand holds the seashell in front of the mirror's reflective surface, which shows an abstract spiral in the mirror. Frame 2 shows a curved arrow indicating the mirror being rotated sideways. In frame 3, viewed from the side, a hand holds the seashell on the left, while the abstract spiral appears in the open air on the right side. In frame 4, the mirror has a physical glass of tea on its left, with an abstract representation (a sphere with rings) floating on its right.}
\end{center}
\vspace{-0.8\baselineskip}
\subsubsection*{Design}
In front of a reflective surface, people tend to shift and tilt to perceive different perspectives. Here, the mirror creates a familiar yet altered world that leverages this instinctive drive. When a viewer holds an object before the mirror, the reflection remains identical in form, yet the object appears as its abstract representation, seemingly grasped by the mirrored hand. When rotated sideways, the abstraction appears in the open air, ready to be taken and used in subsequent interactions. The mirror sustains two worlds at once, with its rotation making the transition between them tangible and participatory.

\begin{center}
  \includegraphics[width=0.999\linewidth]{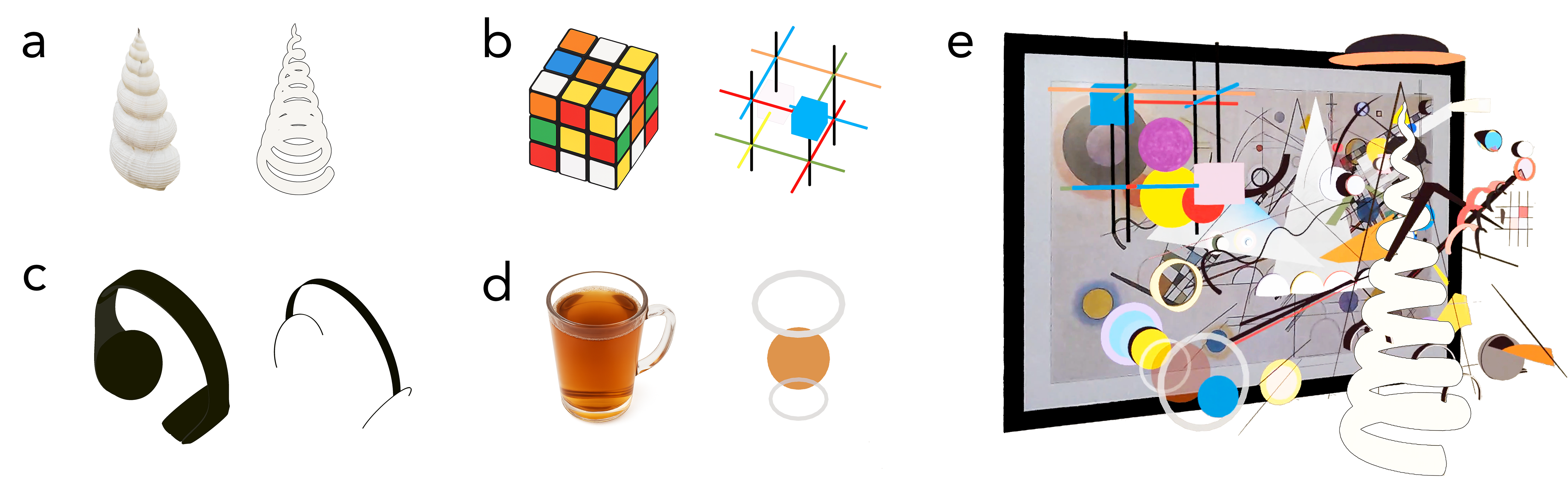}
\Description{A collection of five visual groupings labeled ``a'' through ``e''. Groupings ``a'' through ``d'' display a physical object paired alongside an abstract geometric counterpart: ``a'' pairs a seashell with a wireframe spiral, ``b'' pairs a Rubik's cube with a lattice of colored lines, ``c'' pairs black headphones with three black arcs, and ``d'' pairs a glass of tea with a colored sphere surrounded by white rings. Grouping ``e'' shows the fully pulled-out artwork with these extracted abstract geometric shapes transformed and integrated into it.}
\end{center}

Each mirror reflection reimagines an object in elemental form (a–d), which can then be transformed and incorporated into the painting (e). The seashell (a) becomes a spiral, distilling layered geometry into pure motions of growth and rotation. The Rubik's cube (b) transforms into a lattice of colored lines crossed with small cubes, capturing chromatic richness and modular structure while balancing order and play in a Bauhaus-like palette. Headphones (c) reduce to three arches---a bold arc and two lighter echoes---expressing how a single curve, repeated at different scales, suggests the whole without literal detail. Finally, a glass of tea (d) abstracts into a central sphere of color flanked by concentric rings, evoking the hues of the liquid and the glass; the translucency allows the composition to float with lightness rather than with heavy blocks of color.

\vspace{-0.3\baselineskip}
\noindent
\begin{tabular}{@{} p{0.8\linewidth} @{\hspace{0.01\linewidth}} p{0.19\linewidth} @{}}
  \vspace{0pt} 
\subsubsection*{Composition}
(1) Front-facing mirror: Human \dtag{Direct} manipulation of \dtag{Digital} (seashell, Rubik's cube, headphones) or \dtag{Physical} (cup) objects moving \dtag{Non-contact} relative to a \dtag{Physical}, \dtag{Transformative} boundary (the rotatable mirror), with \dtag{Reflective} visibility and \dtag{Unilateral} perception. For clarity, we illustrate only the \dtag{Digital} object scenario.
&
  \vspace{0.5\baselineskip}
\centering
\includegraphics[width=\linewidth]{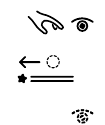}
\Description{A solid hand and an eye symbol are at the top, with a dotted circle and a leftward arrow positioned above a double horizontal line marked with a star. A dotted-outline eye symbol is situated below the boundary line.}
\end{tabular}

\vspace{-0.5\baselineskip}
\setlength{\mySavedIndent}{\parindent}
\noindent
\begin{tabular}{@{} p{0.8\linewidth} @{\hspace{0.01\linewidth}} p{0.19\linewidth} @{}}
  \vspace{0pt} 
\setlength{\parindent}{\mySavedIndent}
\vspace{0pt}
(2) Side-facing mirror (rotated by hand): Identical attributes to (1), but with \dtag{Bilateral} perception (represented by a half-solid, half-dashed eye symbol). Illustrated here with the \dtag{Physical} object scenario.
&
  \vspace{-0.25\baselineskip}
\centering
\includegraphics[width=\linewidth]{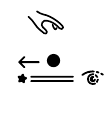}
\Description{Shows a solid hand symbol at the top. Below it, a solid black circle appears with a leftward-pointing arrow beside it, positioned above a double horizontal line marked with a star on the left. To the right of the boundary line is an eye symbol with a half-solid, half-dotted outline.}
\end{tabular}
\vspace{-1.6\baselineskip}
\subsubsection*{Implementation} 
We applied an Inverse Cloak material (\Cref{fig:materials}c), so mirrored objects were visible only through the mirror. The material was implemented as a thin cylinder covering the mirror. Positioning was determined as in earlier examples. Each frame, we achieved one-to-one mapping by reflecting object $A$'s position ($p$) across the mirror plane (normal $n$, point $p_0$): $p' = p - 2n[(p - p_0)\cdot n]$. Orientation was mirrored by reflecting $A$'s forward and up axes across $n$ to reconstruct the rotation. When the mirror was horizontal, the mirrored object switched to a standard material and appeared on the corresponding side using the same mapping. Upon grabbing a mirrored object, the original was disabled. Physical objects like the cup followed the hand's transform instead. Regarding abstractions, the shell was adapted from a spiral asset~\cite{spiral_fab}, reshaped and recolored to match the shell. The cube was constructed from elongated cylinders forming the colored lines, with two cubes at the front and back, using colors inspired by Kandinsky. The headphone arch and teacup (two tori and a central sphere) were modeled using Unity's ProBuilder.

\section{Expert Feedback} \label{sec:expert-sessions}
To examine the usefulness of Unbounded for MR interaction design, as well as its clarity, potential extensions, and limitations, we conducted semi-structured expert feedback sessions. We used our design space (\Cref{fig:design-space}d) and examples (\Cref{sec:demonstrations}) as design probes~\cite{wallace_making_2013} to elicit critique, reflection, and insights.

\subsubsection*{Participants} 
We invited six experts for one-on-one sessions (\Cref{tab:experts}), selected purposively for their senior backgrounds in MR design, research, and industry. Each had at least four years of experience, some over a decade. Their expertise covered MR interaction design, media art, tangible and haptic interaction, and theoretical frameworks, ensuring feedback reflected both academic and industry perspectives across technical, creative, and theoretical dimensions.

\begin{table*}[!htbp]
  \centering
    \caption{Background and Expertise of the Experts}
    \vspace{-3mm}
  \includegraphics[width=\textwidth]{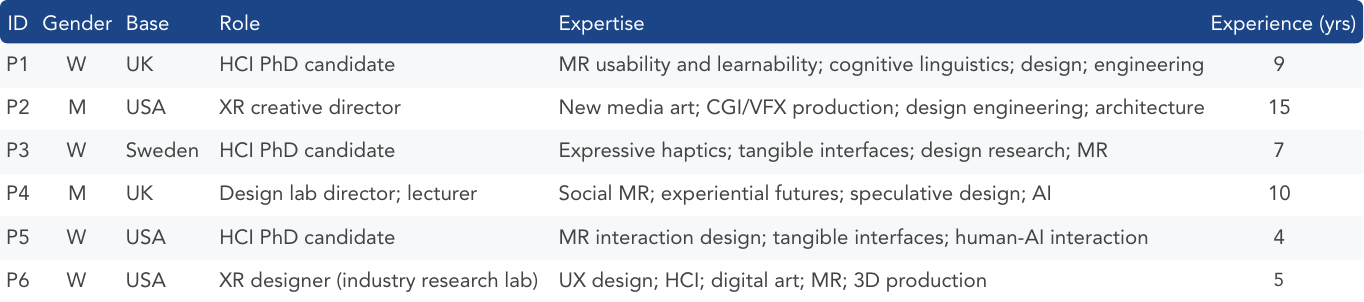}
  \Description{A data table organized into six columns and six rows, detailing the demographics and professional backgrounds of the participating experts. The columns sequentially list each participant's identifier (P1 through P6), self-reported gender, geographic base (UK, USA, or Sweden), current occupational role spanning academia and industry, specific areas of research or design expertise, and total years of professional experience ranging from four to fifteen years.}
  \label{tab:experts} \\
  \vspace{-3.7mm}
  {\footnotesize Note. ``Gender'' denotes self-reported gender identity (Woman [W], Man [M], Non-binary, Self-described, Prefer not to say).}
\end{table*}

\subsubsection*{Procedure}\label{para:procedure}
\begin{figure*}
\centering
  \includegraphics[width=\textwidth]{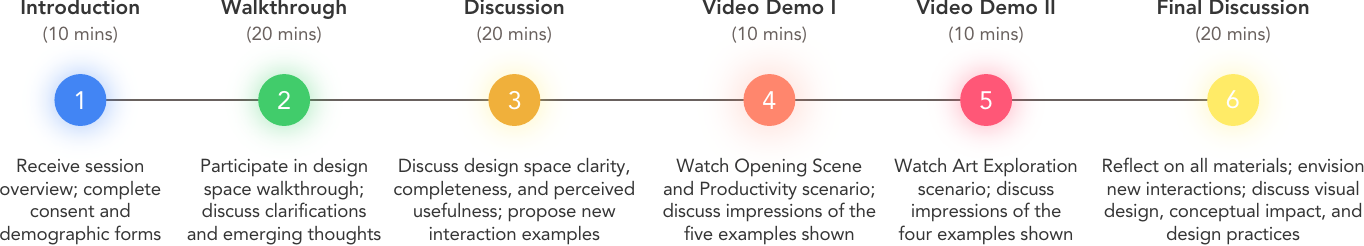}
  \vspace{-1.5em}
\caption{Procedure for each one-on-one expert feedback session. The semi-structured interview questions we used are provided in the supplementary material.}
\Description{A horizontal timeline diagram illustrating a six-step progression. Six sequentially numbered and color-coded circles are connected by a single continuous horizontal line. Above each circle, text indicates the phase name and its time duration, ranging from ten to twenty minutes. Below each circle, brief bullet-point-style text blocks outline the specific activities conducted during that phase, moving from an initial introduction and walkthrough, through discussion and two video viewing sessions, and concluding with a final reflection.}
  \label{fig:procedure}
\end{figure*}

Sessions were conducted and recorded via Zoom to allow international participation. Each session began with a session overview, followed by the completion of informed consent and demographic forms. Following a semi-structured format, we first walked participants through the design space using slides, encouraging them to interrupt for clarification or to share thoughts. Discussion focused on clarity, completeness, and usefulness, and participants were invited to propose new interaction examples. Next, participants viewed the video (from \Cref{sec:demonstrations}) in two parts: first, the Opening Scene and Productivity; second, Art Exploration. After each part, they shared impressions and perceived usefulness of the examples. Finally, they synthesized all materials, envisioned new interactions, discussed conceptual impacts, and shared their design practices to help reflect on the broader relevance of our framework. Each session lasted 90--120 minutes (\Cref{fig:procedure}).

\subsubsection*{Analysis}
We conducted a reflexive thematic analysis~\cite{braun_using_2006} using transcripts in ATLAS.ti, with audio and video aligned to each sentence to ensure that any ambiguous text could be interpreted with full context. We generated codes and themes both inductively (bottom up) and deductively (top down)~\cite{liu_reality_2025}. Deductively, we coded for aspects directly related to the design space and examples that addressed our guiding questions on clarity, completeness, usefulness, value, and conceptual impact (\Cref{fig:procedure} and supplementary material). Inductively, because the sessions were semi-structured and sometimes expanded into broader discussions (e.g., detailed critiques of other designers' work), we created additional codes for the parts that meaningfully related to our design space. The first author conducted the coding and iteratively refined the codebook (e.g., merging codes), and all codes (N=265) were revisited after a full coding cycle to ensure internal coherence. 

To derive themes, we used hierarchical clustering in ATLAS.ti. We first clustered conceptually related codes (e.g., codes on specific elements) into code groups and then organized them into broader categories (e.g., those related to expanding definitions or bridging digital$/$physical realities). Deductively, we focused the theme development on areas central to our research goals, including how the framework and examples could be used or improved (\Cref{sec:situating-abstraction,sec:reimagining-boundaries,sec:how-to-use}) and experts' design practices (\Cref{sec:design-practice-that-resonates}). Inductively, additional themes emerged unexpectedly (\Cref{sec:reframing-human,sec:toward-deeper-connections}). Through iterative comparison and abstraction, seven themes were consolidated as the highest level that preserved conceptual clarity and distinctiveness. Each theme includes subthemes that capture the finer structure of the underlying categories. 

\subsubsection*{Findings} Below we report the findings, which arise from expert feedback specifically on our design space and examples. While they are not positioned as general theoretical claims, and some of the social intuitions they draw on resonate with prior work, their reinterpretation within cross-boundary MR interaction points toward broader implications that we elaborate in \Cref{sec:discussion}. Although we invited experts to propose new interaction examples using our design space, most responses related closely to their own work or to familiar prior work, which we cite throughout the findings. Quotes were translated where participants mixed Mandarin with English; clicking a quotation (QX) links to the corresponding location in the \Cref{full-quotes}.

\myFinding{Situating Abstraction}{sec:situating-abstraction}
\mySubFinding[nospace]{Expressively Grounding the Design Space}{sec:general-need-context} 
Seeing the design space alone, experts recognized a tension between the framework's conceptual clarity and its expressive resonance. While they found the structure clear and valuable (P2), with a ``manageable level of information''~\q{clarity}, some found it overly general and ``perhaps diluted that sense of spirit''~(P4), suggesting the need to clarify motivation, purpose, and use contexts~(\qs{motivation}, \qs{teleology}, \qs{goal}, \qs{usecase}) and to ground the framework in scenarios, such as factory inspection, home decoration, or art-making, to make it actionable (P4, P5).

After seeing specific scenarios (\Cref{sec:demonstrations}), experts began to understand the framework and shifted from critique to engagement \q{now-understand} and imagination \q{hard-imagine}, especially when interactions were woven into a cohesive vignette, as in Art Exploration~\q{art-bundle}. Interactive Painting and Magic Mirror stood out for their metaphorical clarity~\q{mirror} and expressive potential~\q{fun}, as in the idea that ``behind a mirror lies another world''~\q{mirror-intuitive}. By contrast, examples like Trash Can felt ``created solely to illustrate this interaction''~(P2) and less grounded in everyday practice~\q{remove}, though their realism in implementation was praised~(\qs{marker-real}, \qs{marker-fake-real}).  

A forward design stance emerged (P2, P3, P4): MR should do what reality cannot while remaining logically legible~\q{impossible}. Pure replication risks creative stagnation~\q{haptics}. With AI making content generation more accessible, P4 argued, ``MR's edge for utility alone is not enough \ldots{} I would rather make MR do things that are impractical but delightful, things only MR can do''~(\qs{mr-playful}, \qs{mr-fake}). 

\mySubFinding{Cueing Affordances in Examples}{sec:examples-affordance}
Across the eight examples (\Cref{sec:demonstrations}), experts surfaced a different manifestation of the tension noted in \Cref{sec:general-need-context}: while minimal and ambient aesthetics were valued~\q{minimal-shredder}, minimalism risked suppressing the perceptual cues through which users recognize action possibilities. As one expert noted, ``I can understand this interaction because of your illustration, but visually it does not yet convey an affordance''~(P3). Others similarly asked how interaction availability could be communicated ``in a natural, non-intrusive way''~(P1). This highlighted a need for lightweight legibility: subtle cues that maintain aesthetic restraint while enabling discoverability~\q{quiet-mechanisms}, grounding ambiguity~\cite{gaver_ambiguity_2003} in familiarity and keeping interactions creative yet recognizable~\q{ambiguous-yet-familiar}.

Such legibility operates along a spectrum, from simple static hints to richer micro-behaviors, and some examples already enacted it: the drawer knob signaled hidden content~\q{drawer-handle}; everyday priors made the pairing of a marker and a blank surface naturally suggest writing~\q{prior-marker-wall}; and the rim of the Trash Can read as a passage (P1, P2, P6). Yet experts suggested additional cues: Paper Shredder and Interactive Painting would benefit from visual cues for interactivity (P3), and the seam of the Acoustic Partition could faintly glow to suggest liftability (P1, P3). For penetrable boundaries specifically, meeting expectations is important: when an object is expected to pass through, the boundary should briefly shift in behavior or appearance~\q{bouncing}, such as a drop-and-regrow effect~\q{surface-drop-grow} or a swirling pattern that suggests an inward pull~\q{swirl}.

\mySubFinding{Articulating Affordances in Elements}{sec:elements-affordance}
Experts identified another manifestation of the tension (\Cref{sec:general-need-context}): while they appreciated the categorical value of design space elements, they noted a lack of guidance about their affordances, purposes, or functional potentials~(\qs{framework}, \qs{functionality-highdim}), asking what the elements do in interaction: ``Is the boundary constraining movement, or acting as a display surface?''~(P5, see \Cref{sec:boundary-function}). The Drawer illustrates this need: it naturally implies hiding and revealing, yet experts felt the framework did not articulate this layer of meaning~\q{drawer-hiding-aff}. P4 echoed this, arguing that the significance of ``pull[ing] a drawer from a wall'' lies in an affordance grounded in physical experience~\q{embodied-affordance}.

\begin{figure*}
\centering
\includegraphics[width=\textwidth]{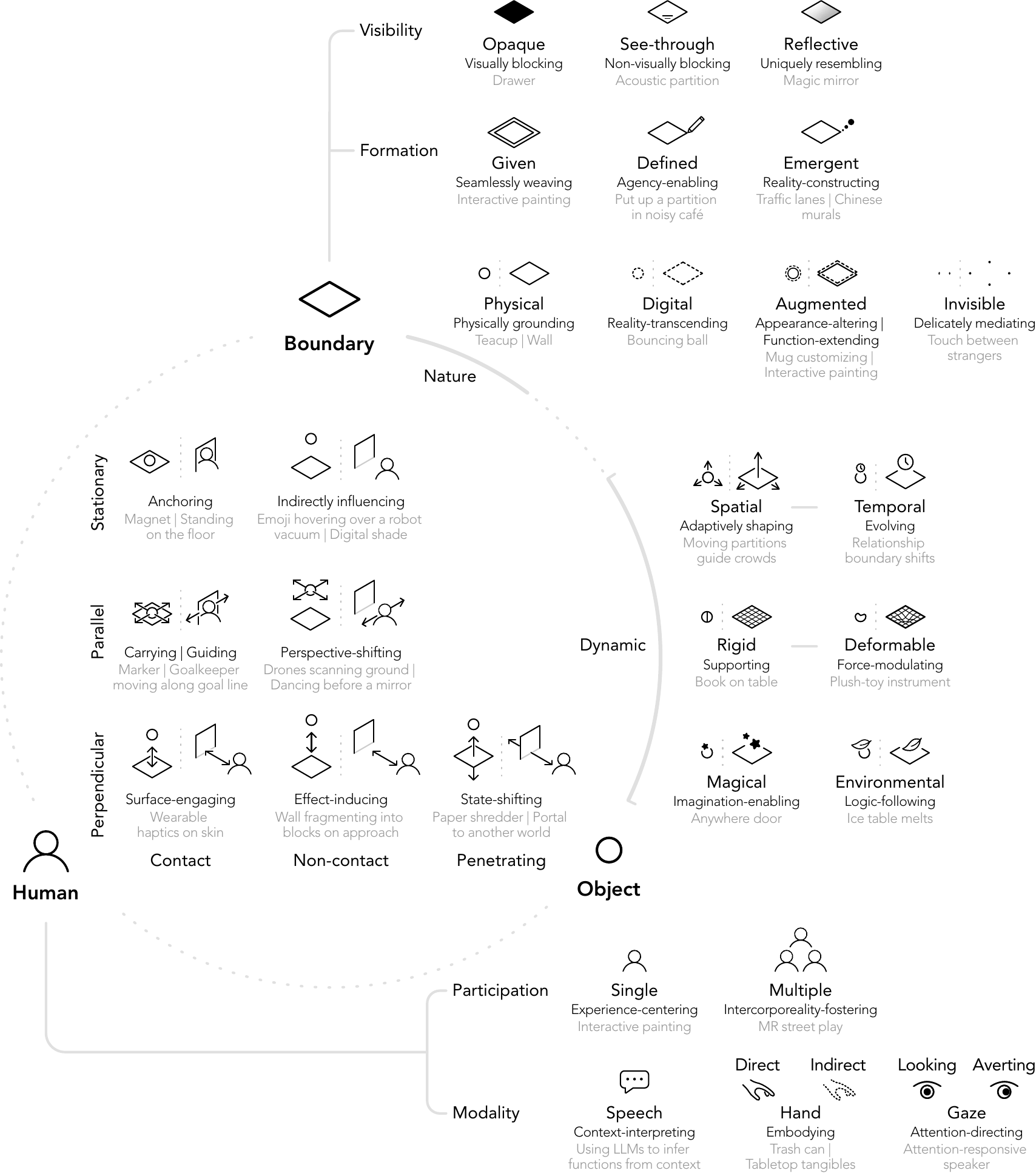}
       \caption{Final design space refined through expert feedback. Following \Cref{sec:situating-abstraction}, each element is presented at three abstraction levels. \Cref{sec:reimagining-boundaries} informs boundary affordances, as well as the inclusion of \dtag{Invisible}, \dtag{Deformable}, and \dtag{Formation}. \Cref{sec:reframing-human} introduces \dtag{Participation}. To reflect \Cref{sec:beyond-visible-ephemeral}, we expanded \dtag{Dynamic} to include \dtag{Spatial}, \dtag{Temporal}, and \dtag{Environmental}, and recategorized movement patterns. \dtag{Speech} and \dtag{Gaze} were added based on \Cref{sec:toward-deeper-connections}. \Cref{sec:design-practice-that-resonates,sec:how-to-use} guide usage.}
        \Description{A structural diagram organizing the design space around three main conceptual nodes: Human (bottom left), Object (bottom right), and Boundary (connecting them via a large dashed arc). The space is visually categorized into several thematic groups. Branching upwards from the top are Visibility (Opaque, See-through, Reflective) and Formation (Given, Defined, Emergent). Positioned to the upper right under ``Nature'' are Physical, Digital, Augmented, and Invisible. Below this, a grid under ``Dynamic'' lists Spatial, Temporal, Rigid, Deformable, Magical, and Environmental. Branching downwards from the Human node are Participation (Single and Multiple) and Modality (Speech, Hand, and Gaze). The center-left occupies a three-by-three matrix detailing relative movement patterns, mapping human/object movement trajectory (Stationary, Parallel, Perpendicular) against boundary interaction states (Contact, Non-contact, Penetrating). Throughout the diagram, every single element illustrates the three abstraction levels mentioned in the caption: a minimalist icon, an abstract functional gerund (e.g., ``Reality-constructing'', ``Agency-enabling''), and concrete contextual examples (e.g., ``Chinese murals'', ``Interactive painting'').}
\label{fig:final-design-space}
\end{figure*}

At the same time, experts emphasized that such functions and purposes are context-dependent~\q{boundary-broad-concept}. To remain generative, what we provide should therefore stay abstract rather than scenario-bound~\q{writing-abstract}, though not so abstract as to lose grounding (see the opening of \Cref{sec:general-need-context}). Experts also cautioned against exhaustively enumerating affordances, as doing so risks fixation~\q{affordance-fixation}. As a middle ground, P1 suggested offering affordances as an optional add-on (\Cref{fig:three-abstractions}), expandable when needed: ``If someone has difficulty generating or applying one, they could expand that attribute to see a non-exhaustive list of possible affordances''~\q{optional-affordance}. 
\begin{center}
    \includegraphics[width=0.404762\textwidth]{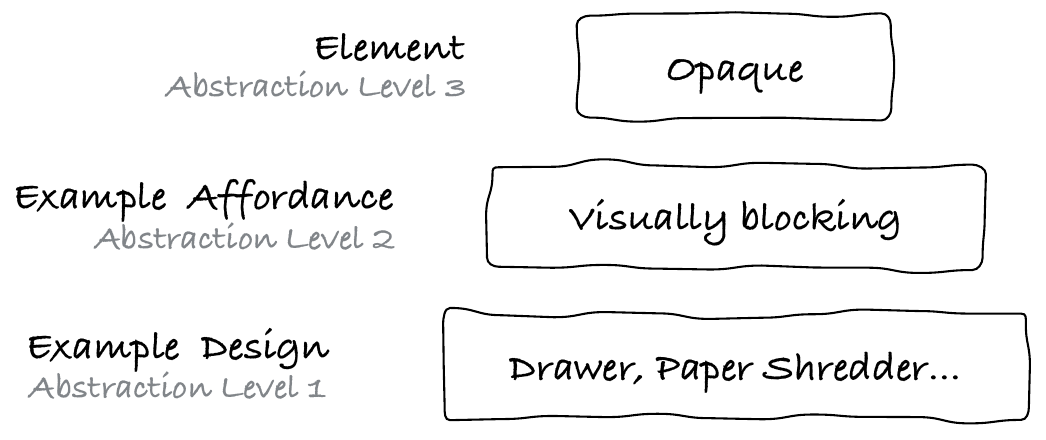}
  \vspace{-1em}
    \captionof{figure}{Three linked abstraction levels illustrated by P1 (using \dtag{Opaque} as an example) to help designers utilize the design space elements. We adopted this structure across the final space (\Cref{fig:final-design-space}).}
    \Description{A conceptual diagram displaying a three-tiered vertical hierarchy using a handwritten font style and loosely drawn rectangular boxes. Descriptive text labels are aligned on the left, paired with corresponding examples enclosed in the boxes on the right. From top to bottom: the highest tier is labeled ``Element'' and ``Abstraction Level 3,'' paired with a box containing the word ``Opaque''; the middle tier is labeled ``Example Affordance'' and ``Abstraction Level 2,'' paired with a box reading ``Visually blocking''; the lowest tier is labeled ``Example Design'' and ``Abstraction Level 1,'' paired with a box reading ``Drawer, Paper Shredder...''.}
  \label{fig:three-abstractions}
\end{center}


\myFinding{Reimagining Boundaries}{sec:reimagining-boundaries}
\mySubFinding[nospace]{Expanding Boundary Definitions}{sec:invisible-main} Boundaries manifest not merely as visible forms but as invisible social implications and the perceptual interplay between the seen and unseen.

\emph{Visible boundaries span diverse forms and materialities.\,}\label[finding]{sec:visble-forms-material}
They extend beyond fixed physical partitions and emerge from how material, spatial, and dynamic conditions shape perception. They vary not only in form (e.g., curved surfaces such as sculptural skins;~\qs{sculptural-surface}) but also in material behaviors, such as deforming (e.g., soft spheres that open when approached~\cite{LiuXin2022PhygitalSupermarket}, or human skin that receives touch;~\qs{skin-haptics}), or transient states such as melting~\q{ice-melts}. They can also cross realities, such as physical boundaries retargeted to provide haptics for digital ones~(P5).

\emph{Invisible boundaries carry social implications.\,}\label[finding]{sec:human-boundary}
They do not merely separate bodies in space but shape how people form, negotiate, and maintain social relationships. The spatial organization of personal space can invite interaction or enforce separation~\q{restaurant-privacy}. Looser boundaries may encourage mingling, yet they also introduce tensions around privacy and trust---particularly in large shared VR or dark environments where people cannot see yet can touch~\q{privacy-touch}. These boundaries are dynamic rather than fixed: ``Between two strangers the invisible boundary is large, but as conversation increases, that boundary shrinks, and people are willing to move closer''~(P5).

\emph{Visible entities can construct invisible boundaries; visible boundaries can also be made perceptually absent.\,}\label[finding]{sec:invisible-visible-sub}
Some marks and objects operate less as direct obstacles than as perceptual cues from which people infer invisible boundaries. Lines on desks (P5) or floors~\q{2d-wall} and the presence of a carpet~\q{carpet-zone} visually generate implicit separations that guide behavior despite offering no physical resistance. 3D structures can also generate fuzzy influence zones~\q{fuzzy-objects} that nudge behavior: ``If I want to dance in a room, the wall influences my behavior even if I never touch it''~(P5). Conversely, physical boundaries may remain but be perceptually absent, such as being visually suppressed in XR~\q{reduced-physical}, or hidden by illusions like Head-on-a-Platter~\cite{ncsm_head_on_a_platter_2022}, where mirrors ``make people believe the space is empty or passable, when it is not''~\q{mirror-illusion}.

\mySubFinding{Unfolding Boundary Functions}{sec:boundary-function}
Beyond restricting~\q{restriction-term}, supporting objects placed on them (P1), and serving as display surfaces for projected content (P5), boundaries can regulate, reveal, and transform.

\emph{Regulating movement and attention.\,}
Boundaries orchestrate how people move, gather, and focus. In dining or exhibition spaces, partitions shape privacy and flow~\q{virtual-wall}. On small surfaces, they guide interaction, such as a tablet that ``guides your hand to slide on this plane''~(P5). Traffic lanes channel movement~(P2,~P5), while a focused beam of light creates a luminous boundary that frames attention~(P5). Boundaries can even embody ``imagined forces''~(P2), like making an object hover~\cite{LiuXin2022PhygitalSupermarket}.

\emph{Concealing and staging information.\,}
Boundaries can act as ``containers''~(P1) that hide or stage content for later retrieval: for instance, in cluttered scenes, people may want to tuck things away and bring them back later~\q{stash-restore}. Alternatively, Paper Shredder-like behaviors suggest ways to permanently remove information from view~\q{old-info}.

\emph{Transforming states across sides.\,}\label[finding]{sec:transform-across-state}
Boundaries can enact any transformation~\q{cross-space}: ``[Place] the object on one side of the boundary and get the transformed result on the other''~(P1). They also define states: whole on one side, fragmented on the other (Paper Shredder, \qs{binary-state}); existent above, absent below (Trash Can, P5); concrete on one side, abstract on the other (Magic Mirror, \qs{trash-mirror}).

\mySubFinding{Constructing Boundaries by Need}{sec:constructing-boundaries} Beyond their forms and functions, the agency to create boundaries is itself an important design dimension: ``I like the idea of creating boundaries [in Acoustic Partition] \ldots{} unlike the other examples where you rely on existing ones''~(P1). Users may thus construct them differently depending on context~\q{define-boundary}, such as drawing lines to constrain movement or divide space~\q{lines-divide}, and sketching a circle to toss objects into~\q{define-throw-trash}.

\myFinding{Reframing Human Experience}{sec:reframing-human}
\mySubFinding[nospace]{Co-Presence}{sec:copresence} We focused on single-person interactions, yet shared agency can heighten impact and reshape meaning. Co-presence introduces intercorporeality~\cite{meyer_intercorporeality_2017}, where seeing and acting together intensify perception and expectation alignment: ``imagine you open and close a drawer, and the person next to you can also open and close it. That shared action becomes very strong. But if the person next to you is looking at a different thing and opens something you did not expect, and you cannot see what they opened, that becomes frustrating''~\q{shared-intensity}. Therefore, design must account for fractured realities: what one perceives as present may not be shared by another in the same scene~\q{fractured-reality}. Spatial heterogeneity~\cite{wong_spatial_2025} across environments compounds this: ``In VR everything is synthetic, so the designer has complete control. In MR the real part is in the user's home, which the designer cannot control''~\q{VR-MR-different}.

\mySubFinding{Culture Imprinting}{sec:culture-imprinting}
Beyond co-presence, societal narratives scaffold comprehension. Global cultural phenomena like Harry Potter normalize magical logics and can be invoked as design material to ease understanding of speculative interactions~\cite{hu_towards_2025}: ``if I give you a wand \ldots{} the only thing you naturally do is swing it down''~\q{magic-wand}.

\mySubFinding{Prior Exposure}{sec:through-not-reasonable}
Experts with extensive experience in TUIs and haptics (P3, P5) showed low tolerance for violations of everyday physics. For the Paper Shredder, P5 read it not as a surface allowing passage but as a ``state division.'' A wall revealing a drawer felt ``inexplicable''~\q{inexplicable} or ``a bit strange''~(P5). When the Interactive Painting expanded into 3D, P5 treated the painting and the extracted digital form as ``two different objects''~\q{two-objects} and questioned how a boundary could ``produce an object''~\q{boundary-produces}. As noted in \Cref{sec:examples-affordance}, affordance cues can clarify these cases: disposal in the Trash Can was acceptable if the surface first ``drops'' to create an opening~\q{surface-drop-grow} or if transitions followed familiar processes such as melting or percolation~\q{ice-melts-holes}.

\mySubFinding{More-Than-Human}{sec:more-than-human}
Although our framework centers on human-led action, experts imagined objects and boundaries acting with their own agency. For example, ``wind moves a curtain that knocks over a flowerpot''~\q{object-agency}. P4 asked what an object might think or do, whether a stone~\cite{hu_autonomous_2025} or a creek~\cite{yang_being_2025}. Boundaries were also envisioned as active, such as deforming to absorb objects~\cite{LiuXin2022PhygitalSupermarket}.

\myFinding{Beyond the Visible and Ephemeral}{sec:beyond-visible-ephemeral}
\mySubFinding[nospace]{Perception Beyond Seeing}{sec:perception-beyond-seeing}
Our initial framing of \dtag{Perception} (\Cref{sec:dimensions}) emphasized fixed positional relations between \dtag{Human} and \dtag{Boundary}, yet feedback highlighted that perception is multi-layered. In terms of gaze, it depends on whether a person is looking and in which direction (P6); as P2 noted, ``looking from one side [of the boundary] reveals a world, but from the other side reveals a different one.'' Spatially, perception extends to whole-body movement~\q{MR-body-movement}: content that ``fades in as I approach and fades out as I leave''~\q{ambient}, floating floor advertisements that move aside when walked through~\cite{Matsuda2010AugmentedCity}, boundaries that appear elusive from afar but reveal portals up close~\cite{deathstranding2_plate_gate_2025}, and even surrounding space and structures that shift with movement~\cite{LiuXin2023ChoreographyArchitecture}. When boundaries are invisible (\Cref{sec:human-boundary}), vision is less relevant~\q{boundary-people-perception} and perception extends to haptics, sound, or scent~\cite{YeseulInvisibleSculptures}.

\mySubFinding{Augmentation Beyond Appearance}{sec:aug-beyond-appearance}
The definition of \dtag{Augmented} extends beyond digital overlays and includes multiple forms: appearance, function, and invisible constructs. Augmentation may be purely visual, as when ``half physical and half digital together form a complete object''~(P6); or functional, as in ``a menu or interface projected on a wall''~(P6). It may also be invisible, often long predating XR: ``In Chinese architecture, murals or painted beams were forms of augmentation. They created a feeling of clouds swirling, of complexity and movement, without needing to construct them physically.'' Western traditions similarly used visual augmentation to intensify spatial experience, aligning frescoes with architectural geometry~\cite{racioppi_st_ignatius_fresco_2026} to evoke divine presence~\q{western-church}.

\mySubFinding{Dynamics Beyond Space and Time}{sec:space-temporal}
\dtag{Dynamic} can be spatial, temporal, or a combination of both~(P5); it can also take deformable (\Cref{sec:visble-forms-material}), magical, or environmental forms. Temporally, boundaries can shift over time, as in human-human boundaries that evolve with contact (\Cref{sec:human-boundary}). Spatially, our framework used only parallel and perpendicular movement (P2, P3), which ``feels a bit two-dimensional''~\q{a-bit-2D}. Experts encouraged richer patterns: angled movement~(P2), ``moving along a curve''~(P1), or varying speeds~(P1). They also highlighted spatio-temporal combinations~\q{chaotically}, as motion often begins from an initial relation that changes over time~\q{initial-motion}. Touch, in particular, differentiates meanings in combined movements: ``Touching the boundary and then moving [the object] in mid-air, versus not touching and directly moving it in mid-air, are two different concepts''~(P3). Finally, P3 noted that \dtag{Dynamic} behaviors may follow environmental processes such as ``ice melting into water, wind causing change, or sunlight shifting over time'' or speculative imaginations, such as the Anywhere Door~\cite{doraemon_anywhere_door_wiki}.

\myFinding{Toward Deeper Connections}{sec:toward-deeper-connections}
\mySubFinding[nospace]{Bridging Physical and Digital Worlds}{sec:physical-digital-worlds}
Experts consistently emphasized that the most compelling interactions emerged when the physical and digital were not merely combined, but mutually implicated (P1, P3--P6). They praised how we ``play with the real--virtual relationship''~(P4) and present interactions in a ``hybrid'' way that feels especially clever~(P3). The most evocative moment was ``pulling a painting out of its physical frame so that the physical and digital spaces begin to interact''~(P6). They valued that this mixing avoids literal mimicry of reality while sustaining presence~\q{mix-presence}.

A key reason this resonance occurs is the anchoring role of everyday physical objects and boundaries. They restore the certainty that mid-air interactions lack through tactile feedback~\q{haptics-value}, making tasks such as drawing or typing more precise~\q{surface-haptics}. Crucially, this value extends beyond tangibility. Repurposing existing surfaces in the environment can translate familiar physical structure into meaningful virtual functionality~\q{repurpose-planes}. They carry familiar spatial logic and continuity that pure virtual surfaces cannot replicate.

Experts also emphasized coherent cross-influences: factors affecting the physical should influence the digital to preserve unified logic~\q{coherent-logic}. They encouraged designs that enact such influences~(P4, P6), such as reducing noise as in Acoustic Partition, extracting real-world materials as augmented textures for chairs~\cite{BaoZhangLiuXin2024UnrealRealm}, or sampling colors from the environment to set smart lamp hues~\cite{wu_megereality_2020}. As P2 observed, ``the biggest issue in XR is that everything is unconstrained, which conflicts with how people expect things to influence one another. Designers need to build in those influences.''

\mySubFinding{Bridging Modalities and Dimensions}{sec:modalities-and-dimensions}
Experts noted that the value of our examples lies in how they preserve embodied, tactile, and logically coherent forms of manipulation. Our focus on hands and direct manipulation was appreciated for experiential richness: ``for {Paper Shredder} the value is in watching the paper gradually shred rather than pressing a delete button \ldots{} which matters wherever the experience matters''~(P1). Experts described this as ``a more embodied interaction''~(P3), valuing gestures that recreate familiar real-world logic~\q{use-hands-real} and emphasizing gesture--interaction coherence, where the motion logic matches people's expectations for the interaction~\q{logic-consistency}. Language input arose in relation to the popularity of large language models (LLMs); while acknowledging their power, experts noted this modality is often inefficient for spatial tasks, where hands remain superior for intuitive positioning and direct control~\q{language-vs-hands}. 

Meanwhile, experts acknowledged that embodied actions are not universally optimal; while delightful and intuitive in one-off cases (e.g., tossing items into the {Trash Can};~\qs{trash-intuitive}), they may become tedious in long-term use, where shortcuts~\q{trash-longterm} or gaze~(\qs{prefer-gaze}, \qs{gaze-tedious}) might be more efficient or less fatiguing. Similarly, some functions, such as noise reduction in the Acoustic Partition, may be better handled automatically by the XR device rather than through explicit interaction~\q{noise-by-device}.

These findings highlight a broader tension: while MR is spatial and inherently 3D, its current interfaces remain dominated by 2D surfaces~\q{buttons-3D-thickness-plane}, which continue to be familiar and easy to learn~\q{planar-familiar}. Yet experts noted that our examples point toward a hybrid future where 3D object interactions and 2D information layers co-exist, with design focusing on orchestrating when to privilege each~\q{2d-and-3d}.

\myFinding{Design Practice that Resonates}{sec:design-practice-that-resonates}
To contextualize the relevance of our framework, we asked experts to reflect on their design practice by describing prior experiences with similar interactions (after viewing our examples) and how they typically approach such designs. These are not necessarily mutually exclusive practices, but situational ways of thinking that surface in different design contexts.
%
\mySubFinding{Relational Abstractions and Self-Consistency}{sec:relational-abs-cons}
Experts reason about interactions as relationships with clear ontologies and consequences. For example, P1 reasoned with image-schematic metaphors~\cite{li_benefits_2024}, asking: ``Is it [the boundary] a container? Is it a surface?'' A container might imply functions such as concealing, while a surface implies displaying~\q{container-surface}.

P2 extended this stance toward topological thinking, asking what a ``pizza-like'' topology might imply in the real world~\q{pizza-topology}. Affirming that a structured design space like ours helps organize these possibilities, the expert noted maintaining a similar mental categorization that considers interaction possibilities among humans, objects, and the environment~\q{relational-system}.

Causality and self-consistency also matter in making these abstractions legible (P2, P3). P3 suggested that novel interaction design should be ``ambiguous but not strange'', and that plausible causes grounded in everyday physics can increase acceptance and creative traction (\Cref{sec:through-not-reasonable,sec:examples-affordance}). P2 further illustrated self-consistency through a piece~\cite{LiuXin2022PhygitalSupermarket} whose motion renders guiding axes, enabling viewers to intuit both path and envelope~\q{elevator-axes}; this consistency enables parametric thinking, where inputs and procedural operations yield legible, logically derived outputs~\q{parametric-io}.

\mySubFinding{Starting from Systems, Constraints, and States}{sec:starts-from-system-and-needs}
Design work sometimes begins from concrete system constraints and functional states rather than abstract frameworks. For example, when working on a specific XR glasses platform that primarily supports overlays (rather than more complex passthrough, as in Meta Quest~3), P6 focused on what was feasible within those capabilities~\q{constrained-explore}. Similarly, in interaction design using a specific headset, P4 said ``I extend my hands to probe the outer limits of its hand-tracking range \ldots{} I identify where the device's tracking begins to break down, then design back from those limits''~\q{design-around-constraint}. Experts cautioned against framework-driven thinking, preferring a needs-first approach over a ``hammer looking for nails''~(\qs{hammer-nail}, \qs{hammer-nail-needs}).

For systems with multiple states or functionalities---such as a smart speaker that reveals different UI layers depending on user distance---P6 emphasized defining these states upfront: ``what states the object has and how they are triggered, and then what each state looks like visually.'' Interaction is then iteratively explored to determine what feels engaging and appropriate. Similarly, when designing interactions for TUI systems, P5 began with functionalities and mapped them to real-world interactions~\cite{wang_push-that-there_2024}, suggesting that functionality guidance in our framework could further accelerate ideation~\q{function-to-usecase}, echoing \Cref{sec:elements-affordance}.

\mySubFinding{Imagination First, Framework Second}{sec:inspiration-instead-of-framework}
For art-led designers (P2, P3, P4, P6), strong creative ideas often emerge from lived observation, media, and embodied experience, with the framework aiding analysis and critique (\qs{practice-based}, \qs{post-hoc-fit}). P2 draws from everyday scenes and synesthetic imagination:
{\begin{quote}
    \textit{When listening to music while walking down the street, I often envision spontaneous visual effects in my surroundings. For example, if all the window lights of a building could turn on and off in sync with the rhythm, it would create a spectacular show. However, given real-world constraints, the most cost-effective way to realize this vision is in XR \ldots{}}
\end{quote}}
Cultural imaginaries (such as magic wands;~\qs{magic-wand}) and embodied metaphors~\q{haptics-metaphor} also serve as inspiration and shared language for interaction design~\cite{hu_towards_2025,zhou_shape-kit_2025}. Experts thus cautioned about the framework's generative power, noting that imagination needs freedom~\q{extreme-expression} and resists rigid taxonomies~\q{synesthetic-imagination}; over-general frameworks can dull the spark that makes interactions compelling~\q{spark-and-sanding}. Great interactions can be decomposed and classified post hoc, but the process is rarely generative in reverse~\q{decompose-not-synthesize}. P4 cited \emph{The Weather Project}~\cite{Eliasson2003WeatherProject} as an example where a work opens vast analytical space, yet analysis cannot algorithmically yield the work~\q{eliasson-reading}.

\myFinding{From Clarity to Composability}{sec:how-to-use}
Experts noted that the design space brought clarity by surfacing structure and sharpening attention to \dtag{Object}--\dtag{Boundary}--\dtag{Human} relations, revealing states~\q{clarity-states} and functionalities~\q{additional-functionality} that had previously remained tacit. For instance, it highlighted motion unfolding along a surface rather than only through collision~\q{along-surface}, and framed \dtag{Crossing} a boundary as a distinct, designable interaction~\q{trash-pass-through}. This clarity, in turn, made the design space usable as compositional material for design.

\mySubFinding{Reinterpret and combine elements}{sec:reinterpret-combine}
\dtag{Boundary} can act as \dtag{Object}~(P4); \dtag{Human} can engage \dtag{Boundary} just as \dtag{Object} does~(P1); and \dtag{Object} itself may embed \dtag{Boundary} in its surface---whether soft, hard, or shape-changeable~\q{object-surface-boundary}. \dtag{Reflective} boundaries can produce \dtag{See-through} illusions (\Cref{sec:invisible-visible-sub}), and different elements (e.g., different motions;~\Cref{sec:space-temporal}) can be combined to form more complex interactions. This combinability enabled designers to rapidly generate concepts: ``By randomly selecting a few elements from each category, you can quickly start brainstorming''~(P3); any two could spark exploration across different focal points~\q{connect-categories}.

\mySubFinding{Keep the framework extensible}{sec:keep-frame-extensible}
Experts suggested that the framework remain extensible so designers can add elements they care about~\q{add-elements} or incorporate new modalities that emerge as technologies mature; gaze, for example, became actionable only once hardware made it reliable~\q{gaze-apple}. At the same time, the scope should remain manageable so the framework stays conceptually tractable~\q{too-complete}.


\mySubFinding{An index of interactions}{para:index}
Experts valued the framework as an index, helping designers anticipate experiential qualities and locate inspirational exemplars: ``Few people think from the perspective of the boundary \ldots{} you offer a valuable index around it. When someone wants to design a similar interaction, they can use your index to understand what experiences it might evoke and check for your existing demos [for inspiration]''~(P1). The index can help designers experiment with variations on a design choice, such as shifting from \dtag{Opaque} to \dtag{See-through} or \dtag{Reflective} to play with different possibilities~\q{index-2}.

\mySubFinding{Examples spark new examples}{sec:cafe-partition}
Magic Mirror inspired auditory transformations: ``On the left side [of the mirror] is ordinary music; on the right, it becomes punk''~\q{magic-mirror-sound}. Acoustic Partition was envisioned for café or studio scenarios where users need quiet focus~(P1, P5), and as a visual ``digital shade'' for users who need to work in low-light environments~\q{digital-shade}. The Trash Can could be applied to deleting unwanted parts in 3D modeling~(P6), while the Marker inspired everyday reminders: ``When leaving home to buy milk, I could leave a note at the door to remind myself not to forget''~(P1). As P1 summarized, ``For each example I can imagine useful scenarios.''

\mySubFinding{Computational support for ideation}{sec:tooling}
Experts envisioned the design space as a foundation for future design tools~\q{ai-function}, describing a system that suggests OBI possibilities from context: ``It does not need to be an authoring tool \ldots{} [but] if I choose factors that I care about, it generates options and proposes interaction choices in a room''~(P3).

\section{Discussion and Conclusion} \label{sec:discussion}
\begin{figure*}
  \centering
\includegraphics[width=0.79\textwidth]{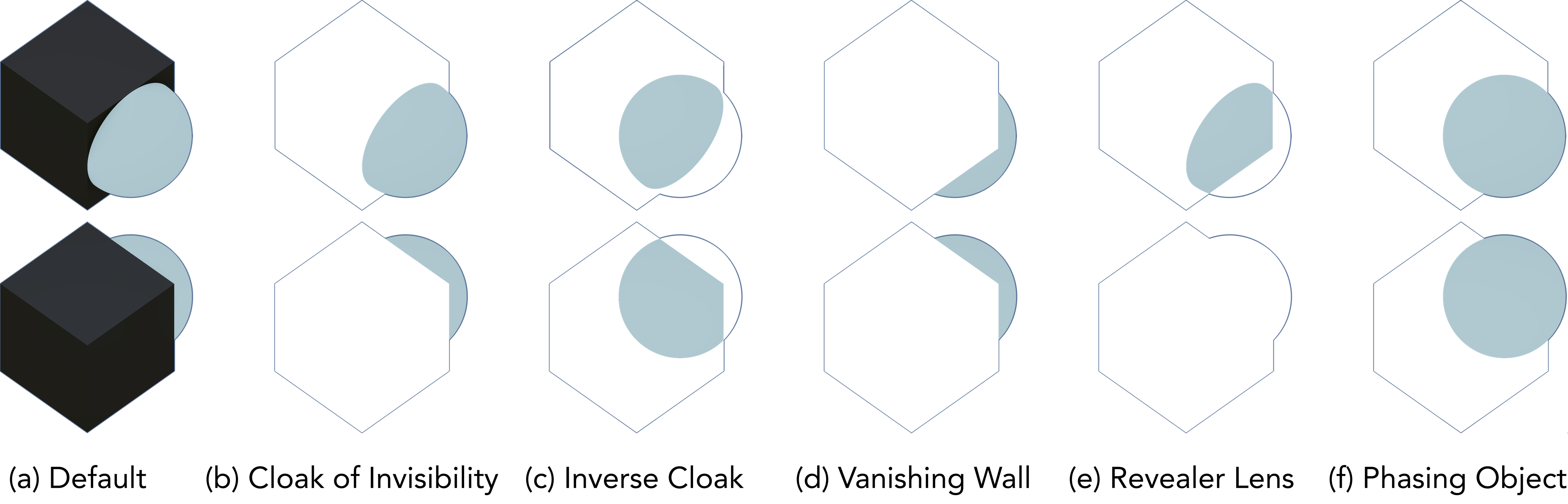}
\vspace{-0.5em}
\caption{Six material effects in Unity, created by varying depth/stencil tests and shader order. Black cube and blue sphere illustrate two intersecting objects; wireframe outlines are shown only for clarity, and blank regions indicate full transparency. (a) embedded portion is invisible; (b) cube is transparent but embedded portion disappears; (c) only embedded portion is visible; (d) cube erases overlapping foreground objects; (e) sphere reveals otherwise invisible cube while leaving normal objects unaffected; (f) object phases through boundaries without occlusion. Examples in \Cref{sec:demonstrations} use (a-c, f). This extends prior work~\cite{elmqvist_taxonomy_2008} by exploring material dimensions alongside geometric relations.}
\Description{A visual sequence demonstrating six different occlusion and visibility effects between two intersecting geometric shapes: a cube and a sphere. The first configuration (a) shows a fully rendered 3D black cube intersecting a light blue sphere. The remaining five configurations (b through f) depict different rendering behaviors, such as making only the embedded volume visible, erasing overlapping sections, or allowing the shapes to visually phase through one another without interference.}
\label{fig:materials}
\end{figure*}

Prior design spaces are often tied to specific technical implementations~\cite{nakagaki_disappearables_2022, follmer_inform_2013, kari_scene_2023} or aim for exhaustive coverage of prior art~\cite{liu_datadancing_2023, rasmussen_shape-changing_2012, wong_spatial_2025}. In contrast, we propose {Unbounded} as a system-agnostic framework focused on interactions that offer experiential richness. We invite the community to explore the space using diverse technical means beyond our prototype and to exercise designerly intentionality by challenging and extending the framework with elements they care about. To facilitate this ongoing dialogue, we share our tools (\Cref{fig:materials,fig:tools}) and the refined space (\Cref{fig:final-design-space}).

\myDiscussion{From Seamless Practicality to Seamful Imagination}
Prior MR and AR work pursues seamlessness through illusions that alter the physical world~\cite{cheng_towards_2022, lee_diminishar_2025, lindlbauer_remixed_2018}, such as masking a chair~\cite{kari_scene_2023} or robot~\cite{kari_reality_2025} with shaders so users perceive them as absent. Suppressing physical reality, however, can disrupt presence~\cite{chandio_investigating_2024} and be ``uncomfortable and disorienting''~\cite{cheng_towards_2022}, as it introduces a tension: humans know two realities exist, while the experience asserts only one. This tension can accumulate over time and undermine trust, as reflected in our more seamless examples: in Paper Shredder and Marker, experts asked whether the paper or pen was real and expressed concern that people may eventually struggle to distinguish the digital from the physical.

Seamful~\cite{chalmers_seamful_2004} MR offers an alternative (\Cref{fig:seamfulMR}). It starts with physical reality and adds digital tightly on top of it, a similar philosophy behind architectural patterns in Chinese temples and Western churches (\Cref{sec:beyond-visible-ephemeral}). Subtle deviations from physical reality (e.g., glow) help maintain this distinction (\Cref{sec:situating-abstraction}) and mitigate the cognitive dissonance caused by violations of physical laws~\cite{ogawa_you_2020} (\Cref{sec:through-not-reasonable}). As current illusions remain detectable and future technology may make them nearly perfect, fully seamless MR---one that collapses realities---risks creating a scenario akin to a ``brain-in-a-vat''~\cite{putnam_reason_1981}, recalling the prisoners in Plato's cave~\cite{grube_republic_1992}. Preserving the integrity of the physical world therefore remains essential.

A recurring tension in MR is whether to prioritize practicality or artistic imagination. Our findings favor the latter, positioning MR less as a tool and more as a medium for speculative exploration (\Cref{sec:general-need-context}). This challenges the dominant framing~\cite{kaeder_working_2024,qiu_marginalia_2025,wang_explainmr_2025} and echoes calls for design as cultural and experiential inquiry~\cite{hu_towards_2025,yang_being_2025,benford_tangles_2025}. Yet, experts cautioned that speculative design cannot disregard usability: even in ``useless'' scenarios, affordances must remain minimal yet sufficient for legibility (\Cref{sec:examples-affordance}). While influenced by our experts' artistic orientation, these insights suggest that MR's value lies less in competing with productivity tools (PCs, phones) than in enabling expressive, non-instrumental experiences that preserve interaction clarity.
\begin{center}
  \includegraphics[width=0.47618\textwidth]{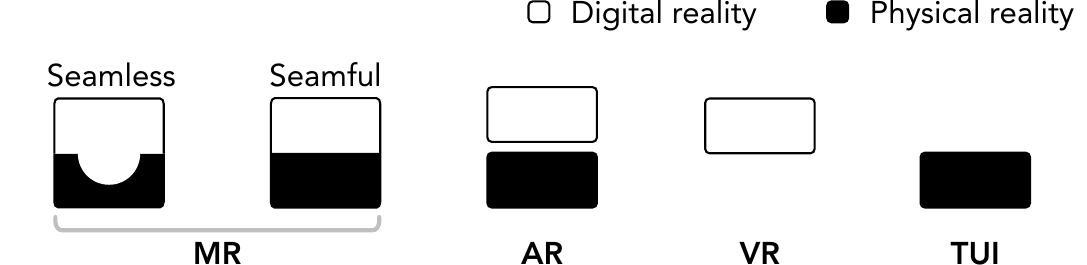}
\captionof{figure}{Seamful MR preserves physical reality, while Seamless MR replaces parts of it with digital reality. Other paradigms are shown through the same lens.}
      \Description{A conceptual diagram using simple geometric shapes to illustrate different reality paradigms. A legend defines outlined white shapes as digital reality and solid black shapes as physical reality. Five configurations are displayed side-by-side. ``Seamless MR'' is depicted as a white shape merging into a black shape with a curved cut-out boundary carved from the black region. ``Seamful MR'' shows the white shape stacked directly on top of the black shape with a strict, flat horizontal dividing line. ``AR'' shows a white rectangle hovering with a distinct spatial gap over a black rectangle. ``VR'' is represented solely by a white rectangle, and ``TUI'' solely by a solid black rectangle.}
  \label{fig:seamfulMR}
\end{center}

\myDiscussion{Invisible Entities and Constructed Realities}
MR has largely emphasized what is visually present or tangible~\cite{ihara_holobots_2023,kari_scene_2023,liu_reality_2025}, a focus we shared. Our findings extend this by foregrounding entities that shape behavior without being visible (\Cref{sec:invisible-main,sec:aug-beyond-appearance}) and by shifting attention from constructing geometry to constructing perception.

\Cref{sec:reimagining-boundaries} indicates that weak boundaries (e.g., a line, a carpet, or a subtle texture) can exert strong behavioral influence. This suggests MR boundaries need not rely on full geometric construction; simple cues (e.g., a digital line on the floor) may suffice. Such weak boundaries may not even take the form of \dtag{Boundary}. Much like how tabletop documents structure personal, group, or storage territories without explicit demarcation~\cite{scott_territoriality_2004}, designers can leverage non-\dtag{Boundary} objects to convey \dtag{Boundary}. In shared MR spaces, users could put up partitions (\Cref{sec:cafe-partition}), but such explicit signaling of blocking access could be socially awkward and cause unintentional exclusion~\cite{onishi_waddlewalls_2022}. Instead, similar to using unrelated hashtags to create online safe spaces~\cite{wan_hashtag_2025}, users might employ uninteresting or unrelated objects; for example, by surrounding themselves with piles of digital documents to signal ``busy,'' or placing culturally coded artifacts (e.g., akin to a \textit{sekimori-ishi}---a Japanese stone marker signifying a closed path) that implicitly communicate ``do not disturb.'' These strategies construct the perception of \dtag{Boundary} without explicit demarcation and return agency to the user: while everyday boundaries are often predefined spatially (e.g., walls, floors) or semantically (e.g., XR safety zones), our findings (\Cref{sec:constructing-boundaries}) highlight the value of letting users co-author the spatial logic of their experience.

Prior work that miniaturizes collaborators onto a tabletop~\cite{kiuchi_minimates_2025} or switches avatar viewpoints~\cite{gronbaek_blended_2024} treats boundaries as rearrangeable \emph{states}. Our findings suggest that interpersonal boundaries behave more like \emph{processes} that evolve with trust and familiarity (\Cref{sec:human-boundary}). Designers can therefore consider how to scaffold this progression, for example, through shared tasks or protocols (e.g., co-wielding wands to defeat an MR dragon~\cite{hu_towards_2025}) that help dissolve boundaries.

This suggests that designers can treat constructed realities as design materials to anticipate and scaffold interactions. Designers should remain cautious, however, since not all physical-world perceptual effects transfer to XR. For example, the doorway effect, where people forget their intentions when passing through a doorway, does not appear to hold in VR~\cite{van_gemert_doorways_2024}.

\myDiscussion{Designing for Human--Physical--Digital Interplay}
We build on work that emphasizes deepening Human--Physical--Digital (HPD) connections~\cite{kari_scene_2023,kari_reality_2025,dupre_tripad_2024,lyu_objestures_2026}, in which coherent cross-influences enable reimagination (\Cref{sec:physical-digital-worlds}). 

One way to deepen such connections is by dissolving functionality into the environment. This approach echoes Weiser's vision that ``the most profound technologies are those that disappear''~\cite{weiser_computer_1999}. It aligns with prior work on displays embedded in everyday surfaces~\cite{olwal_hidden_2022}, information conveyed through ambient cues~\cite{ishii_ambientroom_1998}, and everyday objects repurposed for interaction~\cite{monteiro_teachable_2023,lyu_objestures_2026}. Our experts valued such embedded affordances, yet pointed to the need for a balance between signaling interaction availability and avoiding cognitive overload (\Cref{sec:examples-affordance}).

The choice of modality also shapes HPD interplay. Hand interactions have been widely adopted~\cite{lyu_objestures_2026,kari_reality_2025}, though experts noted the efficiency of alternatives, such as gaze, in the current era where 2D and 3D elements coexist (\Cref{sec:modalities-and-dimensions}). We argue that hands remain central to HPD interplay~\cite{lyu_objestures_2026}, as they afford an embodied experience (\Cref{sec:modalities-and-dimensions}) that will be essential as XR evolves toward fully spatial interfaces.

Importantly, while concrete manifestations evolve, visionary concepts often persist. In the 1990s, researchers recognized the limitations of VR and desktop computing, proposing projection-based tangible manipulation concepts such as the DigitalDesk~\cite{wellner_interacting_1993,wellner_digitaldesk_1991}, which can be viewed as early precursors to MR. In this sense, our Productivity scenario serves as a contemporary extension of this vision, expanding interaction from the desk to the surrounding environment using metaphors that remain familiar, such as shredders operating top-down and trash cans swallowing items.

We thus advocate vision-driven work that prioritizes enduring HPD interplay, echoing Ishii: ``today's technologies will become obsolete in one year, and today's applications will be replaced in 10 years; true visions \ldots{} can last longer than 100 years''~\cite{ishii_sigchi_2019}.

\myDiscussion{Limitations and Future Work}
Our findings are based on six experts engaging in a slide walkthrough and watching video demonstrations, which constitutes a limitation. Hands-on studies with more participants could deepen experiential and embodied understanding. Within-subject experiments could examine how physical objects shape experience~\cite{lyu_objestures_2026}, for example, by comparing the Paper Shredder implemented with a physical shelf versus a digital shelf. Animation strategies, as in Interactive Painting, raise questions about how animation best supports different interactions~\cite{rau_traversing_2025}. 

While {Unbounded} exhibited clarity and descriptive power, its generative potential was more limited than envisioned (\Cref{sec:inspiration-instead-of-framework}). Between-subject workshops could compare how designers employ Unbounded versus free ideation to better understand its generative power and shed light on additional applications~\cite{jeon_sprayable_2025}. Beyond controlled settings, future studies could also investigate social acceptability in public spaces~\cite{rico_usable_2010}.

Crossing boundaries often evokes surprise and indeterminacy, echoing the Exquisite Corpse~\cite{moma_exquisite_corpse_2024}, where each artist continues a drawing while seeing only the edge of the previous section through a fold line (boundary). Similar MR interactions could be imagined: for example, pulling a painting outward could reveal content shaped by the puller's speed or mood, creating a self-generated Exquisite Corpse. Such boundary-crossing activities invite study of the psychological trajectories they evoke, extending work on related experiences such as collision~\cite{benford_tangles_2025}. Research could further probe multi-user immersive speculative enactments~\cite{simeone_immersive_2022,hu_towards_2025}, from which strong concepts~\cite{hook_strong_2012} may emerge. Finally, following suggestions for more-than-human design (\Cref{sec:more-than-human}), Unbounded could explore how objects and boundaries themselves participate in interaction.

\myDiscussion{In Closing}
This work introduced Unbounded, an RtD inquiry into object--boundary interaction in MR. We articulated a design space as a shared language for OBI and demonstrated its potential through eight examples. Expert reflections confirmed its value and expanded its conceptual scope. Together, they highlight how OBI can enrich everyday interactions, inform new practices, and open future directions for MR design that deepen human--physical--digital connections.


\begin{acks}
The authors thank the Associate Chairs and the anonymous reviewers for their thoughtful comments and guidance, the expert participants for their insightful feedback, and the members of the Intelligent Interactive Systems Group for valuable discussions. The first author was supported by the Trinity-Henry Barlow Scholarship and the Cambridge Trust International Scholarship.
\end{acks}

\bibliographystyle{ACM-Reference-Format}
\bibliography{references,references-2}

@misc{ObiCloth_Unity,
  title        = {Obi Cloth},
  author       = {VM Virtual Method},
  howpublished = {Unity Asset Store},
  year         = {2025},
  url          = {https://assetstore.unity.com/packages/tools/physics/obi-cloth-81333}
}

@misc{patrix_Composition8_2023,
  author       = {ARLOOPA},
  title        = {Composition VIII by Wassily Kandinsky [3D model]},
  howpublished = {Fab},
  year         = {2023},
  month        = feb,
  note         = {Standard License},
  url          = {https://fab.com/s/1d0989d6cd39}
}

@misc{AntijnvanderGun_RubiksCube_2016,
  author       = {Shivansh Singh},
  title        = {Rubik's Cube [3D model]},
  howpublished = {Sketchfab},
  year         = {2022},
  month        = jan,
  note         = {CC Attribution license},
  url          = {https://skfb.ly/o8T8S}
}

@misc{spiral_fab,
  author       = {rylertyan1},
  title        = {Spiral [3D model]},
  howpublished = {Sketchfab},
  year         = {2019},
  month        = jan,
  note         = {CC Attribution license},
  url          = {https://skfb.ly/6GzZq}
}

@misc{NHMWien_AmaeaMagnifica_2016,
  author       = {Paolo Chistè},
  title        = {\#10 Amaea magnifica [3D model]},
  howpublished = {Sketchfab},
  year         = {2015},
  month        = dec,
  note         = {CC Attribution license},
  url          = {https://skfb.ly/JoLo}
}

@misc{Poum_WirelessHeadphones_2018,
  author       = {Octaclee},
  title        = {Wireless Headphones [3D model]},
  howpublished = {Sketchfab},
  year         = {2021},
  month        = feb,
  note         = {CC Attribution license},
  url          = {https://skfb.ly/6YXvR}
}

@misc{Edding300Marker_TheLostSock_2022,
  author       = {The Lost Sock},
  title        = {Edding 300 Marker [3D model]},
  howpublished = {Sketchfab},
  year         = {2022},
  month        = sep,
  note         = {CC Attribution license},
  url          = {https://skfb.ly/oywpF}
}

@misc{LiuXin2022PhygitalSupermarket,
  author       = {Liu Xin and Zhu Yuting},
  title        = {Phygital Supermarket: Responsive Architecture in a Vivid Virtual World},
  howpublished = {\url{https://www.liuxin.com/project/phygital-supermarket}},
  year         = {2022},
  note         = {Accessed: 2025-09-11},
}

@misc{meta_pca_unity_2025,
  author       = {{Meta}},
  title        = {Passthrough Camera API for Unity},
  howpublished = {\url{https://developers.meta.com/horizon/documentation/unity/unity-pca-documentation}},
  year         = {2025},
  note         = {Accessed: 2026-01-26}
}

@misc{racioppi_st_ignatius_fresco_2026,
  author       = {Pier Paolo Racioppi},
  title        = {Church of St. Ignatius: {Triumph of St. Ignatius} Ceiling Fresco},
  howpublished = {\url{https://baroqueart.museumwnf.org/database_item.php?id=monument;BAR;it;Mon11;32;en}},
  year         = {2026},
  note         = {Discover Baroque Art, Museum With No Frontiers (MWNF). Accessed: 2026-01-26}
}

@misc{doraemon_anywhere_door_wiki,
  title        = {Anywhere Door},
  author       = {{Doraemon Wiki contributors}},
  howpublished = {\url{https://doraemon.fandom.com/wiki/Anywhere_Door}},
  year         = {2026},
  note         = {Accessed: 2026-01-26}
}

@misc{moma_exquisite_corpse_2024,
  author       = {{The Museum of Modern Art}},
  title        = {How to See an Exquisite Corpse | Surrealism at 100},
  howpublished = {\url{https://youtu.be/B6GuVwN5Ql8}},
  year         = {2024},
  note         = {Accessed: 2026-01-26}
}

@misc{deathstranding2_plate_gate_2025,
  author       = {{ScereBro PSNU}},
  title        = {Sam passes through the plate gate - Death Stranding 2},
  howpublished = {\url{https://youtu.be/x00ASIpz8tQ?t=132}},
  year         = {2025},
  note         = {Accessed: 2026-01-26}
}

@misc{ncsm_head_on_a_platter_2022,
  author       = {{National Council of Science Museums (NCSM)}},
  title        = {Head on a Platter: Magic or Science},
  howpublished = {\url{https://youtu.be/oCDzOR5lY58}},
  year         = {2022},
  note         = {Accessed: 2026-01-26}
}

@misc{LiuXin2023ChoreographyArchitecture,
  author       = {Liu Xin and Candice Wu},
  title        = {Choreography Architecture: Architecture, Body, and Movement},
  howpublished = {\url{https://www.liuxin.com/project/choreography-architecture}},
  year         = {2023},
  note         = {Accessed: 2025-09-11},
}

@misc{Eliasson2003WeatherProject,
  author       = {Olafur Eliasson},
  title        = {The Weather Project},
  howpublished = {\url{https://olafureliasson.net/artwork/the-weather-project-2003/}},
  year         = {2003},
  note         = {Accessed: 2025-09-11},
}

@misc{BaoZhangLiuXin2024UnrealRealm,
  author       = {Aria Xiying Bao and Davide Zhang and Nix Liu Xin},
  title        = {Unreal Realm},
  howpublished = {\url{https://www.ariax.work/home-project/unreal-realm}},
  year         = {2024},
  note         = {Accessed: 2025-09-11}
}

@misc{Matsuda2010AugmentedCity,
  author       = {Keiichi Matsuda},
  title        = {Augmented City},
  howpublished = {\url{https://vimeo.com/14294054}},
  year         = {2010},
  note         = {Design-fiction video; Bartlett School of Architecture (UCL). Accessed: 2025-09-12},
}

@misc{YeseulInvisibleSculptures,
  author       = {Yeseul Song},
  title        = {Invisible Sculptures 1-6},
  howpublished = {\url{https://yeseul.com/Invisible-Sculptures-1-6}},
  year         = {n.d.},
  note         = {Accessed: 2025-09-12},
}

@misc{huang1350fuchun,
  author       = {Huang, Gongwang},
  title        = {Dwelling in the {Fuchun} {Mountains}},
  year         = {1350},
  howpublished = {Handscroll, ink on paper},
  note         = {\url{https://en.wikipedia.org/wiki/Dwelling_in_the_Fuchun_Mountains}},
  address      = {Yuan Dynasty, China}
}

@inproceedings{han_corobos_2025,
	address = {New York, NY, USA},
	series = {{CHI} '25},
	title = {corobos: {A} {Design} for {Mobile} {Robots} {Enabling} {Cooperative} {Transitions} between {Table} and {Wall} {Surfaces}},
	isbn = {979-8-4007-1394-1},
	shorttitle = {corobos},
	url = {https://dl.acm.org/doi/10.1145/3706598.3713440},
	doi = {10.1145/3706598.3713440},
	urldate = {2026-02-21},
	booktitle = {Proceedings of the 2025 {CHI} {Conference} on {Human} {Factors} in {Computing} {Systems}},
	publisher = {Association for Computing Machinery},
	author = {Han, Changyo and Nakagawa, Yosuke and Naemura, Takeshi},
	month = apr,
	year = {2025},
	pages = {1--16},
}

@book{grube_republic_1992,
	series = {Hackett {Classics} {Series}},
	title = {Republic ({Grube} {Edition})},
	isbn = {978-0-87220-136-1},
	publisher = {Hackett Publishing Company},
	author = {Grube, G.M.A. and Reeve, C.D.C.},
	year = {1992},
}

@inproceedings{katins_ad-blocked_2025,
	address = {New York, NY, USA},
	series = {{CHI} '25},
	title = {Ad-{Blocked} {Reality}: {Evaluating} {User} {Perceptions} of {Content} {Blocking} {Concepts} {Using} {Extended} {Reality}},
	isbn = {979-8-4007-1394-1},
	shorttitle = {Ad-{Blocked} {Reality}},
	url = {https://dl.acm.org/doi/10.1145/3706598.3713230},
	doi = {10.1145/3706598.3713230},
	urldate = {2026-02-09},
	booktitle = {Proceedings of the 2025 {CHI} {Conference} on {Human} {Factors} in {Computing} {Systems}},
	publisher = {Association for Computing Machinery},
	author = {Katins, Christopher and Strecker, Jannis and Hinrichs, Jan and Knierim, Pascal and Pfleging, Bastian and Kosch, Thomas},
	month = apr,
	year = {2025},
	pages = {1--18},
}

@article{braun_using_2006,
	title = {Using thematic analysis in psychology},
	volume = {3},
	url = {https://doi.org/10.1191/1478088706qp063oa},
	doi = {10.1191/1478088706qp063oa},
	number = {2},
	journal = {Qualitative Research in Psychology},
	publisher = {Routledge},
	author = {Braun, Virginia and Clarke, Victoria},
	year = {2006},
	pages = {77--101},
}

@inproceedings{ishii_sigchi_2019,
	address = {New York, NY, USA},
	series = {{CHI} {EA} '19},
	title = {{SIGCHI} {Lifetime} {Research} {Award} {Talk}: {Making} {Digital} {Tangible}},
	isbn = {978-1-4503-5971-9},
	shorttitle = {{SIGCHI} {Lifetime} {Research} {Award} {Talk}},
	url = {https://dl.acm.org/doi/10.1145/3290607.3313769},
	doi = {10.1145/3290607.3313769},
	urldate = {2026-02-03},
	booktitle = {Extended {Abstracts} of the 2019 {CHI} {Conference} on {Human} {Factors} in {Computing} {Systems}},
	publisher = {Association for Computing Machinery},
	author = {Ishii, Hiroshi},
	month = may,
	year = {2019},
	pages = {1--4},
}

@inproceedings{wellner_digitaldesk_1991,
	address = {New York, NY, USA},
	series = {{UIST} '91},
	title = {The {DigitalDesk} calculator: tangible manipulation on a desk top display},
	isbn = {978-0-89791-451-2},
	shorttitle = {The {DigitalDesk} calculator},
	url = {https://dl.acm.org/doi/10.1145/120782.120785},
	doi = {10.1145/120782.120785},
	urldate = {2026-02-03},
	booktitle = {Proceedings of the 4th annual {ACM} symposium on {User} interface software and technology},
	publisher = {Association for Computing Machinery},
	author = {Wellner, Pierre},
	month = nov,
	year = {1991},
	pages = {27--33},
}

@book{meyer_intercorporeality_2017,
	title = {Intercorporeality: {Emerging} {Socialities} in {Interaction}},
	isbn = {978-0-19-021046-5},
	shorttitle = {Intercorporeality},
	url = {https://doi.org/10.1093/acprof:oso/9780190210465.001.0001},
	doi = {10.1093/acprof:oso/9780190210465.001.0001},
	urldate = {2025-09-06},
	publisher = {Oxford University Press},
	editor = {Meyer, Christian and Streeck, J and Jordan, J. Scott},
	month = jul,
	year = {2017},
}

@book{dewey_art_1980,
	address = {New York},
	title = {Art as {Experience}},
	publisher = {Perigee Books},
	author = {Dewey, John},
	year = {1980},
	note = {Originally published in 1934 by Minton, Balch \& Company},
}

@book{merleau-ponty_visible_1968,
	address = {Evanston, IL},
	title = {The {Visible} and the {Invisible}},
	publisher = {Northwestern University Press},
	author = {Merleau-Ponty, Maurice},
	editor = {Lefort, Claude},
	translator = {Lingis, Alphonso},
	year = {1968},
}

@article{hu_towards_2025,
	title = {Towards {Immersive} {Mixed} {Reality} {Street} {Play}: {Understanding} {Co}-located {Bodily} {Play} with {See}-through {Head}-{Mounted} {Displays} in {Public} {Spaces}},
	volume = {9},
	shorttitle = {Towards {Immersive} {Mixed} {Reality} {Street} {Play}},
	url = {https://dl.acm.org/doi/10.1145/3757679},
	doi = {10.1145/3757679},
	number = {7},
	urldate = {2026-01-27},
	journal = {Proc. ACM Hum.-Comput. Interact.},
	author = {Hu, Botao Amber and Lin, Rem RunGu and Tao, Yilan Elan and Laato, Samuli and Li, Yue},
	month = oct,
	year = {2025},
	pages = {CSCW498:1--CSCW498:46},
}

@book{putnam_reason_1981,
	series = {Philosophical {Papers}},
	title = {Reason, {Truth} and {History}},
	isbn = {978-0-521-29776-9},
	number = {v. 3},
	publisher = {Cambridge University Press},
	author = {Putnam, H.},
	year = {1981},
	lccn = {81006126},
}

@inproceedings{lyu_objestures_2026,
	address = {New York, NY, USA},
	series = {{CHI} '26},
	title = {Objestures: {Everyday} {Objects} {Meet} {Mid}-{Air} {Gestures} for {Expressive} {Interaction}},
	isbn = {979-8-4007-2278-3},
	shorttitle = {Objestures},
	url = {https://dl.acm.org/doi/10.1145/3772318.3791742},
	doi = {10.1145/3772318.3791742},
	urldate = {2025-05-24},
	booktitle = {Proceedings of the 2026 {CHI} {Conference} on {Human} {Factors} in {Computing} {Systems}},
	publisher = {Association for Computing Machinery},
	author = {Lyu, Zhuoyue and Kristensson, Per Ola},
	month = apr,
	year = {2026},
}

@inproceedings{kiuchi_minimates_2025,
	address = {New York, NY, USA},
	series = {{CHI} '25},
	title = {{MiniMates}: {Miniature} {Avatars} for {AR} {Remote} {Meetings} within {Limited} {Physical} {Spaces}},
	isbn = {979-8-4007-1394-1},
	shorttitle = {{MiniMates}},
	url = {https://dl.acm.org/doi/10.1145/3706598.3714328},
	doi = {10.1145/3706598.3714328},
	urldate = {2025-11-24},
	booktitle = {Proceedings of the 2025 {CHI} {Conference} on {Human} {Factors} in {Computing} {Systems}},
	publisher = {Association for Computing Machinery},
	author = {Kiuchi, Akihiro and Wieland, Jonathan and Igarashi, Takeo and Lindlbauer, David},
	month = apr,
	year = {2025},
	pages = {1--20},
}

@article{chandio_investigating_2024,
	title = {Investigating the {Correlation} {Between} {Presence} and {Reaction} {Time} in {Mixed} {Reality}},
	volume = {30},
	issn = {1941-0506},
	url = {https://ieeexplore.ieee.org/document/10264199/},
	doi = {10.1109/TVCG.2023.3319563},
	number = {9},
	urldate = {2025-11-24},
	journal = {IEEE Transactions on Visualization and Computer Graphics},
	author = {Chandio, Yasra and Bashir, Noman and Interrante, Victoria and Anwar, Fatima M.},
	month = sep,
	year = {2024},
	pages = {5976--5992},
}

@inproceedings{chalmers_seamful_2004,
	address = {Cambridge MA USA},
	title = {Seamful interweaving: heterogeneity in the theory and design of interactive systems},
	isbn = {978-1-58113-787-3},
	shorttitle = {Seamful interweaving},
	url = {https://dl.acm.org/doi/10.1145/1013115.1013149},
	doi = {10.1145/1013115.1013149},
	language = {en},
	urldate = {2025-11-22},
	booktitle = {Proceedings of the 5th conference on {Designing} interactive systems: processes, practices, methods, and techniques},
	publisher = {ACM},
	author = {Chalmers, Matthew and Galani, Areti},
	month = aug,
	year = {2004},
	pages = {243--252},
}

@inproceedings{zhong_ai-assisted_2024,
	address = {Honolulu HI USA},
	title = {{AI}-{Assisted} {Causal} {Pathway} {Diagram} for {Human}-{Centered} {Design}},
	isbn = {979-8-4007-0330-0},
	url = {https://dl.acm.org/doi/10.1145/3613904.3642179},
	doi = {10.1145/3613904.3642179},
	language = {en},
	urldate = {2025-11-11},
	booktitle = {Proceedings of the {CHI} {Conference} on {Human} {Factors} in {Computing} {Systems}},
	publisher = {ACM},
	author = {Zhong, Ruican and Shin, Donghoon and Meza, Rosemary and Klasnja, Predrag and Colusso, Lucas and Hsieh, Gary},
	month = may,
	year = {2024},
	pages = {1--19},
}

@article{paul_milgram_taxonomy_1994,
	title = {A {Taxonomy} of {Mixed} {Reality} {Visual} {Displays}},
	volume = {E77-D},
	number = {12},
	journal = {IEICE TRANSACTIONS on Information},
	author = {Paul MILGRAM, Fumio KISHINO},
	month = dec,
	year = {1994},
	pages = {1321--1329},
}

@article{norman_affordance_1999,
	title = {Affordance, conventions, and design},
	volume = {6},
	issn = {1072-5520},
	url = {https://dl.acm.org/doi/10.1145/301153.301168},
	doi = {10.1145/301153.301168},
	number = {3},
	urldate = {2025-09-12},
	journal = {interactions},
	author = {Norman, Donald A.},
	month = may,
	year = {1999},
	pages = {38--43},
}

@inproceedings{rico_usable_2010,
	address = {New York, NY, USA},
	series = {{CHI} '10},
	title = {Usable gestures for mobile interfaces: evaluating social acceptability},
	isbn = {978-1-60558-929-9},
	shorttitle = {Usable gestures for mobile interfaces},
	url = {https://dl.acm.org/doi/10.1145/1753326.1753458},
	doi = {10.1145/1753326.1753458},
	urldate = {2025-09-09},
	booktitle = {Proceedings of the {SIGCHI} {Conference} on {Human} {Factors} in {Computing} {Systems}},
	publisher = {Association for Computing Machinery},
	author = {Rico, Julie and Brewster, Stephen},
	month = apr,
	year = {2010},
	pages = {887--896},
}

@inproceedings{ishii_ambientroom_1998,
	address = {Los Angeles California USA},
	title = {{ambientROOM}: integrating ambient media with architectural space},
	isbn = {978-1-58113-028-7},
	shorttitle = {{ambientROOM}},
	url = {https://dl.acm.org/doi/10.1145/286498.286652},
	doi = {10.1145/286498.286652},
	language = {en},
	urldate = {2025-09-09},
	booktitle = {{CHI} 98 {Conference} {Summary} on {Human} {Factors} in {Computing} {Systems}},
	publisher = {ACM},
	author = {Ishii, Hiroshi and Wisneski, Craig and Brave, Scott and Dahley, Andrew and Gorbet, Matt and Ullmer, Brygg and Yarin, Paul},
	month = apr,
	year = {1998},
	pages = {173--174},
}

@article{weiser_computer_1999,
	title = {The computer for the 21st century},
	volume = {3},
	issn = {1559-1662},
	url = {https://dl.acm.org/doi/10.1145/329124.329126},
	doi = {10.1145/329124.329126},
	number = {3},
	urldate = {2025-09-09},
	journal = {SIGMOBILE Mob. Comput. Commun. Rev.},
	author = {Weiser, Mark},
	month = jul,
	year = {1999},
	pages = {3--11},
}

@inproceedings{riche_ai-instruments_2025,
	address = {New York, NY, USA},
	series = {{CHI} '25},
	title = {{AI}-{Instruments}: {Embodying} {Prompts} as {Instruments} to {Abstract} \& {Reflect} {Graphical} {Interface} {Commands} as {General}-{Purpose} {Tools}},
	isbn = {979-8-4007-1394-1},
	shorttitle = {{AI}-{Instruments}},
	url = {https://dl.acm.org/doi/10.1145/3706598.3714259},
	doi = {10.1145/3706598.3714259},
	urldate = {2025-09-08},
	booktitle = {Proceedings of the 2025 {CHI} {Conference} on {Human} {Factors} in {Computing} {Systems}},
	publisher = {Association for Computing Machinery},
	author = {Riche, Nathalie and Offenwanger, Anna and Gmeiner, Frederic and Brown, David and Romat, Hugo and Pahud, Michel and Marquardt, Nicolai and Inkpen, Kori and Hinckley, Ken},
	month = apr,
	year = {2025},
	pages = {1--18},
}

@inproceedings{wan_hashtag_2025,
	address = {New York, NY, USA},
	series = {{CHI} '25},
	title = {Hashtag {Re}-{Appropriation} for {Audience} {Control} on {Recommendation}-{Driven} {Social} {Media} {Xiaohongshu} (rednote)},
	isbn = {979-8-4007-1394-1},
	url = {https://dl.acm.org/doi/10.1145/3706598.3713379},
	doi = {10.1145/3706598.3713379},
	urldate = {2025-09-08},
	booktitle = {Proceedings of the 2025 {CHI} {Conference} on {Human} {Factors} in {Computing} {Systems}},
	publisher = {Association for Computing Machinery},
	author = {Wan, Ruyuan and Tong, Lingbo and Knearem, Tiffany and Li, Toby Jia-Jun and Huang, Ting-Hao 'Kenneth' and Wu, Qunfang},
	month = apr,
	year = {2025},
	pages = {1--25},
}

@inproceedings{ihara_holobots_2023,
	address = {San Francisco CA USA},
	title = {{HoloBots}: {Augmenting} {Holographic} {Telepresence} with {Mobile} {Robots} for {Tangible} {Remote} {Collaboration} in {Mixed} {Reality}},
	isbn = {979-8-4007-0132-0},
	shorttitle = {{HoloBots}},
	url = {https://dl.acm.org/doi/10.1145/3586183.3606727},
	doi = {10.1145/3586183.3606727},
	language = {en},
	urldate = {2025-09-08},
	booktitle = {Proceedings of the 36th {Annual} {ACM} {Symposium} on {User} {Interface} {Software} and {Technology}},
	publisher = {ACM},
	author = {Ihara, Keiichi and Faridan, Mehrad and Ichikawa, Ayumi and Kawaguchi, Ikkaku and Suzuki, Ryo},
	month = oct,
	year = {2023},
	pages = {1--12},
}

@inproceedings{wang_explainmr_2025,
	address = {New York, NY, USA},
	series = {{CHI} '25},
	title = {{eXplainMR}: {Generating} {Real}-time {Textual} and {Visual} {eXplanations} to {Facilitate} {UltraSonography} {Learning} in {MR}},
	isbn = {979-8-4007-1394-1},
	shorttitle = {{eXplainMR}},
	url = {https://dl.acm.org/doi/10.1145/3706598.3714015},
	doi = {10.1145/3706598.3714015},
	urldate = {2025-09-07},
	booktitle = {Proceedings of the 2025 {CHI} {Conference} on {Human} {Factors} in {Computing} {Systems}},
	publisher = {Association for Computing Machinery},
	author = {Wang, Jingying and Zhang, Jingjing and Capizzano, Juana Nicoll and Sigakis, Matthew and Wang, Xu and Popov, Vitaliy},
	month = apr,
	year = {2025},
	pages = {1--18},
}

@inproceedings{qiu_marginalia_2025,
	address = {New York, NY, USA},
	series = {{CHI} '25},
	title = {{MaRginalia}: {Enabling} {In}-person {Lecture} {Capturing} and {Note}-taking {Through} {Mixed} {Reality}},
	isbn = {979-8-4007-1394-1},
	shorttitle = {{MaRginalia}},
	url = {https://dl.acm.org/doi/10.1145/3706598.3714065},
	doi = {10.1145/3706598.3714065},
	urldate = {2025-09-07},
	booktitle = {Proceedings of the 2025 {CHI} {Conference} on {Human} {Factors} in {Computing} {Systems}},
	publisher = {Association for Computing Machinery},
	author = {Qiu, Leping and Kim, Erin Seongyoon and Suh, Sangho and Sidenmark, Ludwig and Grossman, Tovi},
	month = apr,
	year = {2025},
	pages = {1--15},
}

@inproceedings{kaeder_working_2024,
	title = {Working with {Mixed} {Reality} in {Public}: {Effects} of {Virtual} {Display} {Layouts} on {Productivity}, {Feeling} of {Safety}, and {Social} {Acceptability}},
	isbn = {979-8-3315-1647-5},
	shorttitle = {Working with {Mixed} {Reality} in {Public}},
	url = {https://www.computer.org/csdl/proceedings-article/ismar/2024/164700a740/22f0jeTKA5W},
	doi = {10.1109/ISMAR62088.2024.00089},
	language = {English},
	urldate = {2025-09-08},
	publisher = {IEEE Computer Society},
	author = {Kaeder, Janne and Vergari, Maurizio and Biener, Verena and Kojić, Tanja and Grubert, Jens and Möller, Sebastian and Antons, Jan-Niklas Voigt},
	month = oct,
	year = {2024},
	pages = {740--748},
}

@inproceedings{kaptelinin_affordances_2012,
	address = {New York, NY, USA},
	series = {{CHI} '12},
	title = {Affordances in {HCI}: toward a mediated action perspective},
	isbn = {978-1-4503-1015-4},
	shorttitle = {Affordances in {HCI}},
	url = {https://dl.acm.org/doi/10.1145/2207676.2208541},
	doi = {10.1145/2207676.2208541},
	urldate = {2025-09-07},
	booktitle = {Proceedings of the {SIGCHI} {Conference} on {Human} {Factors} in {Computing} {Systems}},
	publisher = {Association for Computing Machinery},
	author = {Kaptelinin, Victor and Nardi, Bonnie},
	month = may,
	year = {2012},
	pages = {967--976},
}

@inproceedings{speicher_what_2019,
	address = {New York, NY, USA},
	series = {{CHI} '19},
	title = {What is {Mixed} {Reality}?},
	isbn = {978-1-4503-5970-2},
	url = {https://dl.acm.org/doi/10.1145/3290605.3300767},
	doi = {10.1145/3290605.3300767},
	urldate = {2025-09-07},
	booktitle = {Proceedings of the 2019 {CHI} {Conference} on {Human} {Factors} in {Computing} {Systems}},
	publisher = {Association for Computing Machinery},
	author = {Speicher, Maximilian and Hall, Brian D. and Nebeling, Michael},
	month = may,
	year = {2019},
	pages = {1--15},
}

@inproceedings{wang_push-that-there_2024,
	address = {New York, NY, USA},
	series = {{DIS} '24},
	title = {“{Push}-{That}-{There}”: {Tabletop} {Multi}-robot {Object} {Manipulation} via {Multimodal} '{Object}-level {Instruction}'},
	isbn = {979-8-4007-0583-0},
	shorttitle = {“{Push}-{That}-{There}”},
	url = {https://dl.acm.org/doi/10.1145/3643834.3661542},
	doi = {10.1145/3643834.3661542},
	urldate = {2025-09-07},
	booktitle = {Proceedings of the 2024 {ACM} {Designing} {Interactive} {Systems} {Conference}},
	publisher = {Association for Computing Machinery},
	author = {Wang, Keru and Wang, Zhu and Nakagaki, Ken and Perlin, Ken},
	month = jul,
	year = {2024},
	pages = {2497--2513},
}

@inproceedings{liu_reality_2025,
	address = {Busan, Republic of Korea},
	title = {Reality {Proxy}: {Fluid} {Interactions} with {Real}-{World} {Objects} in {MR} via {Abstract} {Representations}},
	shorttitle = {Reality {Proxy}},
	url = {https://doi.org/10.1145/3746059.3747709},
	doi = {10.1145/3746059.3747709},
	language = {en},
	urldate = {2025-08-05},
	booktitle = {Proceedings of the 38th {Annual} {ACM} {Symposium} on {User} {Interface} {Software} and {Technology}},
	publisher = {ACM},
	author = {Liu, Xiaoan and Jia, Difan and Liu, Xianhao Carton and Gonzalez-Franco, Mar and Zhu-Tian, Chen},
	month = jul,
	year = {2025},
	pages = {1--16},
}

@inproceedings{kari_reality_2025,
	address = {Busan, Republic of Korea},
	title = {Reality {Promises}: {Virtual}-{Physical} {Decoupling} {Illusions} in {Mixed} {Reality} via {Invisible} {Mobile} {Robots}},
	url = {https://doi.org/10.1145/3746059.3747660},
	doi = {10.1145/3746059.3747660},
	language = {en},
	booktitle = {Proceedings of the 38th {Annual} {ACM} {Symposium} on {User} {Interface} {Software} and {Technology}},
	publisher = {ACM},
	author = {Kari, Mohamed},
	year = {2025},
	pages = {1--17},
}

@inproceedings{simeone_immersive_2022,
	address = {New York, NY, USA},
	series = {{CHI} '22},
	title = {Immersive {Speculative} {Enactments}: {Bringing} {Future} {Scenarios} and {Technology} to {Life} {Using} {Virtual} {Reality}},
	isbn = {978-1-4503-9157-3},
	shorttitle = {Immersive {Speculative} {Enactments}},
	url = {https://dl.acm.org/doi/10.1145/3491102.3517492},
	doi = {10.1145/3491102.3517492},
	urldate = {2025-09-05},
	booktitle = {Proceedings of the 2022 {CHI} {Conference} on {Human} {Factors} in {Computing} {Systems}},
	publisher = {Association for Computing Machinery},
	author = {Simeone, Adalberto L. and Cools, Robbe and Depuydt, Stan and Gomes, João Maria and Goris, Piet and Grocott, Joseph and Esteves, Augusto and Gerling, Kathrin},
	month = apr,
	year = {2022},
	pages = {1--20},
}

@inproceedings{wu_megereality_2020,
	address = {Honolulu HI USA},
	title = {"{Megereality}": {Leveraging} {Physical} {Affordances} for {Multi}-{Device} {Gestural} {Interaction} in {Augmented} {Reality}},
	isbn = {978-1-4503-6819-3},
	shorttitle = {"{Megereality}"},
	url = {https://dl.acm.org/doi/10.1145/3334480.3383170},
	doi = {10.1145/3334480.3383170},
	language = {en},
	urldate = {2025-09-03},
	booktitle = {Extended {Abstracts} of the 2020 {CHI} {Conference} on {Human} {Factors} in {Computing} {Systems}},
	publisher = {ACM},
	author = {Wu, Shengzhi and Byrne, Daragh and Steenson, Molly Wright},
	month = apr,
	year = {2020},
	pages = {1--4},
}

@article{benford_tangles_2025,
	title = {Tangles: {Unpacking} {Extended} {Collision} {Experiences} with {Soma} {Trajectories}},
	volume = {32},
	issn = {1073-0516, 1557-7325},
	shorttitle = {Tangles},
	url = {https://dl.acm.org/doi/10.1145/3723875},
	doi = {10.1145/3723875},
	language = {en},
	number = {4},
	urldate = {2025-09-02},
	journal = {ACM Transactions on Computer-Human Interaction},
	author = {Benford, Steve and Garrett, Rachael and Li, Christine and Tennent, Paul and Núñez-Pacheco, Claudia and Kucukyilmaz, Ayse and Tsaknaki, Vasiliki and Höök, Kristina and Caleb-Solly, Praminda and Marshall, Joe and Schneiders, Eike and Popova, Kristina and Afana, Jude},
	month = aug,
	year = {2025},
	pages = {1--34},
}

@inproceedings{beaudouin-lafon_designing_2004,
	address = {New York, NY, USA},
	series = {{AVI} '04},
	title = {Designing interaction, not interfaces},
	isbn = {978-1-58113-867-2},
	url = {https://dl.acm.org/doi/10.1145/989863.989865},
	doi = {10.1145/989863.989865},
	urldate = {2025-08-27},
	booktitle = {Proceedings of the working conference on {Advanced} visual interfaces},
	publisher = {Association for Computing Machinery},
	author = {Beaudouin-Lafon, Michel},
	month = may,
	year = {2004},
	pages = {15--22},
}

@article{kirby_future_2010,
	title = {The {Future} is {Now}: {Diegetic} {Prototypes} and the {Role} of {Popular} {Films} in {Generating} {Real}-world {Technological} {Development}},
	volume = {40},
	issn = {0306-3127},
	shorttitle = {The {Future} is {Now}},
	url = {https://doi.org/10.1177/0306312709338325},
	doi = {10.1177/0306312709338325},
	language = {EN},
	number = {1},
	urldate = {2025-08-28},
	journal = {Social Studies of Science},
	publisher = {SAGE Publications Ltd},
	author = {Kirby, David},
	month = feb,
	year = {2010},
	pages = {41--70},
}

@article{narain_folding_2013,
	title = {Folding and crumpling adaptive sheets},
	volume = {32},
	issn = {0730-0301},
	url = {https://dl.acm.org/doi/10.1145/2461912.2462010},
	doi = {10.1145/2461912.2462010},
	number = {4},
	urldate = {2025-08-26},
	journal = {ACM Trans. Graph.},
	author = {Narain, Rahul and Pfaff, Tobias and O'Brien, James F.},
	month = jul,
	year = {2013},
	pages = {51:1--51:8},
}

@inproceedings{wong_spatial_2025,
	address = {Yokohama Japan},
	title = {Spatial {Heterogeneity} in {Distributed} {Mixed} {Reality} {Collaboration}},
	isbn = {979-8-4007-1394-1},
	url = {https://dl.acm.org/doi/10.1145/3706598.3714033},
	doi = {10.1145/3706598.3714033},
	language = {en},
	urldate = {2025-08-21},
	booktitle = {Proceedings of the 2025 {CHI} {Conference} on {Human} {Factors} in {Computing} {Systems}},
	publisher = {ACM},
	author = {Wong, Emily and Genay, Adélaïde and Grønbæk, Jens Emil Sloth and Velloso, Eduardo},
	month = apr,
	year = {2025},
	pages = {1--19},
}

@article{hook_strong_2012,
	title = {Strong concepts: {Intermediate}-level knowledge in interaction design research},
	volume = {19},
	issn = {1073-0516, 1557-7325},
	shorttitle = {Strong concepts},
	url = {https://dl.acm.org/doi/10.1145/2362364.2362371},
	doi = {10.1145/2362364.2362371},
	language = {en},
	number = {3},
	urldate = {2025-08-21},
	journal = {ACM Transactions on Computer-Human Interaction},
	author = {Höök, Kristina and Löwgren, Jonas},
	month = oct,
	year = {2012},
	pages = {1--18},
}

@inproceedings{yang_being_2025,
	address = {New York, NY, USA},
	series = {{CHI} '25},
	title = {Being {The} {Creek}: {Mobile} {Augmented} {Reality} {Experience} as an {Invitation} for {Exploring} {More}-{Than}-{Human} {Perspectives}},
	isbn = {979-8-4007-1394-1},
	shorttitle = {Being {The} {Creek}},
	url = {https://dl.acm.org/doi/10.1145/3706598.3713713},
	doi = {10.1145/3706598.3713713},
	urldate = {2025-08-20},
	booktitle = {Proceedings of the 2025 {CHI} {Conference} on {Human} {Factors} in {Computing} {Systems}},
	publisher = {Association for Computing Machinery},
	author = {Yang, Yangyang and Ryokai, Kimiko},
	month = apr,
	year = {2025},
	pages = {1--19},
}

@inproceedings{hu_autonomous_2025,
	address = {Aarhus N Denmark},
	title = {Autonomous {Realities}: {A} {Journey} into {Protocolizing} {Digital} {Object} {Permanence} in a {Future} of {Many} {Mixed} {Realities}},
	isbn = {979-8-4007-2003-1},
	shorttitle = {Autonomous {Realities}},
	url = {https://dl.acm.org/doi/10.1145/3744169.3744197},
	doi = {10.1145/3744169.3744197},
	language = {en},
	urldate = {2025-08-20},
	booktitle = {Proceedings of the sixth decennial {Aarhus} conference: {Computing} {X} {Crisis}},
	publisher = {ACM},
	author = {Hu, Botao Amber},
	month = aug,
	year = {2025},
	pages = {290--302},
}

@inproceedings{ma_sensequins_2022,
	address = {Bend OR USA},
	title = {{SenSequins}: {Smart} {Textile} {Using} {3D} {Printed} {Conductive} {Sequins}},
	isbn = {978-1-4503-9320-1},
	shorttitle = {{SenSequins}},
	url = {https://dl.acm.org/doi/10.1145/3526113.3545688},
	doi = {10.1145/3526113.3545688},
	language = {en},
	urldate = {2025-08-17},
	booktitle = {Proceedings of the 35th {Annual} {ACM} {Symposium} on {User} {Interface} {Software} and {Technology}},
	publisher = {ACM},
	author = {Ma, Hua and Yamaoka, Junichi},
	month = oct,
	year = {2022},
	pages = {1--13},
}

@inproceedings{han_parametric_2023,
	address = {New York, NY, USA},
	series = {{UIST} '23},
	title = {Parametric {Haptics}: {Versatile} {Geometry}-based {Tactile} {Feedback} {Devices}},
	isbn = {979-8-4007-0132-0},
	shorttitle = {Parametric {Haptics}},
	url = {https://dl.acm.org/doi/10.1145/3586183.3606766},
	doi = {10.1145/3586183.3606766},
	urldate = {2025-08-17},
	booktitle = {Proceedings of the 36th {Annual} {ACM} {Symposium} on {User} {Interface} {Software} and {Technology}},
	publisher = {Association for Computing Machinery},
	author = {Han, Violet Yinuo and Boadi-Agyemang, Abena and Lin, Yuyu and Lindlbauer, David and Ion, Alexandra},
	month = oct,
	year = {2023},
	pages = {1--13},
}

@inproceedings{yamamura_social_2023,
	address = {Hamburg Germany},
	title = {Social {Digital} {Cyborgs}: {The} {Collaborative} {Design} {Process} of {JIZAI} {ARMS}},
	isbn = {978-1-4503-9421-5},
	shorttitle = {Social {Digital} {Cyborgs}},
	url = {https://dl.acm.org/doi/10.1145/3544548.3581169},
	doi = {10.1145/3544548.3581169},
	language = {en},
	urldate = {2025-08-15},
	booktitle = {Proceedings of the 2023 {CHI} {Conference} on {Human} {Factors} in {Computing} {Systems}},
	publisher = {ACM},
	author = {Yamamura, Nahoko and Uriu, Daisuke and Muramatsu, Mitsuru and Kamiyama, Yusuke and Kashino, Zendai and Sakamoto, Shin and Tanaka, Naoki and Tanigawa, Toma and Onishi, Akiyoshi and Yoshida, Shigeo and Yamanaka, Shunji and Inami, Masahiko},
	month = apr,
	year = {2023},
	pages = {1--19},
}

@inproceedings{lee_diminishar_2025,
	address = {New York, NY, USA},
	series = {{CHI} '25},
	title = {{DiminishAR}: {Diminishing} {Visual} {Distractions} via {Holographic} {AR} {Displays}},
	isbn = {979-8-4007-1394-1},
	shorttitle = {{DiminishAR}},
	url = {https://dl.acm.org/doi/10.1145/3706598.3713415},
	doi = {10.1145/3706598.3713415},
	urldate = {2025-08-10},
	booktitle = {Proceedings of the 2025 {CHI} {Conference} on {Human} {Factors} in {Computing} {Systems}},
	publisher = {Association for Computing Machinery},
	author = {Lee, JangHyeon and Kim, Lawrence H.},
	month = apr,
	year = {2025},
	pages = {1--16},
}

@book{dunne_speculative_2013,
	address = {Cambridge},
	series = {The {MIT} {Press} {Ser}},
	title = {Speculative {Everything}: {Design}, {Fiction}, and {Social} {Dreaming}},
	isbn = {978-0-262-01984-2 978-0-262-31850-1},
	shorttitle = {Speculative {Everything}},
	language = {en},
	publisher = {MIT Press},
	author = {Dunne, Anthony},
	collaborator = {Raby, Fiona},
	year = {2013},
}

@inproceedings{zhou_shape-kit_2025,
	address = {Yokohama Japan},
	title = {Shape-{Kit}: {A} {Design} {Toolkit} for {Crafting} {On}-{Body} {Expressive} {Haptics}},
	isbn = {979-8-4007-1394-1},
	shorttitle = {Shape-{Kit}},
	url = {https://dl.acm.org/doi/10.1145/3706598.3713981},
	doi = {10.1145/3706598.3713981},
	language = {en},
	urldate = {2025-08-01},
	booktitle = {Proceedings of the 2025 {CHI} {Conference} on {Human} {Factors} in {Computing} {Systems}},
	publisher = {ACM},
	author = {Zhou, Ran and Ding, Jianru and Gao, Chenfeng and Qian, Wanli and Erickson, Benjamin and Balaam, Madeline and Leithinger, Daniel and Nakagaki, Ken},
	month = apr,
	year = {2025},
	pages = {1--26},
}

@inproceedings{odom_design_2018,
	address = {Montreal QC Canada},
	title = {On the {Design} of {OLO} {Radio}: {Investigating} {Metadata} as a {Design} {Material}},
	copyright = {https://www.acm.org/publications/policies/copyright\_policy\#Background},
	shorttitle = {On the {Design} of {OLO} {Radio}},
	url = {https://dl.acm.org/doi/10.1145/3173574.3173678},
	doi = {10.1145/3173574.3173678},
	urldate = {2025-07-12},
	booktitle = {Proceedings of the 2018 {CHI} {Conference} on {Human} {Factors} in {Computing} {Systems}},
	publisher = {ACM},
	author = {Odom, William and Duel, Tijs},
	month = apr,
	year = {2018},
	pages = {1--9},
}

@inproceedings{uriu_designing_2025,
	address = {New York, NY, USA},
	series = {{CHI} '25},
	title = {Designing {Virtual} {Funerals} as a {Design} {Fiction}: {A} {Film}-{Based} {Exploration} of {Near}-{Future} {Memorial} {Rituals}},
	isbn = {979-8-4007-1394-1},
	shorttitle = {Designing {Virtual} {Funerals} as a {Design} {Fiction}},
	url = {https://dl.acm.org/doi/10.1145/3706598.3713399},
	doi = {10.1145/3706598.3713399},
	urldate = {2025-07-12},
	booktitle = {Proceedings of the 2025 {CHI} {Conference} on {Human} {Factors} in {Computing} {Systems}},
	publisher = {Association for Computing Machinery},
	author = {Uriu, Daisuke and Arima, Shun},
	month = apr,
	year = {2025},
	pages = {1--19},
}

@inproceedings{follmer_inform_2013,
	address = {New York, NY, USA},
	series = {{UIST} '13},
	title = {{inFORM}: dynamic physical affordances and constraints through shape and object actuation},
	isbn = {978-1-4503-2268-3},
	shorttitle = {{inFORM}},
	url = {https://dl.acm.org/doi/10.1145/2501988.2502032},
	doi = {10.1145/2501988.2502032},
	urldate = {2025-06-30},
	booktitle = {Proceedings of the 26th annual {ACM} symposium on {User} interface software and technology},
	publisher = {Association for Computing Machinery},
	author = {Follmer, Sean and Leithinger, Daniel and Olwal, Alex and Hogge, Akimitsu and Ishii, Hiroshi},
	month = oct,
	year = {2013},
	pages = {417--426},
}

@inproceedings{wallace_making_2013,
	address = {Paris France},
	title = {Making design probes work},
	isbn = {978-1-4503-1899-0},
	url = {https://dl.acm.org/doi/10.1145/2470654.2466473},
	doi = {10.1145/2470654.2466473},
	language = {en},
	urldate = {2025-06-27},
	booktitle = {Proceedings of the {SIGCHI} {Conference} on {Human} {Factors} in {Computing} {Systems}},
	publisher = {ACM},
	author = {Wallace, Jayne and McCarthy, John and Wright, Peter C. and Olivier, Patrick},
	month = apr,
	year = {2013},
	pages = {3441--3450},
}

@inproceedings{holmquist_bits_2023,
	address = {New York, NY, USA},
	series = {{CHI} {EA} '23},
	title = {Bits are {Cheap}, {Atoms} are {Expensive}: {Critiquing} the {Turn} {Towards} {Tangibility} in {HCI}},
	isbn = {978-1-4503-9422-2},
	shorttitle = {Bits are {Cheap}, {Atoms} are {Expensive}},
	url = {https://dl.acm.org/doi/10.1145/3544549.3582744},
	doi = {10.1145/3544549.3582744},
	urldate = {2025-06-11},
	booktitle = {Extended {Abstracts} of the 2023 {CHI} {Conference} on {Human} {Factors} in {Computing} {Systems}},
	publisher = {Association for Computing Machinery},
	author = {Holmquist, Lars Erik},
	month = apr,
	year = {2023},
	pages = {1--8},
}

@book{michotte_perception_2017,
	address = {London},
	title = {The {Perception} of {Causality}},
	isbn = {978-1-315-51905-0},
	doi = {10.4324/9781315519050},
	publisher = {Routledge},
	author = {Michotte, Albert},
	month = mar,
	year = {2017},
}

@book{piaget_construction_2013,
	address = {London},
	title = {The {Construction} {Of} {Reality} {In} {The} {Child}},
	isbn = {978-1-315-00965-0},
	doi = {10.4324/9781315009650},
	publisher = {Routledge},
	author = {Piaget, Jean},
	month = jul,
	year = {2013},
}

@article{li_benefits_2025,
	title = {On the {Benefits} of {Sensorimotor} {Regularities} as {Design} {Constraints} for {Superpower} {Interactions} in {Mixed} {Reality}},
	volume = {31},
	issn = {1941-0506},
	url = {https://ieeexplore.ieee.org/document/10919211/},
	doi = {10.1109/TVCG.2025.3549876},
	number = {5},
	urldate = {2025-06-11},
	journal = {IEEE Transactions on Visualization and Computer Graphics},
	author = {Li, Jingyi and Kristensson, Per Ola},
	month = may,
	year = {2025},
	pages = {2568--2578},
}

@inproceedings{abtahi_im_2019,
	address = {New York, NY, USA},
	series = {{CHI} '19},
	title = {I'm a {Giant}: {Walking} in {Large} {Virtual} {Environments} at {High} {Speed} {Gains}},
	isbn = {978-1-4503-5970-2},
	shorttitle = {I'm a {Giant}},
	url = {https://dl.acm.org/doi/10.1145/3290605.3300752},
	doi = {10.1145/3290605.3300752},
	urldate = {2025-06-11},
	booktitle = {Proceedings of the 2019 {CHI} {Conference} on {Human} {Factors} in {Computing} {Systems}},
	publisher = {Association for Computing Machinery},
	author = {Abtahi, Parastoo and Gonzalez-Franco, Mar and Ofek, Eyal and Steed, Anthony},
	month = may,
	year = {2019},
	pages = {1--13},
}

@inproceedings{pohl_poros_2021,
	address = {New York, NY, USA},
	series = {{CHI} '21},
	title = {Poros: {Configurable} {Proxies} for {Distant} {Interactions} in {VR}},
	isbn = {978-1-4503-8096-6},
	shorttitle = {Poros},
	url = {https://dl.acm.org/doi/10.1145/3411764.3445685},
	doi = {10.1145/3411764.3445685},
	urldate = {2025-06-11},
	booktitle = {Proceedings of the 2021 {CHI} {Conference} on {Human} {Factors} in {Computing} {Systems}},
	publisher = {Association for Computing Machinery},
	author = {Pohl, Henning and Lilija, Klemen and McIntosh, Jess and Hornbæk, Kasper},
	month = may,
	year = {2021},
	pages = {1--12},
}

@inproceedings{onishi_waddlewalls_2022,
	address = {New York, NY, USA},
	series = {{UIST} '22},
	title = {{WaddleWalls}: {Room}-scale {Interactive} {Partitioning} {System} using a {Swarm} of {Robotic} {Partitions}},
	isbn = {978-1-4503-9320-1},
	shorttitle = {{WaddleWalls}},
	url = {https://dl.acm.org/doi/10.1145/3526113.3545615},
	doi = {10.1145/3526113.3545615},
	urldate = {2025-05-28},
	booktitle = {Proceedings of the 35th {Annual} {ACM} {Symposium} on {User} {Interface} {Software} and {Technology}},
	publisher = {Association for Computing Machinery},
	author = {Onishi, Yuki and Takashima, Kazuki and Higashiyama, Shoi and Fujita, Kazuyuki and Kitamura, Yoshifumi},
	month = oct,
	year = {2022},
	pages = {1--15},
}

@inproceedings{pohl_integrated_2024,
	address = {New York, NY, USA},
	series = {{DIS} '24},
	title = {Integrated {Calculators}: {Moving} {Calculation} into the {World}},
	isbn = {979-8-4007-0583-0},
	shorttitle = {Integrated {Calculators}},
	url = {https://dl.acm.org/doi/10.1145/3643834.3661523},
	doi = {10.1145/3643834.3661523},
	urldate = {2025-05-28},
	booktitle = {Proceedings of the 2024 {ACM} {Designing} {Interactive} {Systems} {Conference}},
	publisher = {Association for Computing Machinery},
	author = {Pohl, Henning and Hornbæk, Kasper},
	month = jul,
	year = {2024},
	pages = {343--355},
}

@inproceedings{olwal_hidden_2022,
	address = {New York, NY, USA},
	series = {{CHI} '22},
	title = {Hidden {Interfaces} for {Ambient} {Computing}: {Enabling} {Interaction} in {Everyday} {Materials} through {High}-brightness {Visuals} on {Low}-cost {Matrix} {Displays}},
	isbn = {978-1-4503-9157-3},
	shorttitle = {Hidden {Interfaces} for {Ambient} {Computing}},
	url = {https://dl.acm.org/doi/10.1145/3491102.3517674},
	doi = {10.1145/3491102.3517674},
	urldate = {2025-05-26},
	booktitle = {Proceedings of the 2022 {CHI} {Conference} on {Human} {Factors} in {Computing} {Systems}},
	publisher = {Association for Computing Machinery},
	author = {Olwal, Alex and Dementyev, Artem},
	month = apr,
	year = {2022},
	pages = {1--20},
}

@inproceedings{rau_traversing_2025,
	address = {New York, NY, USA},
	series = {{CHI} '25},
	title = {Traversing {Dual} {Realities}: {Investigating} {Techniques} for {Transitioning} {3D} {Objects} between {Desktop} and {Augmented} {Reality} {Environments}},
	isbn = {979-8-4007-1394-1},
	shorttitle = {Traversing {Dual} {Realities}},
	url = {https://dl.acm.org/doi/10.1145/3706598.3713949},
	doi = {10.1145/3706598.3713949},
	urldate = {2025-05-24},
	booktitle = {Proceedings of the 2025 {CHI} {Conference} on {Human} {Factors} in {Computing} {Systems}},
	publisher = {Association for Computing Machinery},
	author = {Rau, Tobias and Isenberg, Tobias and Koehn, Andreas and Sedlmair, Michael and Lee, Benjamin},
	month = apr,
	year = {2025},
	pages = {1--16},
}

@inproceedings{jeon_sprayable_2025,
	address = {New York, NY, USA},
	series = {{CHI} '25},
	title = {Sprayable {Sound}: {Exploring} the {Experiential} and {Design} {Potential} of {Physically} {Spraying} {Sound} {Interaction}},
	isbn = {979-8-4007-1394-1},
	shorttitle = {Sprayable {Sound}},
	url = {https://dl.acm.org/doi/10.1145/3706598.3713786},
	doi = {10.1145/3706598.3713786},
	urldate = {2025-04-28},
	booktitle = {Proceedings of the 2025 {CHI} {Conference} on {Human} {Factors} in {Computing} {Systems}},
	publisher = {Association for Computing Machinery},
	author = {Jeon, Jongik and Lee, Chang Hee},
	month = apr,
	year = {2025},
	pages = {1--17},
}

@inproceedings{van_gemert_doorways_2024,
	address = {New York, NY, USA},
	series = {{CHI} '24},
	title = {Doorways {Do} {Not} {Always} {Cause} {Forgetting}: {Studying} the {Effect} of {Locomotion} {Technique} and {Doorway} {Visualization} in {Virtual} {Reality}},
	isbn = {979-8-4007-0330-0},
	shorttitle = {Doorways {Do} {Not} {Always} {Cause} {Forgetting}},
	url = {https://dl.acm.org/doi/10.1145/3613904.3642879},
	doi = {10.1145/3613904.3642879},
	urldate = {2025-04-05},
	booktitle = {Proceedings of the 2024 {CHI} {Conference} on {Human} {Factors} in {Computing} {Systems}},
	publisher = {Association for Computing Machinery},
	author = {van Gemert, Thomas and Chew, Sean and Kalaitzoglou, Yiannis and Bergström, Joanna},
	month = may,
	year = {2024},
	pages = {1--13},
}

@inproceedings{ogawa_you_2020,
	address = {New York, NY, USA},
	series = {{CHI} '20},
	title = {Do {You} {Feel} {Like} {Passing} {Through} {Walls}?: {Effect} of {Self}-{Avatar} {Appearance} on {Facilitating} {Realistic} {Behavior} in {Virtual} {Environments}},
	isbn = {978-1-4503-6708-0},
	shorttitle = {Do {You} {Feel} {Like} {Passing} {Through} {Walls}?},
	url = {https://dl.acm.org/doi/10.1145/3313831.3376562},
	doi = {10.1145/3313831.3376562},
	urldate = {2025-04-05},
	booktitle = {Proceedings of the 2020 {CHI} {Conference} on {Human} {Factors} in {Computing} {Systems}},
	publisher = {Association for Computing Machinery},
	author = {Ogawa, Nami and Narumi, Takuji and Kuzuoka, Hideaki and Hirose, Michitaka},
	month = apr,
	year = {2020},
	pages = {1--14},
}

@inproceedings{muller_baselase_2015,
	address = {New York, NY, USA},
	series = {{CHI} '15},
	title = {{BaseLase}: {An} {Interactive} {Focus}+{Context} {Laser} {Floor}},
	isbn = {978-1-4503-3145-6},
	shorttitle = {{BaseLase}},
	url = {https://dl.acm.org/doi/10.1145/2702123.2702246},
	doi = {10.1145/2702123.2702246},
	urldate = {2025-04-05},
	booktitle = {Proceedings of the 33rd {Annual} {ACM} {Conference} on {Human} {Factors} in {Computing} {Systems}},
	publisher = {Association for Computing Machinery},
	author = {Müller, Jörg and Eberle, Dieter and Schmidt, Constantin},
	month = apr,
	year = {2015},
	pages = {3869--3878},
}

@inproceedings{dollinger_are_2023,
	address = {New York, NY, USA},
	series = {{CHI} '23},
	title = {Are {Embodied} {Avatars} {Harmful} to our {Self}-{Experience}? {The} {Impact} of {Virtual} {Embodiment} on {Body} {Awareness}},
	isbn = {978-1-4503-9421-5},
	shorttitle = {Are {Embodied} {Avatars} {Harmful} to our {Self}-{Experience}?},
	url = {https://dl.acm.org/doi/10.1145/3544548.3580918},
	doi = {10.1145/3544548.3580918},
	urldate = {2025-04-05},
	booktitle = {Proceedings of the 2023 {CHI} {Conference} on {Human} {Factors} in {Computing} {Systems}},
	publisher = {Association for Computing Machinery},
	author = {Döllinger, Nina and Wolf, Erik and Botsch, Mario and Latoschik, Marc Erich and Wienrich, Carolin},
	month = apr,
	year = {2023},
	pages = {1--14},
}

@inproceedings{dupre_tripad_2024,
	address = {New York, NY, USA},
	series = {{CHI} '24},
	title = {{TriPad}: {Touch} {Input} in {AR} on {Ordinary} {Surfaces} with {Hand} {Tracking} {Only}},
	isbn = {979-8-4007-0330-0},
	url = {https://doi.org/10.1145/3613904.3642323},
	doi = {10.1145/3613904.3642323},
	booktitle = {Proceedings of the {CHI} {Conference} on {Human} {Factors} in {Computing} {Systems}},
	publisher = {Association for Computing Machinery},
	author = {Dupré, Camille and Appert, Caroline and Rey, Stéphanie and Saidi, Houssem and Pietriga, Emmanuel},
	year = {2024},
}

@inproceedings{jacob_reality-based_2008,
	address = {New York, NY, USA},
	series = {{CHI} '08},
	title = {Reality-based interaction: a framework for post-{WIMP} interfaces},
	isbn = {978-1-60558-011-1},
	url = {https://doi.org/10.1145/1357054.1357089},
	doi = {10.1145/1357054.1357089},
	booktitle = {Proceedings of the {SIGCHI} {Conference} on {Human} {Factors} in {Computing} {Systems}},
	publisher = {Association for Computing Machinery},
	author = {Jacob, Robert J.K. and Girouard, Audrey and Hirshfield, Leanne M. and Horn, Michael S. and Shaer, Orit and Solovey, Erin Treacy and Zigelbaum, Jamie},
	year = {2008},
	pages = {201--210},
}

@inproceedings{ishii_tangible_1997,
	address = {New York, NY, USA},
	series = {{CHI} '97},
	title = {Tangible bits: towards seamless interfaces between people, bits and atoms},
	isbn = {0-89791-802-9},
	url = {https://doi.org/10.1145/258549.258715},
	doi = {10.1145/258549.258715},
	booktitle = {Proceedings of the {ACM} {SIGCHI} {Conference} on {Human} {Factors} in {Computing} {Systems}},
	publisher = {Association for Computing Machinery},
	author = {Ishii, Hiroshi and Ullmer, Brygg},
	year = {1997},
	pages = {234--241},
}

@inproceedings{gong_affordance-based_2023,
	address = {New York, NY, USA},
	series = {{DIS} '23},
	title = {Affordance-{Based} and {User}-{Defined} {Gestures} for {Spatial} {Tangible} {Interaction}},
	isbn = {978-1-4503-9893-0},
	url = {https://doi.org/10.1145/3563657.3596032},
	doi = {10.1145/3563657.3596032},
	booktitle = {Proceedings of the 2023 {ACM} {Designing} {Interactive} {Systems} {Conference}},
	publisher = {Association for Computing Machinery},
	author = {Gong, Weilun and Santosa, Stephanie and Grossman, Tovi and Glueck, Michael and Clarke, Daniel and Lai, Frances},
	year = {2023},
	pages = {1500--1514},
}

@inproceedings{gaver_ambiguity_2003,
	address = {New York, NY, USA},
	series = {{CHI} '03},
	title = {Ambiguity as a resource for design},
	isbn = {978-1-58113-630-2},
	url = {https://dl.acm.org/doi/10.1145/642611.642653},
	doi = {10.1145/642611.642653},
	urldate = {2025-03-21},
	booktitle = {Proceedings of the {SIGCHI} {Conference} on {Human} {Factors} in {Computing} {Systems}},
	publisher = {Association for Computing Machinery},
	author = {Gaver, William W. and Beaver, Jacob and Benford, Steve},
	month = apr,
	year = {2003},
	pages = {233--240},
}

@inproceedings{lin_throwio_2023,
	address = {New York, NY, USA},
	series = {{CHI} '23},
	title = {{ThrowIO}: {Actuated} {TUIs} that {Facilitate} “{Throwing} and {Catching}” {Spatial} {Interaction} with {Overhanging} {Mobile} {Wheeled} {Robots}},
	isbn = {978-1-4503-9421-5},
	shorttitle = {{ThrowIO}},
	url = {https://dl.acm.org/doi/10.1145/3544548.3581267},
	doi = {10.1145/3544548.3581267},
	urldate = {2025-02-26},
	booktitle = {Proceedings of the 2023 {CHI} {Conference} on {Human} {Factors} in {Computing} {Systems}},
	publisher = {Association for Computing Machinery},
	author = {Lin, Ting-Han and Yang, Willa Yunqi and Nakagaki, Ken},
	month = apr,
	year = {2023},
	pages = {1--17},
}

@inproceedings{yu_aerorigui_2023,
	address = {New York, NY, USA},
	series = {{CHI} '23},
	title = {{AeroRigUI}: {Actuated} {TUIs} for {Spatial} {Interaction} using {Rigging} {Swarm} {Robots} on {Ceilings} in {Everyday} {Space}},
	isbn = {978-1-4503-9421-5},
	shorttitle = {{AeroRigUI}},
	url = {https://dl.acm.org/doi/10.1145/3544548.3581437},
	doi = {10.1145/3544548.3581437},
	urldate = {2025-02-26},
	booktitle = {Proceedings of the 2023 {CHI} {Conference} on {Human} {Factors} in {Computing} {Systems}},
	publisher = {Association for Computing Machinery},
	author = {Yu, Lilith and Gao, Chenfeng and Wu, David and Nakagaki, Ken},
	month = apr,
	year = {2023},
	pages = {1--18},
}

@inproceedings{lindlbauer_remixed_2018,
	address = {New York, NY, USA},
	series = {{CHI} '18},
	title = {Remixed {Reality}: {Manipulating} {Space} and {Time} in {Augmented} {Reality}},
	isbn = {978-1-4503-5620-6},
	shorttitle = {Remixed {Reality}},
	url = {https://dl.acm.org/doi/10.1145/3173574.3173703},
	doi = {10.1145/3173574.3173703},
	urldate = {2025-02-24},
	booktitle = {Proceedings of the 2018 {CHI} {Conference} on {Human} {Factors} in {Computing} {Systems}},
	publisher = {Association for Computing Machinery},
	author = {Lindlbauer, David and Wilson, Andy D.},
	month = apr,
	year = {2018},
	pages = {1--13},
}

@inproceedings{zhu_bishare_2020,
	address = {New York, NY, USA},
	series = {{CHI} '20},
	title = {{BISHARE}: {Exploring} {Bidirectional} {Interactions} {Between} {Smartphones} and {Head}-{Mounted} {Augmented} {Reality}},
	isbn = {978-1-4503-6708-0},
	shorttitle = {{BISHARE}},
	url = {https://dl.acm.org/doi/10.1145/3313831.3376233},
	doi = {10.1145/3313831.3376233},
	urldate = {2025-02-15},
	booktitle = {Proceedings of the 2020 {CHI} {Conference} on {Human} {Factors} in {Computing} {Systems}},
	publisher = {Association for Computing Machinery},
	author = {Zhu, Fengyuan and Grossman, Tovi},
	month = apr,
	year = {2020},
	pages = {1--14},
}

@inproceedings{liu_datadancing_2023,
	address = {New York, NY, USA},
	series = {{CHI} '23},
	title = {{DataDancing}: {An} {Exploration} of the {Design} {Space} {For} {Visualisation} {View} {Management} for {3D} {Surfaces} and {Spaces}},
	isbn = {978-1-4503-9421-5},
	shorttitle = {{DataDancing}},
	url = {https://dl.acm.org/doi/10.1145/3544548.3580827},
	doi = {10.1145/3544548.3580827},
	urldate = {2025-02-15},
	booktitle = {Proceedings of the 2023 {CHI} {Conference} on {Human} {Factors} in {Computing} {Systems}},
	publisher = {Association for Computing Machinery},
	author = {Liu, Jiazhou and Ens, Barrett and Prouzeau, Arnaud and Smiley, Jim and Nixon, Isobel Kara and Goodwin, Sarah and Dwyer, Tim},
	month = apr,
	year = {2023},
	pages = {1--17},
}

@inproceedings{kari_scene_2023,
	address = {San Francisco CA USA},
	title = {Scene {Responsiveness} for {Visuotactile} {Illusions} in {Mixed} {Reality}},
	isbn = {979-8-4007-0132-0},
	url = {https://dl.acm.org/doi/10.1145/3586183.3606825},
	doi = {10.1145/3586183.3606825},
	language = {en},
	urldate = {2025-02-14},
	booktitle = {Proceedings of the 36th {Annual} {ACM} {Symposium} on {User} {Interface} {Software} and {Technology}},
	publisher = {ACM},
	author = {Kari, Mohamed and Schütte, Reinhard and Sodhi, Raj},
	month = oct,
	year = {2023},
	pages = {1--15},
}

@inproceedings{monteiro_teachable_2023,
	address = {New York, NY, USA},
	series = {{CHI} '23},
	title = {Teachable {Reality}: {Prototyping} {Tangible} {Augmented} {Reality} with {Everyday} {Objects} by {Leveraging} {Interactive} {Machine} {Teaching}},
	isbn = {978-1-4503-9421-5},
	shorttitle = {Teachable {Reality}},
	url = {https://dl.acm.org/doi/10.1145/3544548.3581449},
	doi = {10.1145/3544548.3581449},
	urldate = {2025-02-13},
	booktitle = {Proceedings of the 2023 {CHI} {Conference} on {Human} {Factors} in {Computing} {Systems}},
	publisher = {Association for Computing Machinery},
	author = {Monteiro, Kyzyl and Vatsal, Ritik and Chulpongsatorn, Neil and Parnami, Aman and Suzuki, Ryo},
	month = apr,
	year = {2023},
	pages = {1--15},
}

@inproceedings{tsai_gait_2024,
	address = {New York, NY, USA},
	series = {{UIST} '24},
	title = {Gait {Gestures}: {Examining} {Stride} and {Foot} {Strike} {Variation} as an {Input} {Method} {While} {Walking}},
	isbn = {979-8-4007-0628-8},
	shorttitle = {Gait {Gestures}},
	url = {https://dl.acm.org/doi/10.1145/3654777.3676342},
	doi = {10.1145/3654777.3676342},
	urldate = {2025-02-13},
	booktitle = {Proceedings of the 37th {Annual} {ACM} {Symposium} on {User} {Interface} {Software} and {Technology}},
	publisher = {Association for Computing Machinery},
	author = {Tsai, Ching-Yi and Yen, Ryan and Kim, Daekun and Vogel, Daniel},
	month = oct,
	year = {2024},
	pages = {1--16},
}

@inproceedings{gronbaek_blended_2024,
	address = {New York, NY, USA},
	series = {{CHI} '24},
	title = {Blended {Whiteboard}: {Physicality} and {Reconfigurability} in {Remote} {Mixed} {Reality} {Collaboration}},
	isbn = {979-8-4007-0330-0},
	shorttitle = {Blended {Whiteboard}},
	url = {https://dl.acm.org/doi/10.1145/3613904.3642293},
	doi = {10.1145/3613904.3642293},
	urldate = {2025-02-13},
	booktitle = {Proceedings of the 2024 {CHI} {Conference} on {Human} {Factors} in {Computing} {Systems}},
	publisher = {Association for Computing Machinery},
	author = {Grønbæk, Jens Emil Sloth and Sánchez Esquivel, Juan and Leiva, Germán and Velloso, Eduardo and Gellersen, Hans and Pfeuffer, Ken},
	month = may,
	year = {2024},
	pages = {1--16},
}

@inproceedings{hendriks_undertable_2024,
	address = {New York, NY, USA},
	series = {{DIS} '24},
	title = {The {Undertable}: {A} {Design} {Remake} of the {Mediated} {Body}},
	isbn = {979-8-4007-0583-0},
	shorttitle = {The {Undertable}},
	url = {https://dl.acm.org/doi/10.1145/3643834.3660698},
	doi = {10.1145/3643834.3660698},
	urldate = {2025-02-12},
	booktitle = {Proceedings of the 2024 {ACM} {Designing} {Interactive} {Systems} {Conference}},
	publisher = {Association for Computing Machinery},
	author = {Hendriks, Sjoerd and Gamboa, Mafalda and Obaid, Mohammad},
	month = jul,
	year = {2024},
	pages = {2591--2610},
}

@inproceedings{rasmussen_shape-changing_2012,
	address = {New York, NY, USA},
	series = {{CHI} '12},
	title = {Shape-changing interfaces: a review of the design space and open research questions},
	isbn = {978-1-4503-1015-4},
	shorttitle = {Shape-changing interfaces},
	url = {https://dl.acm.org/doi/10.1145/2207676.2207781},
	doi = {10.1145/2207676.2207781},
	urldate = {2025-02-12},
	booktitle = {Proceedings of the {SIGCHI} {Conference} on {Human} {Factors} in {Computing} {Systems}},
	publisher = {Association for Computing Machinery},
	author = {Rasmussen, Majken K. and Pedersen, Esben W. and Petersen, Marianne G. and Hornbæk, Kasper},
	month = may,
	year = {2012},
	pages = {735--744},
}

@inproceedings{jones_illumiroom_2013,
	address = {New York, NY, USA},
	series = {{CHI} '13},
	title = {{IllumiRoom}: peripheral projected illusions for interactive experiences},
	isbn = {978-1-4503-1899-0},
	shorttitle = {{IllumiRoom}},
	url = {https://dl.acm.org/doi/10.1145/2470654.2466112},
	doi = {10.1145/2470654.2466112},
	urldate = {2025-02-10},
	booktitle = {Proceedings of the {SIGCHI} {Conference} on {Human} {Factors} in {Computing} {Systems}},
	publisher = {Association for Computing Machinery},
	author = {Jones, Brett R. and Benko, Hrvoje and Ofek, Eyal and Wilson, Andrew D.},
	month = apr,
	year = {2013},
	pages = {869--878},
}

@inproceedings{liu_virtual_2022,
	address = {New York, NY, USA},
	series = {{CHI} '22},
	title = {Virtual {Transcendent} {Dream}: {Empowering} {People} through {Embodied} {Flying} in {Virtual} {Reality}},
	isbn = {978-1-4503-9157-3},
	shorttitle = {Virtual {Transcendent} {Dream}},
	url = {https://dl.acm.org/doi/10.1145/3491102.3517677},
	doi = {10.1145/3491102.3517677},
	urldate = {2025-02-10},
	booktitle = {Proceedings of the 2022 {CHI} {Conference} on {Human} {Factors} in {Computing} {Systems}},
	publisher = {Association for Computing Machinery},
	author = {Liu, Pinyao and Stepanova, Ekaterina R. and Kitson, Alexandra and Schiphorst, Thecla and Riecke, Bernhard E.},
	month = apr,
	year = {2022},
	pages = {1--18},
}

@inproceedings{rixen_exploring_2021,
	address = {New York, NY, USA},
	series = {{CHI} '21},
	title = {Exploring {Augmented} {Visual} {Alterations} in {Interpersonal} {Communication}},
	isbn = {978-1-4503-8096-6},
	url = {https://dl.acm.org/doi/10.1145/3411764.3445597},
	doi = {10.1145/3411764.3445597},
	urldate = {2025-02-10},
	booktitle = {Proceedings of the 2021 {CHI} {Conference} on {Human} {Factors} in {Computing} {Systems}},
	publisher = {Association for Computing Machinery},
	author = {Rixen, Jan Ole and Hirzle, Teresa and Colley, Mark and Etzel, Yannick and Rukzio, Enrico and Gugenheimer, Jan},
	month = may,
	year = {2021},
	pages = {1--11},
}

@article{elmqvist_taxonomy_2008,
	title = {A {Taxonomy} of {3D} {Occlusion} {Management} for {Visualization}},
	volume = {14},
	issn = {1941-0506},
	url = {https://ieeexplore.ieee.org/document/4483791/?arnumber=4483791},
	doi = {10.1109/TVCG.2008.59},
	number = {5},
	urldate = {2025-02-09},
	journal = {IEEE Transactions on Visualization and Computer Graphics},
	author = {Elmqvist, Niklas and Tsigas, Philippas},
	month = sep,
	year = {2008},
	note = {Conference Name: IEEE Transactions on Visualization and Computer Graphics},
	pages = {1095--1109},
}

@inproceedings{lilija_augmented_2019,
	address = {New York, NY, USA},
	series = {{CHI} '19},
	title = {Augmented {Reality} {Views} for {Occluded} {Interaction}},
	isbn = {978-1-4503-5970-2},
	url = {https://dl.acm.org/doi/10.1145/3290605.3300676},
	doi = {10.1145/3290605.3300676},
	urldate = {2025-02-09},
	booktitle = {Proceedings of the 2019 {CHI} {Conference} on {Human} {Factors} in {Computing} {Systems}},
	publisher = {Association for Computing Machinery},
	author = {Lilija, Klemen and Pohl, Henning and Boring, Sebastian and Hornbæk, Kasper},
	month = may,
	year = {2019},
	pages = {1--12},
}

@article{wellner_interacting_1993,
	title = {Interacting with paper on the {DigitalDesk}},
	volume = {36},
	issn = {0001-0782},
	url = {https://dl.acm.org/doi/10.1145/159544.159630},
	doi = {10.1145/159544.159630},
	number = {7},
	urldate = {2025-02-09},
	journal = {Commun. ACM},
	author = {Wellner, Pierre},
	month = jul,
	year = {1993},
	pages = {87--96},
}

@inproceedings{fitzmaurice_bricks_1995,
	address = {USA},
	series = {{CHI} '95},
	title = {Bricks: laying the foundations for graspable user interfaces},
	isbn = {978-0-201-84705-5},
	shorttitle = {Bricks},
	url = {https://dl.acm.org/doi/10.1145/223904.223964},
	doi = {10.1145/223904.223964},
	urldate = {2025-02-09},
	booktitle = {Proceedings of the {SIGCHI} {Conference} on {Human} {Factors} in {Computing} {Systems}},
	publisher = {ACM Press/Addison-Wesley Publishing Co.},
	author = {Fitzmaurice, George W. and Ishii, Hiroshi and Buxton, William A. S.},
	month = may,
	year = {1995},
	pages = {442--449},
}

@inproceedings{wigdor_lucid_2007,
	address = {New York, NY, USA},
	series = {{UIST} '07},
	title = {Lucid touch: a see-through mobile device},
	isbn = {978-1-59593-679-0},
	shorttitle = {Lucid touch},
	url = {https://dl.acm.org/doi/10.1145/1294211.1294259},
	doi = {10.1145/1294211.1294259},
	urldate = {2025-02-09},
	booktitle = {Proceedings of the 20th annual {ACM} symposium on {User} interface software and technology},
	publisher = {Association for Computing Machinery},
	author = {Wigdor, Daniel and Forlines, Clifton and Baudisch, Patrick and Barnwell, John and Shen, Chia},
	month = oct,
	year = {2007},
	pages = {269--278},
}

@inproceedings{cheng_towards_2022,
	address = {New York, NY, USA},
	series = {{CHI} '22},
	title = {Towards {Understanding} {Diminished} {Reality}},
	isbn = {978-1-4503-9157-3},
	url = {https://dl.acm.org/doi/10.1145/3491102.3517452},
	doi = {10.1145/3491102.3517452},
	urldate = {2025-02-09},
	booktitle = {Proceedings of the 2022 {CHI} {Conference} on {Human} {Factors} in {Computing} {Systems}},
	publisher = {Association for Computing Machinery},
	author = {Cheng, Yi Fei and Yin, Hang and Yan, Yukang and Gugenheimer, Jan and Lindlbauer, David},
	month = apr,
	year = {2022},
	pages = {1--16},
}

@inproceedings{jacobs_performative_2019,
	address = {New York, NY, USA},
	series = {{CHI} '19},
	title = {The {Performative} {Mirror} {Space}},
	isbn = {978-1-4503-5970-2},
	url = {https://dl.acm.org/doi/10.1145/3290605.3300630},
	doi = {10.1145/3290605.3300630},
	urldate = {2025-02-08},
	booktitle = {Proceedings of the 2019 {CHI} {Conference} on {Human} {Factors} in {Computing} {Systems}},
	publisher = {Association for Computing Machinery},
	author = {Jacobs, Rachel and Schnädelbach, Holger and Jäger, Nils and Leal, Silvia and Shackford, Robin and Benford, Steve and Patel, Roma},
	month = may,
	year = {2019},
	pages = {1--14},
}

@inproceedings{rajcic_mirror_2020,
	address = {New York, NY, USA},
	series = {{CHI} '20},
	title = {Mirror {Ritual}: {An} {Affective} {Interface} for {Emotional} {Self}-{Reflection}},
	isbn = {978-1-4503-6708-0},
	shorttitle = {Mirror {Ritual}},
	url = {https://dl.acm.org/doi/10.1145/3313831.3376625},
	doi = {10.1145/3313831.3376625},
	urldate = {2025-02-08},
	booktitle = {Proceedings of the 2020 {CHI} {Conference} on {Human} {Factors} in {Computing} {Systems}},
	publisher = {Association for Computing Machinery},
	author = {Rajcic, Nina and McCormack, Jon},
	month = apr,
	year = {2020},
	pages = {1--13},
}

@inproceedings{zhou_here_2023,
	address = {New York, NY, USA},
	series = {{CHI} '23},
	title = {Here and {Now}: {Creating} {Improvisational} {Dance} {Movements} with a {Mixed} {Reality} {Mirror}},
	isbn = {978-1-4503-9421-5},
	shorttitle = {Here and {Now}},
	url = {https://dl.acm.org/doi/10.1145/3544548.3580666},
	doi = {10.1145/3544548.3580666},
	urldate = {2025-02-08},
	booktitle = {Proceedings of the 2023 {CHI} {Conference} on {Human} {Factors} in {Computing} {Systems}},
	publisher = {Association for Computing Machinery},
	author = {Zhou, Qiushi and Grebel, Louise and Irlitti, Andrew and Minaai, Julie Ann and Goncalves, Jorge and Velloso, Eduardo},
	month = apr,
	year = {2023},
	pages = {1--16},
}

@inproceedings{soler-dominguez_arcadia_2024,
	address = {New York, NY, USA},
	series = {{CHI} '24},
	title = {{ARCADIA}: {A} {Gamified} {Mixed} {Reality} {System} for {Emotional} {Regulation} and {Self}-{Compassion}},
	isbn = {979-8-4007-0330-0},
	shorttitle = {{ARCADIA}},
	url = {https://dl.acm.org/doi/10.1145/3613904.3642123},
	doi = {10.1145/3613904.3642123},
	urldate = {2025-02-08},
	booktitle = {Proceedings of the 2024 {CHI} {Conference} on {Human} {Factors} in {Computing} {Systems}},
	publisher = {Association for Computing Machinery},
	author = {Soler-Dominguez, Jose Luis and Navas-Medrano, Samuel and Pons, Patricia},
	month = may,
	year = {2024},
	pages = {1--17},
}

@inproceedings{kim_perspective_2023,
	address = {New York, NY, USA},
	series = {{CHI} '23},
	title = {Perspective and {Geometry} {Approaches} to {Mouse} {Cursor} {Control} in {Spatial} {Augmented} {Reality}},
	isbn = {978-1-4503-9421-5},
	url = {https://dl.acm.org/doi/10.1145/3544548.3580849},
	doi = {10.1145/3544548.3580849},
	urldate = {2025-02-08},
	booktitle = {Proceedings of the 2023 {CHI} {Conference} on {Human} {Factors} in {Computing} {Systems}},
	publisher = {Association for Computing Machinery},
	author = {Kim, Daekun and Joshi, Nikhita and Vogel, Daniel},
	month = apr,
	year = {2023},
	pages = {1--19},
}

@inproceedings{ishii_pingpongplus_1999,
	address = {New York, NY, USA},
	series = {{CHI} '99},
	title = {{PingPongPlus}: design of an athletic-tangible interface for computer-supported cooperative play},
	isbn = {978-0-201-48559-2},
	shorttitle = {{PingPongPlus}},
	url = {https://dl.acm.org/doi/10.1145/302979.303115},
	doi = {10.1145/302979.303115},
	urldate = {2025-02-07},
	booktitle = {Proceedings of the {SIGCHI} conference on {Human} {Factors} in {Computing} {Systems}},
	publisher = {Association for Computing Machinery},
	author = {Ishii, Hiroshi and Wisneski, Craig and Orbanes, Julian and Chun, Ben and Paradiso, Joe},
	month = may,
	year = {1999},
	pages = {394--401},
}

@inproceedings{branzel_gravityspace_2013,
	address = {New York, NY, USA},
	series = {{CHI} '13},
	title = {{GravitySpace}: tracking users and their poses in a smart room using a pressure-sensing floor},
	isbn = {978-1-4503-1899-0},
	shorttitle = {{GravitySpace}},
	url = {https://dl.acm.org/doi/10.1145/2470654.2470757},
	doi = {10.1145/2470654.2470757},
	urldate = {2025-02-07},
	booktitle = {Proceedings of the {SIGCHI} {Conference} on {Human} {Factors} in {Computing} {Systems}},
	publisher = {Association for Computing Machinery},
	author = {Bränzel, Alan and Holz, Christian and Hoffmann, Daniel and Schmidt, Dominik and Knaust, Marius and Lühne, Patrick and Meusel, René and Richter, Stephan and Baudisch, Patrick},
	month = apr,
	year = {2013},
	pages = {725--734},
}

@inproceedings{grosse-puppendahl_exploring_2016,
	address = {New York, NY, USA},
	series = {{UIST} '16},
	title = {Exploring the {Design} {Space} for {Energy}-{Harvesting} {Situated} {Displays}},
	isbn = {978-1-4503-4189-9},
	url = {https://dl.acm.org/doi/10.1145/2984511.2984513},
	doi = {10.1145/2984511.2984513},
	urldate = {2025-02-07},
	booktitle = {Proceedings of the 29th {Annual} {Symposium} on {User} {Interface} {Software} and {Technology}},
	publisher = {Association for Computing Machinery},
	author = {Grosse-Puppendahl, Tobias and Hodges, Steve and Chen, Nicholas and Helmes, John and Taylor, Stuart and Scott, James and Fromm, Josh and Sweeney, David},
	month = oct,
	year = {2016},
	pages = {41--48},
}

@inproceedings{lee_icy_2024,
	address = {New York, NY, USA},
	series = {{DIS} '24},
	title = {{ICY} {Interfaces}: {Exploration} of {Ice}’s {Ephemeral} {Features} for {Digital} {Game} {User} {Experience}},
	isbn = {979-8-4007-0583-0},
	shorttitle = {{ICY} {Interfaces}},
	url = {https://dl.acm.org/doi/10.1145/3643834.3660725},
	doi = {10.1145/3643834.3660725},
	urldate = {2025-02-07},
	booktitle = {Proceedings of the 2024 {ACM} {Designing} {Interactive} {Systems} {Conference}},
	publisher = {Association for Computing Machinery},
	author = {Lee, Yoonji and Lee, Chang Hee},
	month = jul,
	year = {2024},
	pages = {2107--2124},
}

@article{benford_understanding_1998,
	title = {Understanding and constructing shared spaces with mixed-reality boundaries},
	volume = {5},
	issn = {1073-0516},
	url = {https://dl.acm.org/doi/10.1145/292834.292836},
	doi = {10.1145/292834.292836},
	number = {3},
	urldate = {2025-02-08},
	journal = {ACM Trans. Comput.-Hum. Interact.},
	author = {Benford, Steve and Greenhalgh, Chris and Reynard, Gail and Brown, Chris and Koleva, Boriana},
	month = sep,
	year = {1998},
	pages = {185--223},
}

@inproceedings{scott_territoriality_2004,
	address = {New York, NY, USA},
	series = {{CSCW} '04},
	title = {Territoriality in collaborative tabletop workspaces},
	isbn = {978-1-58113-810-8},
	url = {https://dl.acm.org/doi/10.1145/1031607.1031655},
	doi = {10.1145/1031607.1031655},
	urldate = {2025-02-07},
	booktitle = {Proceedings of the 2004 {ACM} conference on {Computer} supported cooperative work},
	publisher = {Association for Computing Machinery},
	author = {Scott, Stacey D. and Carpendale, M. Sheelagh T. and Inkpen, Kori},
	month = nov,
	year = {2004},
	pages = {294--303},
}

@article{zhou_reflected_2024,
	title = {Reflected {Reality}: {Augmented} {Reality} through the {Mirror}},
	volume = {7},
	shorttitle = {Reflected {Reality}},
	url = {https://dl.acm.org/doi/10.1145/3631431},
	doi = {10.1145/3631431},
	number = {4},
	urldate = {2025-02-08},
	journal = {Proc. ACM Interact. Mob. Wearable Ubiquitous Technol.},
	author = {Zhou, Qiushi and Syiem, Brandon Victor and Li, Beier and Goncalves, Jorge and Velloso, Eduardo},
	month = jan,
	year = {2024},
	pages = {202:1--202:28},
}

@inproceedings{li_benefits_2024,
	address = {Honolulu HI USA},
	title = {On the {Benefits} of {Image}-{Schematic} {Metaphors} when {Designing} {Mixed} {Reality} {Systems}},
	isbn = {979-8-4007-0330-0},
	url = {https://dl.acm.org/doi/10.1145/3613904.3642925},
	doi = {10.1145/3613904.3642925},
	language = {en},
	urldate = {2025-02-06},
	booktitle = {Proceedings of the {CHI} {Conference} on {Human} {Factors} in {Computing} {Systems}},
	publisher = {ACM},
	author = {Li, Jingyi and Kristensson, Per Ola},
	month = may,
	year = {2024},
	pages = {1--20},
}

@inproceedings{nakagaki_disappearables_2022,
	address = {New Orleans LA USA},
	title = {({Dis}){Appearables}: {A} {Concept} and {Method} for {Actuated} {Tangible} {UIs} to {Appear} and {Disappear} based on {Stages}},
	isbn = {978-1-4503-9157-3},
	shorttitle = {({Dis}){Appearables}},
	url = {https://dl.acm.org/doi/10.1145/3491102.3501906},
	doi = {10.1145/3491102.3501906},
	language = {en},
	urldate = {2025-02-04},
	booktitle = {{CHI} {Conference} on {Human} {Factors} in {Computing} {Systems}},
	publisher = {ACM},
	author = {Nakagaki, Ken and Tappa, Jordan L and Zheng, Yi and Forman, Jack and Leong, Joanne and Koenig, Sven and Ishii, Hiroshi},
	month = apr,
	year = {2022},
	pages = {1--13},
}

@inproceedings{zimmerman_research_2007,
	address = {San Jose California USA},
	title = {Research through design as a method for interaction design research in {HCI}},
	isbn = {978-1-59593-593-9},
	url = {https://dl.acm.org/doi/10.1145/1240624.1240704},
	doi = {10.1145/1240624.1240704},
	language = {en},
	urldate = {2025-02-04},
	booktitle = {Proceedings of the {SIGCHI} {Conference} on {Human} {Factors} in {Computing} {Systems}},
	publisher = {ACM},
	author = {Zimmerman, John and Forlizzi, Jodi and Evenson, Shelley},
	month = apr,
	year = {2007},
	pages = {493--502},
}


%

\appendix
\counterwithin{table}{section}
\counterwithin{figure}{section}
\begingroup
\footnotesize

\captionsetup{
  font=footnotesize,
  labelfont={bf,footnotesize}
}

\myAppSection{Representative Prior Work}{list-papers}
This appendix presents representative and inspirational prior works that informed the dimensions of the design space, curated through iterative search and snowballing as described in \Cref{sec:dimensions}.
\vspace{0.5em}

\noindent
\begin{minipage}[t]{0.48\linewidth}
\begin{itemize}[leftmargin=*]
 \item[1.] {\citet{benford_understanding_1998}}
 \item[2.] {\citet{branzel_gravityspace_2013}}
 \item[3.] {\citet{cheng_towards_2022}}
 \item[4.] {\citet{dollinger_are_2023}}
 \item[5.] {\citet{han_corobos_2025}}
 \item[6.] {\citet{fitzmaurice_bricks_1995}}
 \item[7.] {\citet{gronbaek_blended_2024}}
 \item[8.] {\citet{grosse-puppendahl_exploring_2016}}
 \item[9.] {\citet{hendriks_undertable_2024}}
 \item[10.] {\citet{ishii_pingpongplus_1999}}
 \item[11.] {\citet{jacobs_performative_2019}}
 \item[12.] {\citet{kim_perspective_2023}}
 \item[13.] {\citet{lee_icy_2024}}
 \item[14.] {\citet{lilija_augmented_2019}}
 \item[15.] {\citet{lin_throwio_2023}}
 \item[16.] {\citet{liu_reality_2025}}
\end{itemize}
\end{minipage}%
\hfill
\begin{minipage}[t]{0.48\linewidth}
\begin{itemize}[leftmargin=*]
 \item[17.] {\citet{lyu_objestures_2026}}
 \item[18.] {\citet{muller_baselase_2015}}
 \item[19.] {\citet{nakagaki_disappearables_2022}}
 \item[20.] {\citet{ogawa_you_2020}}
 \item[21.] {\citet{onishi_waddlewalls_2022}}
 \item[22.] {\citet{pohl_integrated_2024}}
 \item[23.] {\citet{pohl_poros_2021}}
 \item[24.] {\citet{rajcic_mirror_2020}}
 \item[25.] {\citet{rau_traversing_2025}}
 \item[26.] {\citet{riche_ai-instruments_2025}}
 \item[27.] {\citet{soler-dominguez_arcadia_2024}}
 \item[28.] {\citet{van_gemert_doorways_2024}}
 \item[29.] {\citet{wigdor_lucid_2007}}
 \item[30.] {\citet{yu_aerorigui_2023}}
 \item[31.] {\citet{zhou_here_2023}}
\end{itemize}
\end{minipage}
\myAppSection{Design Reflections}{sec:design-implementation}
Reflecting on the process, designing each example required extending the information provided by the framework, which unfolded bidirectionally. For examples such as the Trash Can, Drawer, and Paper Shredder, the concepts predated the framework iterations (\Cref{fig:design-space}a) and were grounded in the distinction between \dtag{Crossing} and \dtag{Non-crossing}, particularly how these relations imply concealment and state transitions (\Cref{sec:boundary-function}). In these cases, the framework acted as a backbone that guided detailed design decisions. For instance, our initial Paper Shredder idea did not consider how users would perceive the experience. The framework prompted us to clarify the \dtag{Perception}: because we wanted users to experience the transition rather than a simple disappearance, Paper Shredder requires \dtag{Unilateral}, whereas the Trash Can uses \dtag{Bilateral}. 

Other examples emerged more generatively: for the Marker, this began with exploring how \dtag{Contact} and \dtag{Along} could be combined in the Productivity scenario. The notion of an object moving along the boundary suggested an affordance of something being carried on the surface, which led to the idea of a stroke carried across it, and the Marker gradually took shape. Examples such as Interactive Painting and Magic Mirror required substantial additions beyond the framework elements and were highly context-dependent. In Magic Mirror, for example, we used abstract representations because the selected artwork (\textit{Composition VIII}) consists of abstract elements. Implementing similar 2D--3D \dtag{Crossing} relations for other types of artwork would require different choices, relying on designers' imagination and interpretation. To support this, \Cref{fig:final-design-space} provides example affordances and designs for each element that may help inspire designers.

\myAppSection{Supplementary Discussion and Development Tools}{appen:diss-tools}
\noindent{\textit{\sffamily From Semantics to Shared Understanding.\,}} 
Interpreting ``boundary'' as restriction caused difficulties in interactions (such as pulling in {Interactive Painting}; \Cref{sec:through-not-reasonable}) until we used ``surface.'' Such semantic challenges are common: ``mixed reality'' lacks a stable definition~\cite{speicher_what_2019}, and ``affordance'' has been repeatedly revisited~\cite{norman_affordance_1999,kaptelinin_affordances_2012}. We suggest frameworks benefit from semantic flexibility. Ambiguity at the outset can mislead, so multiple entry points (``boundary'', ``surface'', etc.) and concrete examples help. Once clarity is reached, ambiguity~\cite{gaver_ambiguity_2003} of coexisting meanings (restriction, support, etc.) enables new interpretations where the definition matters less than the shared understanding.
\begin{center}
  \includegraphics[width=\linewidth]{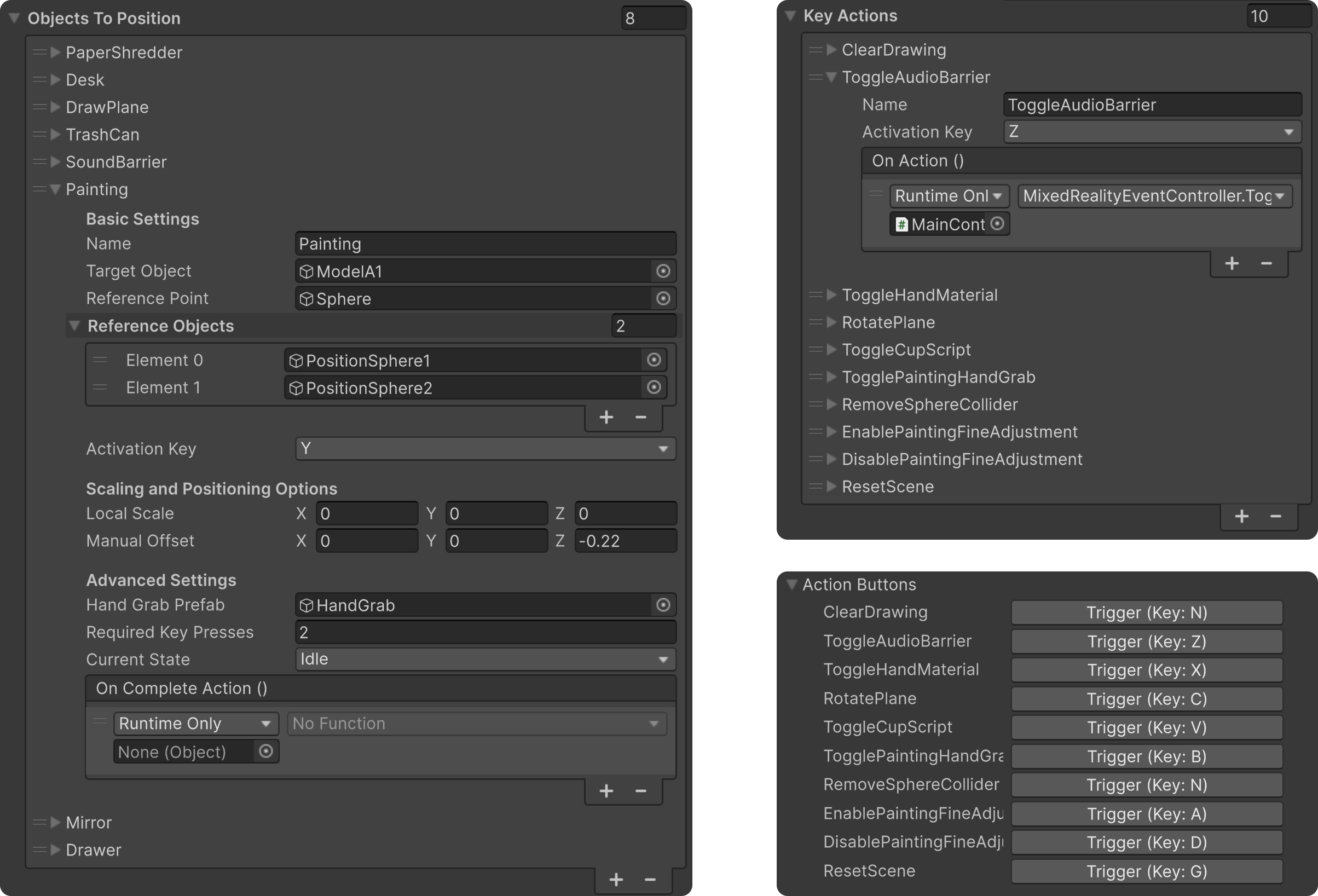}
  \vspace{-2em}
  \captionof{figure}{We implemented commonly used scripts as Unity Editor tools. On the left is a generic positioner that samples one, two, or three reference points on key press to place and align objects. With one point, the object moves to that location; with two points, it is uniformly scaled to match the segment length and translated so the first marker maps to the first point; and with three points, it is placed at the triangle centroid and oriented by the plane normal. Transforms are saved per object in JSON and restored on load, allowing multiple independently configured objects. The tools on the right support debugging: the top one lets developers specify actions, and the script binds them to functions and generates UI elements, shown as buttons in the lower figure. Developers can trigger these during runtime via the UI or keyboard, or through the Quest Developer Hub after the app is built. Our prototype relied on manual positioning, but with Quest video-feed access~\cite{meta_pca_unity_2025}, systems could automate detection and propose environment-specific interactions (\Cref{sec:tooling}).}
  \Description{A screenshot of a dark-themed software configuration interface composed of three distinct panels. The tall panel on the left displays a hierarchical list of objects, with one item expanded to reveal numerous input fields, dropdown menus, and numeric coordinate boxes for adjusting properties like ``Reference Objects,'' ``Local Scale,'' and ``Manual Offset.'' On the right, two smaller panels are stacked vertically. The top right panel features a collapsible list of programmable events, showing an expanded block where a specific action is assigned to a keyboard key. The bottom right panel displays a simple vertical column of rectangular buttons, each explicitly labeled with an action name paired with its corresponding keyboard shortcut trigger.}
  \label{fig:tools}
\end{center}

\vspace{-0.5em}
\myAppSection[nospaceafter]{Extended Quotes}{full-quotes}
\vspace{-0.5em}
\Q{P1}{clarity}{It covers fundamental elements. It is clear now and not too confusing, and it has a manageable level of information for me to work with.}

\Q{P2}{motivation}{What is your motivation, and what are you hoping to achieve?}

\Q{P5}{goal}{What is your design goal? If I have this design space, I do not yet know what to do with it.}

\Q{P4}{teleology}{We have not emphasized purpose. With a clearer purpose, the use of this design space could become less nebulous.}

\Q{P5}{usecase}{I want to understand what use case this design space could serve.}

\Q{P4}{cards}{I know you may hope it works like a deck of cards, where people can combine them to invent new interactions. With the framework feeling abstract so far, this potential is not yet fully realized.}

\Q{P4}{hard-imagine}{Only after seeing your examples did I think of related work; the design space alone made it hard to imagine concrete interactions.}

\Q{P3}{now-understand}{It is quite interesting; now I understand why this topic matters.}

\Q{P2}{western-church}{Western churches also used ceiling frescoes to enhance the sacred quality of space. The paintings aligned with the geometry of arches and domes, and perspective was corrected so that when people looked up, they truly felt the presence of divine figures. In that sense, augmentation has always been psychological as well as visual.}

\Q{P2}{art-bundle}{Very exciting. Compared with the previous Productivity [scenario], the biggest benefit [of the Art Exploration scenario] is that it weaves several interaction types into one small scene, and the actions in that scene relate to each other. The relationships are tighter, each action has ample room for imagination, and it still aligns with your classification framework.}

\Q{P6}{mirror}{I really like how you use the mirror as a metaphor \ldots{} and then abstract an object and place it back into this already abstract painting. The very act of abstracting an object itself is fascinating.}

\Q{P3}{fun}{This is the most fun. After you turn the [Magic Mirror] around, the physical object [placed on one side] turns into something else [on the other side].}

\Q{P1}{optional-affordance}{To make [the design space] a more usable tool, this can be presented as optional. If someone has difficulty coming up with or making use of it, they could expand that attribute to see a non-exhaustive list of possible affordances.}

\Q{P2}{mirror-intuitive}{This feels very intuitive; many people have imagined in real life that behind a mirror lies another world or another version of something. This idea has also been expressed in many films: a false mirror or a sheet of glass, where whatever someone does on this side is mirrored by another performer on the other side doing the same actions. I find that very interesting.}

\Q{P5}{remove}{Rather than saying you are throwing it away, maybe it is just not needed now, so you take it out of the scene.}

\Q{P6}{marker-real}{Was that marker real?}

\Q{P2}{marker-fake-real}{It looked so real that I thought it was a physical marker.}

\Q{P2}{impossible}{If you want the marker's drawing to be understood as XR, it must produce effects impossible with a real pen, for example, strokes that glow, while still following the logic of real-world interaction. That may be what makes the second scenario [Art Exploration] appealing: the Magic Mirror reveals objects clearly impossible in reality, which is where the charm comes from.}

\Q{P3}{haptics}{If you simply replicate reality exactly as it is, it becomes boring. Most work in haptics centers on replication, with everything aimed at achieving realistic touch. From a design perspective, that way of thinking is limiting and ultimately uninteresting.}

\Q{P4}{mr-fake}{MR is perfect for showing fake things. Magic is fake; ghosts are fake. MR makes fake things feel real.}

\Q{P4}{mr-playful}{Nowadays, your phone interface is just a grid of apps. Do you really need those little boxes? If AI continues to advance for a year or two, I believe the concept of separate apps will be unnecessary. MR's edge for utility alone is not enough, so trying to make MR purely useful becomes almost absurd. I would rather make MR do things that are impractical but delightful, things only MR can do.}

\Q{P6}{minimal-shredder}{In the Productivity scenario, the visual representation is very minimal, which I really like. For the Paper Shredder, at first, I did not realize it was functional, but once I saw the paper being shredded, I understood its purpose. Visually, I also like this design.}

\Q{P2}{drawer-handle}{That drawer is also very intuitive. In reality, people naturally think an item would be in a place like a drawer. So on the wall, there is a drawer knob; you pull it, and all the magical things inside are revealed.}

\Q{P3}{ambiguous-yet-familiar}{We want more creative interactions, but for people to understand them, they need to be ambiguous yet not strange. There should be elements that people are already familiar with.}
\Q{P2}{swirl}{If you added a swirling pattern on the circled area for the Trash Can, it would signal that this surface is penetrable. I would then know the rest of the table is safe to place objects on, but this area works like a black hole where I can throw things away.}
\Q{P1}{prior-marker-wall}{Give me a marker and a wall; I will naturally pick up the marker and try to write on the wall.}

\Q{P3}{surface-drop-grow}{In [the Trash Can], the circled area drops down so you can throw your marker in, then grows back again.}

\Q{P3}{bouncing}{If the ball is bouncing and is about to sink through the surface, I would expect the surface to look slightly different, so I can feel this is the moment it should sink.}

\Q{P1}{quiet-mechanisms}{Simple mechanisms indicating interactivity can encourage exploration. A major benefit of your boundary examples is that they can be hidden; they can be made very ``quiet'' and unobtrusive, concealing functionality so you are not overwhelmed by visual clutter from UI elements. You obviously do not want text popping up to tell you what to do when approaching [an interactive area]. But you could consider adding subtle affordances or signals that naturally let users know something is hidden there, [and] what they can do with it.}

\Q{P3}{framework}{In my view, it reads more like a design framework.}

\Q{P5}{functionality-highdim}{The design space you present is very fundamental, yet as an HCI practitioner, I would like to see its higher-dimensional functionality.}

\Q{P1}{drawer-hiding-aff}{A drawer inside a boundary, what does it imply? It implies you can hide things. At the moment, I have to abstract this layer myself. If your design space also specified these affordances, I would find it more usable.}

\Q{P4}{embodied-affordance}{When you pull a drawer from a wall, the essence that the design space does not yet express is the affordance. We have this experience in the physical world; it is embodied, and the current framework does not yet show it.}

\Q{P5}{boundary-broad-concept}{Boundary is a broad concept, and in different situations, its meaning in interaction changes.}

\Q{P1}{writing-abstract}{I think one thing to note is that the affordance should be written in a very abstract way. Otherwise, if you present it as a very concrete idea, like writing, it will fix thinking on that example and prevent imagining other possibilities. So if you add affordances, they need to be added very carefully, at a very abstract level.}

\Q{P1}{affordance-fixation}{I understand. Indeed, if you cannot cover everything, highlighting only part may limit thinking to that part. For example, if you list two affordances, people may focus only on those and not imagine others. So adding affordances has both sides: it can be helpful, but also limiting.}

\Q{P2}{sculptural-surface}{It could be a curved surface, like the surface of a sculpture.}
\Q{P3}{skin-haptics}{From the perspective of haptic devices on the body, the human skin surface can serve as a boundary.}
\Q{P3}{ice-melts}{A boundary made of ice will melt into water.}

\Q{P2}{restaurant-privacy}{In a restaurant, we have tables or private rooms, with gradients of privacy and publicness. A more private table gathers a group, and the bond within that group feels stronger than the bond you feel with the table next to you.}

\Q{P4}{privacy-touch}{When relationships need to mingle, or boundaries are relaxed, privacy and trust become issues. In popular large-space VR, or in my project, where strangers are in the dark and cannot see each other, but can still touch, these questions become very present.}

\Q{P2}{2d-wall}{Through a simple 2D graphic, you instill the idea of a wall in the middle that should not be crossed.}

\Q{P2}{carpet-zone}{In architectural design, placing a rug in the living room creates a small space within the larger space. People often take off their shoes there, or at least feel they have entered a zone that is different from the surrounding tile.}

\Q{P5}{fuzzy-objects}{For example, placing numerous objects in a virtual space can be seen as a way to constrain movement trajectories. Beyond simply limiting the maximum range of motion, it also exerts an influence on a user's most natural behavior; for instance, people do not typically walk with their shoulders glued to the walls just because the room is enclosed by them.}

\Q{P2}{mirror-illusion}{It is like those old magic tricks with mirrors, where the performer hides inside a box, and mirrors reflect the body away. To the audience, it looks as if the person has no head, or as if their head is sitting in a vase. In the 18th and 19th centuries, circuses often used this method: angled mirrors with flowers covering the edges. From the front, it seemed the person's head floated in the vase, when in fact their body was hidden. This reflection makes people believe the space is empty or passable, when it is not, almost the reverse of a drawn line that can be crossed but feels solid.}

\Q{P4}{reduced-physical}{The boundary is physical, but I can choose not to let you see it. It is reduced.}

\Q{P5}{restriction-term}{For example, with the Bouncing Ball, the boundary serves as a restrictive surface.}

\Q{P2}{virtual-wall}{Imagine being in a large space, wanting some privacy with friends, like a private room in a restaurant. You could build a fake wall around you, sometimes semi-transparent, sometimes fully opaque. Visually, it appears as a solid cement wall, and in someone's perception, it is impenetrable, until they touch it and realize it is only made of pixels. Most of the time, people will not actually test it, so they act as if the wall is real. In an exhibition-like space, such virtual walls influence interaction. Even if people could technically walk through, they default to avoiding it.}

\Q{P3}{stash-restore}{With many elements and windows in XR, if you leave everything open, it floats everywhere and becomes annoying. You need reasonable ways to stash things so they appear and disappear appropriately. In 2D, we swipe up to close apps. In 3D, we will need new ways.}

\Q{P1}{old-info}{It is about how to handle information you do not need anymore, to get rid of old information.}

\Q{P6}{cross-space}{It feels cross-space and magical. As paper goes from one side to the other, its form changes.}

\Q{P5}{binary-state}{A boundary can determine an object's state in a binary way. On one side, the state is whole. On the other side, it becomes fragmented.}

\Q{P5}{trash-mirror}{In the Trash Can example, above the boundary the object exists, below it does not. In the Magic Mirror example, the left side is physical, and the right side is abstract.}

\Q{P1}{define-boundary}{Users could define the boundary themselves depending on their purpose, or are you more focused on just using existing physical things as the boundary?}

\Q{P5}{lines-divide}{Drawing lines to limit where things can move, or even deciding how to partition the space, is all possible. Depending on the use case, people will prefer different interaction methods.}

\Q{P4}{shared-intensity}{Two people viewing the same thing is more powerful than one person viewing alone. It is not that a single viewer will forget; it is relative. For example, imagine you open and close a drawer, and the person next to you can also open and close it. That shared action becomes very strong. But if the person next to you is looking at a different thing and opens something you did not expect, and you cannot see what they opened, that becomes frustrating.}

\Q{P4}{fractured-reality}{In the future, the common condition will not be a single shared reality but a fractured one. I saw it; does that mean you saw it? Not necessarily. What I consider real may not be what you can confidently consider real.}

\Q{P4}{VR-MR-different}{VR and MR are very different. In VR everything is synthetic, so the designer has complete control. In MR the real part is in the user's home, which the designer cannot control.}

\Q{P3}{inexplicable}{Pulling a drawer out of a wall to retrieve something felt inexplicable to me.}

\Q{P5}{two-objects}{To me, the physical elements of the painting and the digital element pulled from it are two different objects.}

\Q{P5}{boundary-produces}{I do not understand why a boundary can produce an object.}

\Q{P5}{ice-melts-holes}{Ice melts into water. If the surface has holes, when it is still ice, it rests on the surface; after it melts, it drips through.}

\Q{P3}{object-agency}{Objects have their own logic to move with external forces, not just through human manipulation. For example, wind moves a curtain that knocks over a flowerpot.}

\Q{P4}{magic-wand}{What comes to mind is waving a magic wand. If your hand is empty, any gesture you make usually needs instruction. But if I give you a wand and say it is real, heavy, and solid, the only thing you naturally do is swing it down; you would not do anything else. There is no ambiguity. Part of this comes from decades of cultural background, like Harry Potter, and part from the wand itself: it has weight, a certain length, and it is not long enough to plant on the ground \ldots{} This creates an extremely simple and self-evident form of interaction.}

\Q{P6}{ambient}{I feel your categories are centered on the eyes. I am not sure whether other parts of the body are included. The person's position, the boundary, and the object can have many interaction relationships. For example, something displayed on a wall does not need to remain visible all the time. It could fade in as I approach and fade out as I leave. We call this ambient technology: you do not need everything to be constantly visible and noisy.}

\Q{P5}{define-throw-trash}{In the Trash Can example, it would make sense if the user could define it. The user draws a circle, then throws an object in.}

\Q{P5}{boundary-people-perception}{For example, the boundary between people may not involve this kind of perceptual relationship.}

\Q{P4}{MR-body-movement}{In MR, body movement is naturally involved.}

\Q{P3}{initial-motion}{For example, \dtag{Non-contact} and \dtag{Contact}: besides being parallel movement, the key is that it starts with \dtag{Contact} or \dtag{Non-contact}, an initial relation, and only then does the motion unfold.}

\Q{P3}{a-bit-2D}{It feels a bit two-dimensional, but since it is 3D space, maybe other factors could be included to make motion more interesting.}

\Q{P3}{chaotically}{When you move while holding something, it will inevitably move around chaotically in different directions, distances, and speeds.}

\Q{P3}{mix-presence}{What is interesting is that you mixed the virtual and the real to create a sense of participation in the environment, without merely mimicking real interactions. Your demos maintain a certain degree of presence, which makes sense.}

\Q{P1}{haptics-value}{Leveraging physical boundaries in daily life can provide tactile feedback in interaction, which is largely missing in most mid-air MR interactions and can hinder usability. Smart use of real objects or boundaries can close that gap.}

\Q{P5}{surface-haptics}{When drawing, writing, or typing in VR on a purely virtual surface, there is no haptic feedback, so it is hard to know if I am in contact. If there is a tangible surface, even if the pen is virtual, my real hand feels the surface. That is a strong use case. Many people doing virtual typing place the virtual keyboard on a real surface. Haptics give reliable feedback for motion tasks, whereas vision alone is unclear about contact. Haptics are very useful.}

\Q{P5}{repurpose-planes}{Even if I am not using the tangible aspect, it still makes sense. For example, using a bookshelf plane instead of spawning a virtual object can turn my everyday physical space into something that also has virtual functionality. This holds true even without the tangible element.}

\Q{P5}{coherent-logic}{When a \dtag{Digital} {object} is related to a \dtag{Physical} {object}, it should be influenced by the same factors that affect the \dtag{Physical} one. For example, in your case, sometimes you interact with \dtag{Digital} objects and sometimes with \dtag{Physical} ones. In those moments, people naturally feel that everything is co-existing in a very immersive workspace, without the sense of separation between \dtag{Digital} and \dtag{Physical}.}

\Q{P6}{use-hands-real}{Using hands is engaging. It is exciting to recreate real-world interactions in the digital world.}

\Q{P3}{logic-consistency}{{Paper Shredder} works because the gesture and object logic align. The paper goes down and is gradually shredded. Writing with a marker is parallel to the surface, which feels right. {Magic Mirror} makes sense because facing it yields a direct reflection, while stepping aside produces a virtual image at the expected location.}

\Q{P5}{language-vs-hands}{Language can be different and sometimes cognitively heavy. If I want to place a painting precisely, describing tiny left-right adjustments is cumbersome. With my hand, it is immediate. Moving a cup from here to there by speech is slow; by hand it is trivial.}

\Q{P4}{prefer-gaze}{I would use very precise gaze interaction because it is excellent for fine control. I would not lift a sheet of paper and place it into the shredder if that costs more energy.}

\Q{P6}{gaze-tedious}{Gaze is less tiring. Constantly grabbing can become tedious.}

\Q{P1}{trash-intuitive}{{Trash Can} is efficient and intuitive. Tossing down means it is gone.}

\Q{P6}{trash-longterm}{If I work in the space long term, I might not want to walk over to throw things away each time. A shortcut gesture to delete would be more convenient, though the concept is interesting.}

\Q{P6}{noise-by-device}{With AR glasses, you can reduce noise through the device itself; you do not necessarily need to raise a visible acoustic partition.}

\Q{P6}{buttons-3D-thickness-plane}{You can give buttons some thickness so they look a bit 3D, but they are still essentially planar.}

\Q{P6}{planar-familiar}{Planar interaction is easier and more familiar for people.}

\Q{P6}{2d-and-3d}{What you are demonstrating is not traditional interface interaction; it is more about interacting with objects. Naturally, interacting with objects requires a 3D approach. I believe future MR interfaces will be a hybrid of 3D and 2D elements. The core design challenge will be creating an effective combination of both. Small elements could transition to 3D, such as icons becoming small 3D models instead of flat graphics, while text will inevitably remain 2D.}

\Q{P1}{container-surface}{If it is a container, then it is \dtag{Unilateral} and \dtag{Opaque}, and I can hide information inside it; if it is a surface, then things can be visible on one side and not visible on the other.}

\Q{P2}{pizza-topology}{What real-world phenomenon corresponds to a topology like a pizza?}

\Q{P2}{relational-system}{I also have my own perspective on how entities interact in XR. These different perspectives could be selectively articulated through different projects. In that sense, there would indeed be a framework similar to your diagram \ldots{} I consider the possible forms of interaction, as well as different types of interacting subjects: human--human, human--object, object--object, or interactions with the environment \ldots{} Interaction serves as a connective thread that links the object and its physical, digital, and other attributes into a coherent whole.}

\Q{P2}{elevator-axes}{I render vertical and horizontal ring structures that push [the cabin] along. Technically, I could move the cabin from the upper left to the lower right directly, but that feels unreal. Using real-world structural logic, I show rods like a 3D printer's axes. People then accept how it moves and understand its envelope, rather than it roaming into empty space.}

\Q{P2}{parametric-io}{Parametric design means I give inputs, a series of shape computations happens, and I get outputs, rather than sculpting the output directly like a static sculpture.}

\Q{P6}{constrained-explore}{Within constrained hardware, we explore what the glasses can do across scenarios, from listening to music, taking photos, sending messages, to asking AI what an object is and how to use it.}

\Q{P4}{design-around-constraint}{As MR designers, we treat device limitations as primary. We design around the constraint. I first list all constraints. For Apple Vision Pro, I extend my hands to probe the outer limits of its hand-tracking range, and even ask someone to interfere with the tracking. I identify where the device's tracking begins to break down, then design back from those limits.}

\Q{P4}{hammer-nail}{Start from real scenarios and needs, not a hammer looking for nails.}
\Q{P3}{hammer-nail-needs}{Do not be a hammer looking for nails; let needs lead the design.}

\Q{P5}{function-to-usecase}{Starting from a functionality category, such as layout replication, I can quickly map to real-life use cases. If your framework included functionality and affordance guidance, it would be easier to generate more examples.}

\Q{P4}{magic-culture}{Many people have been educated by the culture of magic through film, so gesture-based effects feel immediately legible.}

\Q{P3}{haptics-metaphor}{When designing haptics, people first experience different tactile sensations and recall similar feelings from the past. They assign metaphors; otherwise, describing touch is difficult.}

\Q{P4}{spark-and-sanding}{Summarizing into a design space can drift upward into abstraction and risk sanding away the spark and playfulness that make an interaction compelling.}

\Q{P4}{synesthetic-imagination}{Interaction design requires a high degree of synesthetic intuition that resists being neatly constrained by a design space.}

\Q{P4}{decompose-not-synthesize}{Given a good interaction, I can decompose and classify it. Going the other way is hard: starting from classifications to synthesize a good interaction is not straightforward.}

\Q{P3}{practice-based}{Personally, I’m not someone who thinks through frameworks; my process is more artistic and inspiration-led. Having a practice-based, art-school background, I find it a bit challenging to generate ideas starting purely from abstract concepts.}

\Q{P3}{post-hoc-fit}{It feels as if the ideas emerged first and were then mapped onto the framework, rather than the ideas being generated from the framework itself.}

\Q{P4}{eliasson-reading}{Take Olafur Eliasson's \emph{The Weather Project}~\cite{Eliasson2003WeatherProject} at Tate Modern: the description is simple, but the design space it opens is vast. You can analyze interactions, symbolism, and collective attention endlessly, yet you cannot derive the piece like a formula.}

\Q{P2}{extreme-expression}{I imagine the interaction, and if it feels possible, I express it in its most extreme form without yet worrying about restrictions.}

\Q{P6}{clarity-states}{The framework really clarified things for me. I saw how objects, boundaries, humans, and motion can each exist in these distinct states. It made the whole concept much clearer.}

\Q{P1}{additional-functionality}{It shaped my perspective on object--boundary relationships. I had rarely thought about them before this framework. We take boundaries for granted. We have physical and virtual boundaries everywhere in the physical world and in MR, so we do not really think about what other functions they can provide or how we can make use of them. This framework made me think about the additional functionality of boundaries beyond just supporting physical objects on them.}

\Q{P2}{along-surface}{I had never really thought of continuous contact with a boundary as a distinct relationship. Previously, I only associated boundary contact with collisions. I had not considered interactions where movement occurs continuously along a surface. Yet in the real world, many things happen exactly this way: a point moving across a plane, with various events unfolding on that surface.}

\Q{P2}{trash-pass-through}{I had never considered direct pass-through as a distinct category of interaction design. It is very different from my previous example of opening a wall. A direct pass-through is like tossing an object through a surface where it simply disappears. Users do not expect to see it again or interact with it further; the opening acts as a portal into a black box. In contrast, when opening a wall, users expect to walk through to the other side and anticipate further actions. With the pass-through, the interaction simply ends. Once the object is dropped in, it is gone.}

\Q{P3}{object-surface-boundary}{An object's own surface can serve as its boundary, whether it is soft, hard, or shape-changeable.}

\Q{P6}{connect-categories}{I found it very inspiring because it made me think of many things. Even just connecting two categories here can open up explorations. For example, in two of my previous projects, one was mainly about the relationship between \dtag{Object} and \dtag{Boundary}, while another was about the relationship between \dtag{Human} states and \dtag{Object}.}

\Q{P1}{add-elements}{People can definitely add the elements that they care about into the design space.}

\Q{P4}{gaze-apple}{Before Apple, it was hard to imagine what gaze interaction could achieve. Coming from the HoloLens 2 era, I have used it for almost seven years; while it always had gaze tracking, it never felt usable for interaction. Once Apple pushed the technology above a certain bar, you suddenly realized it could be actively designed around.}

\Q{P1}{too-complete}{If it tries to cover too many aspects and be too complete, it might become unreadable or even unusable for designers, because we need to keep track of relations and attributes across elements.}

\Q{P2}{index-2}{Perhaps you do not need to define every specific experience. With these categories and demos, designers can infer what fits their context. They might instinctively default to \dtag{Opaque}, but this framework helps them find adjacent options that satisfy their essential requirements. They can explore the space, for instance, by shifting from \dtag{Opaque} to \dtag{See-through} or \dtag{Reflective}, which they might not have considered. This allows them to play with \dtag{visibility} or \dtag{Perception}.}

\Q{P1}{magic-mirror-sound}{For example, on the left side [of the mirror] is ordinary music; on the right, it becomes punk. It can work like a filter and a means of transformation.}

\Q{P6}{digital-shade}{I simply cannot work in brightly lit areas; I have to be in a very dark environment. That is why I desperately need a digital shade. Bright environments cause me a lot of frustration, and I cannot realistically install physical curtains everywhere I go. Having a virtual shade like this would be incredibly useful for me.}

\Q{P4}{ai-function}{You could know which function is useful in which situation, and AI can help determine those situations.}

\endgroup
\end{document}